\documentclass[a4paper,12pt]{article}
\usepackage{a4p}
\usepackage{pennames}
\usepackage{citesort}

\usepackage[final]{graphics}

\newcommand{\RTVaa}    {\ensuremath{1.764}}
\newcommand{\RTVba}    {\ensuremath{1.264}}
\newcommand{\RTVbb}    {\ensuremath{0.2980}}
\newcommand{\RTVbc}    {\ensuremath{0.0942}}
\newcommand{\RTVbd}    {\ensuremath{0.0403}}
\newcommand{\RTAaa}    {\ensuremath{1.720}}
\newcommand{\RTAba}    {\ensuremath{1.240}}
\newcommand{\RTAbb}    {\ensuremath{0.2510}}
\newcommand{\RTAbc}    {\ensuremath{0.1090}}
\newcommand{\RTAbd}    {\ensuremath{0.0518}}

\newcommand{\RTVBRaa}  {\ensuremath{\pm 0.016}}

\newcommand{\RTVESba}  {\ensuremath{\pm 0.004}}
\newcommand{\RTVPSba}  {\ensuremath{\pm 0.001}}
\newcommand{\RTVRGba}  {\ensuremath{\pm 0.001}}
\newcommand{\RTVDAba}  {\ensuremath{\pm 0.005}}
\newcommand{\RTVBRba}  {\ensuremath{\pm 0.010}}
\newcommand{\RTVMCba}  {\ensuremath{\pm 0.004}}
\newcommand{\RTVESbb}  {\ensuremath{\pm 0.0005}}
\newcommand{\RTVPSbb}  {\ensuremath{\pm 0.0002}}
\newcommand{\RTVRGbb}  {\ensuremath{\pm 0.0000}}
\newcommand{\RTVDAbb}  {\ensuremath{\pm 0.0012}}
\newcommand{\RTVBRbb}  {\ensuremath{\pm 0.0031}}
\newcommand{\RTVMCbb}  {\ensuremath{\pm 0.0008}}
\newcommand{\RTVESbc}  {\ensuremath{\pm 0.0007}}
\newcommand{\RTVPSbc}  {\ensuremath{\pm 0.0000}}
\newcommand{\RTVRGbc}  {\ensuremath{\pm 0.0001}}
\newcommand{\RTVDAbc}  {\ensuremath{\pm 0.0006}}
\newcommand{\RTVBRbc}  {\ensuremath{\pm 0.0016}}
\newcommand{\RTVMCbc}  {\ensuremath{\pm 0.0004}}
\newcommand{\RTVESbd}  {\ensuremath{\pm 0.0005}}
\newcommand{\RTVPSbd}  {\ensuremath{\pm 0.0001}}
\newcommand{\RTVRGbd}  {\ensuremath{\pm 0.0001}}
\newcommand{\RTVDAbd}  {\ensuremath{\pm 0.0008}}
\newcommand{\RTVBRbd}  {\ensuremath{\pm 0.0011}}
\newcommand{\RTVMCbd}  {\ensuremath{\pm 0.0005}}

\newcommand{\RTABRaa}  {\ensuremath{\pm 0.017}}

\newcommand{\RTAESba}  {\ensuremath{\pm 0.002}}
\newcommand{\RTAPSba}  {\ensuremath{\pm 0.002}}
\newcommand{\RTARGba}  {\ensuremath{\pm 0.003}}
\newcommand{\RTADAba}  {\ensuremath{\pm 0.004}}
\newcommand{\RTABRba}  {\ensuremath{\pm 0.012}}
\newcommand{\RTAMCba}  {\ensuremath{\pm 0.002}}
\newcommand{\RTAESbb}  {\ensuremath{\pm 0.0003}}
\newcommand{\RTAPSbb}  {\ensuremath{\pm 0.0002}}
\newcommand{\RTARGbb}  {\ensuremath{\pm 0.0004}}
\newcommand{\RTADAbb}  {\ensuremath{\pm 0.0010}}
\newcommand{\RTABRbb}  {\ensuremath{\pm 0.0029}}
\newcommand{\RTAMCbb}  {\ensuremath{\pm 0.0007}}
\newcommand{\RTAESbc}  {\ensuremath{\pm 0.0006}}
\newcommand{\RTAPSbc}  {\ensuremath{\pm 0.0001}}
\newcommand{\RTARGbc}  {\ensuremath{\pm 0.0005}}
\newcommand{\RTADAbc}  {\ensuremath{\pm 0.0008}}
\newcommand{\RTABRbc}  {\ensuremath{\pm 0.0015}}
\newcommand{\RTAMCbc}  {\ensuremath{\pm 0.0005}}
\newcommand{\RTAESbd}  {\ensuremath{\pm 0.0004}}
\newcommand{\RTAPSbd}  {\ensuremath{\pm 0.0002}}
\newcommand{\RTARGbd}  {\ensuremath{\pm 0.0003}}
\newcommand{\RTADAbd}  {\ensuremath{\pm 0.0007}}
\newcommand{\RTABRbd}  {\ensuremath{\pm 0.0008}}
\newcommand{\RTAMCbd}  {\ensuremath{\pm 0.0004}}

\newcommand{\RTVEXaa}  {\ensuremath{\pm 0.016}}
\newcommand{\RTVEXba}  {\ensuremath{\pm 0.012}}
\newcommand{\RTVEXbb}  {\ensuremath{\pm 0.0034}}
\newcommand{\RTVEXbc}  {\ensuremath{\pm 0.0019}}
\newcommand{\RTVEXbd}  {\ensuremath{\pm 0.0016}}
\newcommand{\RTAEXaa}  {\ensuremath{\pm 0.017}}
\newcommand{\RTAEXba}  {\ensuremath{\pm 0.013}}
\newcommand{\RTAEXbb}  {\ensuremath{\pm 0.0032}}
\newcommand{\RTAEXbc}  {\ensuremath{\pm 0.0019}}
\newcommand{\RTAEXbd}  {\ensuremath{\pm 0.0013}}
\newcommand{\CVaaVba}  {\ensuremath{ 72}}
\newcommand{\CVaaVbb}  {\ensuremath{ 87}}
\newcommand{\CVaaVbc}  {\ensuremath{ 74}}
\newcommand{\CVaaVbd}  {\ensuremath{ 53}}
\newcommand{\CVaaAaa}  {\ensuremath{  2}}
\newcommand{\CVaaAba}  {\ensuremath{  9}}
\newcommand{\CVaaAbb}  {\ensuremath{ -5}}
\newcommand{\CVaaAbc}  {\ensuremath{ -8}}
\newcommand{\CVaaAbd}  {\ensuremath{-10}}
\newcommand{\CVbaVbb}  {\ensuremath{ 72}}
\newcommand{\CVbaVbc}  {\ensuremath{ 14}}
\newcommand{\CVbaVbd}  {\ensuremath{-18}}
\newcommand{\CVbaAaa}  {\ensuremath{  9}}
\newcommand{\CVbaAba}  {\ensuremath{  4}}
\newcommand{\CVbaAbb}  {\ensuremath{ -2}}
\newcommand{\CVbaAbc}  {\ensuremath{  3}}
\newcommand{\CVbaAbd}  {\ensuremath{  8}}
\newcommand{\CVbbVbc}  {\ensuremath{ 72}}
\newcommand{\CVbbVbd}  {\ensuremath{ 37}}
\newcommand{\CVbbAaa}  {\ensuremath{  0}}
\newcommand{\CVbbAba}  {\ensuremath{  0}}
\newcommand{\CVbbAbb}  {\ensuremath{ -7}}
\newcommand{\CVbbAbc}  {\ensuremath{  0}}
\newcommand{\CVbbAbd}  {\ensuremath{  3}}
\newcommand{\CVbcVbd}  {\ensuremath{ 90}}
\newcommand{\CVbcAaa}  {\ensuremath{ -8}}
\newcommand{\CVbcAba}  {\ensuremath{  4}}
\newcommand{\CVbcAbb}  {\ensuremath{ -7}}
\newcommand{\CVbcAbc}  {\ensuremath{-12}}
\newcommand{\CVbcAbd}  {\ensuremath{-16}}
\newcommand{\CVbdAaa}  {\ensuremath{ -8}}
\newcommand{\CVbdAba}  {\ensuremath{  8}}
\newcommand{\CVbdAbb}  {\ensuremath{ -4}}
\newcommand{\CVbdAbc}  {\ensuremath{-17}}
\newcommand{\CVbdAbd}  {\ensuremath{-26}}
\newcommand{\CAaaAba}  {\ensuremath{ 85}}
\newcommand{\CAaaAbb}  {\ensuremath{ 79}}
\newcommand{\CAaaAbc}  {\ensuremath{ 64}}
\newcommand{\CAaaAbd}  {\ensuremath{ 51}}
\newcommand{\CAbaAbb}  {\ensuremath{ 56}}
\newcommand{\CAbaAbc}  {\ensuremath{ 22}}
\newcommand{\CAbaAbd}  {\ensuremath{  2}}
\newcommand{\CAbbAbc}  {\ensuremath{ 85}}
\newcommand{\CAbbAbd}  {\ensuremath{ 63}}
\newcommand{\CAbcAbd}  {\ensuremath{ 94}}
\newcommand{\NDAapapz} {\ensuremath{32316}}
\newcommand{\EFVapapz} {\ensuremath{28.7}}
\newcommand{\EFEapapz} {\ensuremath{ 0.1}}
\newcommand{\NBCapapz} {\ensuremath{ 7.7}}
\newcommand{\EBCapapz} {\ensuremath{ 0.2}}
\newcommand{\NBUapapz} {\ensuremath{ 7.9}}
\newcommand{\EBUapapz} {\ensuremath{ 0.1}}
\newcommand{\NDAapbpz} {\ensuremath{13814}}
\newcommand{\EFVapbpz} {\ensuremath{18.8}}
\newcommand{\EFEapbpz} {\ensuremath{ 0.1}}
\newcommand{\NBCapbpz} {\ensuremath{45.0}}
\newcommand{\EBCapbpz} {\ensuremath{ 0.6}}
\newcommand{\NBUapbpz} {\ensuremath{ 8.4}}
\newcommand{\EBUapbpz} {\ensuremath{ 0.1}}
\newcommand{\NDAapcpz} {\ensuremath{ 1738}}
\newcommand{\EFVapcpz} {\ensuremath{ 8.0}}
\newcommand{\EFEapcpz} {\ensuremath{ 0.2}}
\newcommand{\NBCapcpz} {\ensuremath{70.0}}
\newcommand{\EBCapcpz} {\ensuremath{ 2.2}}
\newcommand{\NBUapcpz} {\ensuremath{11.4}}
\newcommand{\EBUapcpz} {\ensuremath{ 0.5}}
\newcommand{\NDAcpopz} {\ensuremath{14321}}
\newcommand{\EFVcpopz} {\ensuremath{34.6}}
\newcommand{\EFEcpopz} {\ensuremath{ 0.1}}
\newcommand{\NBCcpopz} {\ensuremath{ 9.7}}
\newcommand{\EBCcpopz} {\ensuremath{ 0.3}}
\newcommand{\NBUcpopz} {\ensuremath{ 3.8}}
\newcommand{\EBUcpopz} {\ensuremath{ 0.1}}
\newcommand{\NDAcpapz} {\ensuremath{ 2455}}
\newcommand{\EFVcpapz} {\ensuremath{11.0}}
\newcommand{\EFEcpapz} {\ensuremath{ 0.1}}
\newcommand{\NBCcpapz} {\ensuremath{21.3}}
\newcommand{\EBCcpapz} {\ensuremath{ 1.0}}
\newcommand{\NBUcpapz} {\ensuremath{ 6.1}}
\newcommand{\EBUcpapz} {\ensuremath{ 0.3}}
\newcommand{\NDAcpbpz} {\ensuremath{ 1255}}
\newcommand{\EFVcpbpz} {\ensuremath{ 8.3}}
\newcommand{\EFEcpbpz} {\ensuremath{ 0.4}}
\newcommand{\NBCcpbpz} {\ensuremath{82.3}}
\newcommand{\EBCcpbpz} {\ensuremath{ 2.9}}
\newcommand{\NBUcpbpz} {\ensuremath{ 7.1}}
\newcommand{\EBUcpbpz} {\ensuremath{ 0.5}}
\newcommand{\NDAtotal} {\ensuremath{65899}}
\newcommand{\NBUtotal} {\ensuremath{26.6}}

\newcommand{\FOPTASMT} {\ensuremath{ 0.324}}

\newcommand{\FOPTSMMT} {\ensuremath{ 0.002}}

\newcommand{\FOPTTMMT} {\ensuremath{ 0.013}}
\newcommand{\FOPTKFMT} {\ensuremath{ 0.006}}
\newcommand{\FOPTZEMT} {\ensuremath{ 0.009}}
\newcommand{\FOPTDAMT} {\ensuremath{ 0.001}}

\newcommand{\FOPTBRMT} {\ensuremath{ 0.006}}
\newcommand{\FOPTRSMT} {\ensuremath{ 0.009}}
\newcommand{\FOPTcATVL}{\ensuremath{ 0.323}}

\newcommand{\FOPTcATSM}{\ensuremath{ 0.002}}

\newcommand{\FOPTcATTM}{\ensuremath{ 0.014}}

\newcommand{\FOPTcATDA}{\ensuremath{ 0.001}}

\newcommand{\FOPTcATBR}{\ensuremath{ 0.008}}

\newcommand{\FOPTASMZ} {\ensuremath{ 0.1191}}

\newcommand{\FOPTEMMZ} {\ensuremath{ 0.0008}}

\newcommand{\FOPTTMMZ} {\ensuremath{ 0.0013}}
\newcommand{\FOPTKFMZ} {\ensuremath{ 0.0007}}
\newcommand{\FOPTZEMZ} {\ensuremath{ 0.0009}}

\newcommand{\FOPTRSMZ} {\ensuremath{ 0.0005}}
\newcommand{\FOPTcGGVL}{\ensuremath{ 0.017}}

\newcommand{\FOPTcGGSM}{\ensuremath{ 0.003}}

\newcommand{\FOPTcGGTM}{\ensuremath{ 0.010}}

\newcommand{\FOPTcGGDA}{\ensuremath{ 0.003}}

\newcommand{\FOPTcGGBR}{\ensuremath{ 0.004}}

\newcommand{\FOPTGGVL} {\ensuremath{ 0.014}}

\newcommand{\FOPTGGSM} {\ensuremath{ 0.005}}

\newcommand{\FOPTGGTM} {\ensuremath{ 0.013}}

\newcommand{\FOPTGGDA} {\ensuremath{ 0.007}}

\newcommand{\FOPTGGBR} {\ensuremath{ 0.006}}

\newcommand{\FOPTcVSVL}{\ensuremath{ 0.0271}}

\newcommand{\FOPTcVSSM}{\ensuremath{ 0.0018}}

\newcommand{\FOPTcVSTM}{\ensuremath{ 0.0056}}

\newcommand{\FOPTcVSDA}{\ensuremath{ 0.0017}}

\newcommand{\FOPTcVSBR}{\ensuremath{ 0.0025}}

\newcommand{\FOPTSSVL} {\ensuremath{ 0.0028}}

\newcommand{\FOPTSSSM} {\ensuremath{ 0.0030}}

\newcommand{\FOPTSSTM} {\ensuremath{ 0.0068}}

\newcommand{\FOPTSSDA} {\ensuremath{ 0.0034}}

\newcommand{\FOPTSSBR} {\ensuremath{ 0.0034}}

\newcommand{\FOPTcVEVL}{\ensuremath{-0.0085}}

\newcommand{\FOPTcVESM}{\ensuremath{ 0.0005}}

\newcommand{\FOPTcVETM}{\ensuremath{ 0.0012}}

\newcommand{\FOPTcVEDA}{\ensuremath{ 0.0010}}

\newcommand{\FOPTcVEBR}{\ensuremath{ 0.0007}}

\newcommand{\FOPTSEVL} {\ensuremath{-0.0015}}

\newcommand{\FOPTSESM} {\ensuremath{ 0.0014}}

\newcommand{\FOPTSETM} {\ensuremath{ 0.0019}}

\newcommand{\FOPTSEDA} {\ensuremath{ 0.0024}}

\newcommand{\FOPTSEBR} {\ensuremath{ 0.0016}}

\newcommand{\FOPTcASVL}{\ensuremath{-0.0183}}

\newcommand{\FOPTcASSM}{\ensuremath{ 0.0019}}

\newcommand{\FOPTcASTM}{\ensuremath{ 0.0052}}

\newcommand{\FOPTcASDA}{\ensuremath{ 0.0016}}

\newcommand{\FOPTcASBR}{\ensuremath{ 0.0023}}

\newcommand{\FOPTcAEVL}{\ensuremath{ 0.0036}}

\newcommand{\FOPTcAESM}{\ensuremath{ 0.0008}}

\newcommand{\FOPTcAETM}{\ensuremath{ 0.0011}}

\newcommand{\FOPTcAEDA}{\ensuremath{ 0.0011}}

\newcommand{\FOPTcAEBR}{\ensuremath{ 0.0012}}

\newcommand{\FOPTCSMT} {\ensuremath{0.17}}
\newcommand{\FOPTcCHSQ}{\ensuremath{0.62}}
\newcommand{\CORRFOATGG}{\ensuremath{-57}}
\newcommand{\CORRFOATVS}{\ensuremath{-55}}
\newcommand{\CORRFOATAS}{\ensuremath{-61}}
\newcommand{\CORRFOATVE}{\ensuremath{ 41}}
\newcommand{\CORRFOATAE}{\ensuremath{ 42}}
\newcommand{\CORRFOGGVS}{\ensuremath{ 99}}
\newcommand{\CORRFOGGAS}{\ensuremath{ 96}}
\newcommand{\CORRFOGGVE}{\ensuremath{-92}}
\newcommand{\CORRFOGGAE}{\ensuremath{-87}}
\newcommand{\CORRFOVSAS}{\ensuremath{ 96}}
\newcommand{\CORRFOVSVE}{\ensuremath{-90}}
\newcommand{\CORRFOVSAE}{\ensuremath{-86}}
\newcommand{\CORRFOASVE}{\ensuremath{-84}}
\newcommand{\CORRFOASAE}{\ensuremath{-77}}
\newcommand{\CORRFOVEAE}{\ensuremath{ 89}}
\newcommand{\CIPTASMT} {\ensuremath{ 0.348}}

\newcommand{\CIPTSMMT} {\ensuremath{ 0.002}}

\newcommand{\CIPTTMMT} {\ensuremath{ 0.019}}
\newcommand{\CIPTKFMT} {\ensuremath{ 0.012}}
\newcommand{\CIPTZEMT} {\ensuremath{ 0.006}}
\newcommand{\CIPTDAMT} {\ensuremath{ 0.002}}

\newcommand{\CIPTBRMT} {\ensuremath{ 0.009}}
\newcommand{\CIPTRSMT} {\ensuremath{ 0.015}}
\newcommand{\CIPTcATVL}{\ensuremath{ 0.347}}

\newcommand{\CIPTcATSM}{\ensuremath{ 0.002}}

\newcommand{\CIPTcATTM}{\ensuremath{ 0.019}}

\newcommand{\CIPTcATDA}{\ensuremath{ 0.001}}

\newcommand{\CIPTcATBR}{\ensuremath{ 0.012}}

\newcommand{\CIPTKFMZ} {\ensuremath{ 0.0013}}
\newcommand{\CIPTZEMZ} {\ensuremath{ 0.0005}}

\newcommand{\CIPTRSMZ} {\ensuremath{ 0.0009}}
\newcommand{\CIPTcGGVL}{\ensuremath{ 0.001}}

\newcommand{\CIPTcGGSM}{\ensuremath{ 0.003}}

\newcommand{\CIPTcGGTM}{\ensuremath{ 0.004}}

\newcommand{\CIPTcGGDA}{\ensuremath{ 0.003}}

\newcommand{\CIPTcGGBR}{\ensuremath{ 0.006}}

\newcommand{\CIPTGGVL} {\ensuremath{-0.003}}

\newcommand{\CIPTGGSM} {\ensuremath{ 0.006}}

\newcommand{\CIPTGGTM} {\ensuremath{ 0.005}}

\newcommand{\CIPTGGDA} {\ensuremath{ 0.007}}

\newcommand{\CIPTGGBR} {\ensuremath{ 0.007}}

\newcommand{\CIPTcVSVL}{\ensuremath{ 0.0256}}

\newcommand{\CIPTcVSSM}{\ensuremath{ 0.0017}}

\newcommand{\CIPTcVSTM}{\ensuremath{ 0.0006}}

\newcommand{\CIPTcVSDA}{\ensuremath{ 0.0017}}

\newcommand{\CIPTcVSBR}{\ensuremath{ 0.0024}}

\newcommand{\CIPTSSVL} {\ensuremath{ 0.0012}}

\newcommand{\CIPTSSSM} {\ensuremath{ 0.0029}}

\newcommand{\CIPTSSTM} {\ensuremath{ 0.0006}}

\newcommand{\CIPTSSDA} {\ensuremath{ 0.0034}}

\newcommand{\CIPTSSBR} {\ensuremath{ 0.0033}}

\newcommand{\CIPTcVEVL}{\ensuremath{-0.0080}}

\newcommand{\CIPTcVESM}{\ensuremath{ 0.0005}}

\newcommand{\CIPTcVETM}{\ensuremath{ 0.0002}}

\newcommand{\CIPTcVEDA}{\ensuremath{ 0.0010}}

\newcommand{\CIPTcVEBR}{\ensuremath{ 0.0007}}

\newcommand{\CIPTSEVL} {\ensuremath{-0.0010}}

\newcommand{\CIPTSESM} {\ensuremath{ 0.0015}}

\newcommand{\CIPTSETM} {\ensuremath{ 0.0003}}

\newcommand{\CIPTSEDA} {\ensuremath{ 0.0024}}

\newcommand{\CIPTSEBR} {\ensuremath{ 0.0016}}

\newcommand{\CIPTcASVL}{\ensuremath{-0.0197}}

\newcommand{\CIPTcASSM}{\ensuremath{ 0.0019}}

\newcommand{\CIPTcASTM}{\ensuremath{ 0.0010}}

\newcommand{\CIPTcASDA}{\ensuremath{ 0.0016}}

\newcommand{\CIPTcASBR}{\ensuremath{ 0.0022}}

\newcommand{\CIPTcAEVL}{\ensuremath{ 0.0041}}

\newcommand{\CIPTcAESM}{\ensuremath{ 0.0008}}

\newcommand{\CIPTcAETM}{\ensuremath{ 0.0002}}

\newcommand{\CIPTcAEDA}{\ensuremath{ 0.0012}}

\newcommand{\CIPTcAEBR}{\ensuremath{ 0.0013}}

\newcommand{\CIPTCSMT} {\ensuremath{0.16}}
\newcommand{\CIPTcCHSQ}{\ensuremath{0.63}}

\newcommand{\RCPTASMT} {\ensuremath{ 0.306}}

\newcommand{\RCPTSMMT} {\ensuremath{ 0.001}}

\newcommand{\RCPTTMMT} {\ensuremath{ 0.011}}

\newcommand{\RCPTZEMT} {\ensuremath{ 0.011}}
\newcommand{\RCPTDAMT} {\ensuremath{ 0.001}}

\newcommand{\RCPTBRMT} {\ensuremath{ 0.005}}
\newcommand{\RCPTRSMT} {\ensuremath{ 0.000}}
\newcommand{\RCPTcATVL}{\ensuremath{ 0.305}}

\newcommand{\RCPTcATSM}{\ensuremath{ 0.001}}

\newcommand{\RCPTcATTM}{\ensuremath{ 0.011}}

\newcommand{\RCPTcATDA}{\ensuremath{ 0.001}}

\newcommand{\RCPTcATBR}{\ensuremath{ 0.007}}

\newcommand{\RCPTASMZ} {\ensuremath{ 0.1169}}

\newcommand{\RCPTEMMZ} {\ensuremath{ 0.0007}}

\newcommand{\RCPTTMMZ} {\ensuremath{ 0.0015}}

\newcommand{\RCPTZEMZ} {\ensuremath{ 0.0015}}

\newcommand{\RCPTRSMZ} {\ensuremath{ 0.0005}}
\newcommand{\RCPTcGGVL}{\ensuremath{ 0.002}}

\newcommand{\RCPTcGGSM}{\ensuremath{ 0.003}}

\newcommand{\RCPTcGGTM}{\ensuremath{ 0.001}}

\newcommand{\RCPTcGGDA}{\ensuremath{ 0.003}}

\newcommand{\RCPTcGGBR}{\ensuremath{ 0.005}}

\newcommand{\RCPTGGVL} {\ensuremath{-0.002}}

\newcommand{\RCPTGGSM} {\ensuremath{ 0.005}}

\newcommand{\RCPTGGTM} {\ensuremath{ 0.002}}

\newcommand{\RCPTGGDA} {\ensuremath{ 0.007}}

\newcommand{\RCPTGGBR} {\ensuremath{ 0.007}}

\newcommand{\RCPTcVSVL}{\ensuremath{ 0.0202}}

\newcommand{\RCPTcVSSM}{\ensuremath{ 0.0018}}

\newcommand{\RCPTcVSTM}{\ensuremath{ 0.0009}}

\newcommand{\RCPTcVSDA}{\ensuremath{ 0.0018}}

\newcommand{\RCPTcVSBR}{\ensuremath{ 0.0033}}

\newcommand{\RCPTSSVL} {\ensuremath{-0.0047}}

\newcommand{\RCPTSSSM} {\ensuremath{ 0.0032}}

\newcommand{\RCPTSSTM} {\ensuremath{ 0.0011}}

\newcommand{\RCPTSSDA} {\ensuremath{ 0.0036}}

\newcommand{\RCPTSSBR} {\ensuremath{ 0.0040}}

\newcommand{\RCPTcVEVL}{\ensuremath{-0.0075}}

\newcommand{\RCPTcVESM}{\ensuremath{ 0.0005}}

\newcommand{\RCPTcVETM}{\ensuremath{ 0.0002}}

\newcommand{\RCPTcVEDA}{\ensuremath{ 0.0010}}

\newcommand{\RCPTcVEBR}{\ensuremath{ 0.0008}}

\newcommand{\RCPTSEVL} {\ensuremath{-0.0001}}

\newcommand{\RCPTSESM} {\ensuremath{ 0.0015}}

\newcommand{\RCPTSETM} {\ensuremath{ 0.0003}}

\newcommand{\RCPTSEDA} {\ensuremath{ 0.0024}}

\newcommand{\RCPTSEBR} {\ensuremath{ 0.0017}}

\newcommand{\RCPTcASVL}{\ensuremath{-0.0252}}

\newcommand{\RCPTcASSM}{\ensuremath{ 0.0020}}

\newcommand{\RCPTcASTM}{\ensuremath{ 0.0006}}

\newcommand{\RCPTcASDA}{\ensuremath{ 0.0017}}

\newcommand{\RCPTcASBR}{\ensuremath{ 0.0032}}

\newcommand{\RCPTcAEVL}{\ensuremath{ 0.0047}}

\newcommand{\RCPTcAESM}{\ensuremath{ 0.0008}}

\newcommand{\RCPTcAETM}{\ensuremath{ 0.0001}}

\newcommand{\RCPTcAEDA}{\ensuremath{ 0.0012}}

\newcommand{\RCPTcAEBR}{\ensuremath{ 0.0013}}

\newcommand{\RCPTCSMT} {\ensuremath{0.07}}
\newcommand{\RCPTcCHSQ}{\ensuremath{0.61}}

\newcommand{\MEANASVLMT}{\ensuremath{ 0.348}}

\newcommand{\MEANASXMMT}{\ensuremath{ 0.009}}

\newcommand{\MEANASMTMT}{\ensuremath{ 0.019}}

\newcommand{\MEANASVLMZ}{\ensuremath{ 0.1219}}

\newcommand{\MEANASTMMZ}{\ensuremath{ 0.0017}}

\newcommand{\MEANASXMMZ}{\ensuremath{ 0.0010}}

\newcommand{\MEANASMTMZ}{\ensuremath{ 0.0017}}

\mathchardef\im="10
\mathchardef\md="2D
\newcommand{\Sy}        {\scriptstyle}

\newcommand{\p}         {\phantom}
\newcommand{\Order}     {\ensuremath{O}}
\newcommand{\Op}        {\ensuremath{{\cal O}}}
\newcommand{\Ce}        {\ensuremath{{\cal C}}}
\newcommand{\alphas}    {\ensuremath{\alpha_{\rm s}}}
\newcommand{\balphas}   {\boldmath\ensuremath{\alpha_{\rm s}}}
\newcommand{\btau}      {\boldmath\ensuremath{\tau}}
\newcommand{\Rtaus}     {\ensuremath{R_\tau(s_0)}}

\newcommand{\rd}        {\ensuremath{\mathrm{d}}}
\newcommand{\Ph}        {\ensuremath{\mathrm{h}}}

\newcommand{\Pgtm}      {\ensuremath{\mathrm{\tau^-}}}
\newcommand{\Pgtp}      {\ensuremath{\mathrm{\tau^+}}}
\newcommand{\mz}        {\ensuremath{m_\mathrm{Z}}}
\newcommand{\mzsq}      {\ensuremath{m_\mathrm{Z}^2}}
\renewcommand{\Pgo}     {\ensuremath{\mathrm{\omega}}}
\renewcommand{\Pgr}     {\ensuremath{\mathrm{\rho}}}
\renewcommand{\Pai}     {\ensuremath{\mathrm{a_1}}}

\begin{document}
\begin{titlepage}
\begin{center}
{\large EUROPEAN LABORATORY FOR PARTICLE PHYSICS}
\end{center}
\begin{flushright}
CERN-EP/98-102\\
June 22, 1998
\end{flushright}
\vspace{5cm}

\begin{center}
  \LARGE\bfseries Measurement of the Strong Coupling
  Constant~\balphas\ and the Vector and Axial-Vector Spectral
  Functions in Hadronic Tau Decays
\end{center}
\bigskip

\begin{center}
  \Large\bfseries The OPAL Collaboration
\end{center}
\bigskip

\bigskip

\bigskip

\begin{abstract}
\noindent
The spectral functions of the vector current and the axial-vector
current have been measured in hadronic \Pgt\ decays using the OPAL
detector at LEP.  Within the framework of the Operator Product
Expansion a simultaneous determination of the strong coupling constant
\alphas, the non-perturbative operators of dimension 6 and 8 and of
the gluon condensate has been performed.  Different perturbative
descriptions have been compared to the data.  The Contour Improved
Fixed Order Perturbation Theory gives $\alphas(m_\Pgt^2) = \MEANASVLMT
\pm \MEANASXMMT_{\rm exp} \pm \MEANASMTMT_{\rm theo}$ at the \Pgt-mass
scale and $\alphas(\mzsq) = \MEANASVLMZ \pm \MEANASXMMZ_{\rm exp} \pm
\MEANASMTMZ_{\rm theo}$ at the \PZz-mass scale.  The values obtained
for $\alphas(\mzsq)$ using Fixed Order Perturbation Theory or
Renormalon Chain Resummation are $2.3\,\%$ and $4.1\,\%$ smaller,
respectively.  The \lq running\rq\ of the strong coupling between $s_0
\simeq 1.3\,{\rm GeV}^2$ and $s_0 = m_\Pgt^2$ has been tested from
direct fits to the integrated differential hadronic decay rate \Rtaus.
A test of the saturation of QCD sum rules at the \Pgt-mass scale has
been performed.
\end{abstract}

\vfill
\begin{center}
 {\em\large (Submitted to European Physical Journal C)}
\end{center}
\bigskip
\end{titlepage}


\begin{center}{\Large        The OPAL Collaboration
}\end{center}\bigskip
\begin{center}{
K.\thinspace Ackerstaff$^{  8}$,
G.\thinspace Alexander$^{ 23}$,
J.\thinspace Allison$^{ 16}$,
N.\thinspace Altekamp$^{  5}$,
K.J.\thinspace Anderson$^{  9}$,
S.\thinspace Anderson$^{ 12}$,
S.\thinspace Arcelli$^{  2}$,
S.\thinspace Asai$^{ 24}$,
S.F.\thinspace Ashby$^{  1}$,
D.\thinspace Axen$^{ 29}$
% ...
}\end{center}\bigskip
\bigskip
$^{  1}$School of Physics and Astronomy, University of Birmingham,
Birmingham B15 2TT, UK
\newline
$^{  2}$Dipartimento di Fisica dell' Universit{\`a} di Bologna and INFN,
I-40126 Bologna, Italy
\newline
$^{  3}$Physikalisches Institut, Universit{\"a}t Bonn,
D-53115 Bonn, Germany
\newline
$^{  4}$Department of Physics, University of California,
Riverside CA 92521, USA
\newline
$^{  5}$Cavendish Laboratory, Cambridge CB3 0HE, UK
\newline
$^{  6}$Ottawa-Carleton Institute for Physics,
Department of Physics, Carleton University,
Ottawa, Ontario K1S 5B6, Canada
\newline
$^{  7}$Centre for Research in Particle Physics,
Carleton University, Ottawa, Ontario K1S 5B6, Canada
\newline
$^{  8}$CERN, European Organisation for Particle Physics,
CH-1211 Geneva 23, Switzerland
\newline
$^{  9}$Enrico Fermi Institute and Department of Physics,
University of Chicago, Chicago IL 60637, USA
\newline
$^{ 10}$Fakult{\"a}t f{\"u}r Physik, Albert Ludwigs Universit{\"a}t,
D-79104 Freiburg, Germany
\newline
$^{ 11}$Physikalisches Institut, Universit{\"a}t
Heidelberg, D-69120 Heidelberg, Germany
\newline
$^{ 12}$Indiana University, Department of Physics,
Swain Hall West 117, Bloomington IN 47405, USA
\newline
$^{ 13}$Queen Mary and Westfield College, University of London,
London E1 4NS, UK
\newline
$^{ 14}$Technische Hochschule Aachen, III Physikalisches Institut,
Sommerfeldstrasse 26-28, D-52056 Aachen, Germany
\newline
$^{ 15}$University College London, London WC1E 6BT, UK
\newline
$^{ 16}$Department of Physics, Schuster Laboratory, The University,
Manchester M13 9PL, UK
\newline
$^{ 17}$Department of Physics, University of Maryland,
College Park, MD 20742, USA
\newline
$^{ 18}$Laboratoire de Physique Nucl{\'e}aire, Universit{\'e} de Montr{\'e}al,
Montr{\'e}al, Quebec H3C 3J7, Canada
\newline
$^{ 19}$University of Oregon, Department of Physics, Eugene
OR 97403, USA
\newline
$^{ 20}$Rutherford Appleton Laboratory, Chilton,
Didcot, Oxfordshire OX11 0QX, UK
\newline
$^{ 22}$Department of Physics, Technion-Israel Institute of
Technology, Haifa 32000, Israel
\newline
$^{ 23}$Department of Physics and Astronomy, Tel Aviv University,
Tel Aviv 69978, Israel
\newline
$^{ 24}$International Centre for Elementary Particle Physics and
Department of Physics, University of Tokyo, Tokyo 113, and
Kobe University, Kobe 657, Japan
\newline
$^{ 25}$Institute of Physical and Environmental Sciences,
Brunel University, Uxbridge, Middlesex UB8 3PH, UK
\newline
$^{ 26}$Particle Physics Department, Weizmann Institute of Science,
Rehovot 76100, Israel
\newline
$^{ 27}$Universit{\"a}t Hamburg/DESY, II Institut f{\"u}r Experimental
Physik, Notkestrasse 85, D-22607 Hamburg, Germany
\newline
$^{ 28}$University of Victoria, Department of Physics, P O Box 3055,
Victoria BC V8W 3P6, Canada
\newline
$^{ 29}$University of British Columbia, Department of Physics,
Vancouver BC V6T 1Z1, Canada
\newline
$^{ 30}$University of Alberta,  Department of Physics,
Edmonton AB T6G 2J1, Canada
\newline
$^{ 31}$Duke University, Dept of Physics,
Durham, NC 27708-0305, USA
\newline
$^{ 32}$Research Institute for Particle and Nuclear Physics,
H-1525 Budapest, P O  Box 49, Hungary
\newline
$^{ 33}$Institute of Nuclear Research,
H-4001 Debrecen, P O  Box 51, Hungary
\newline
$^{ 34}$Ludwigs-Maximilians-Universit{\"a}t M{\"u}nchen,
Sektion Physik, Am Coulombwall 1, D-85748 Garching, Germany
\newline
\bigskip\newline
$^{  a}$ and at TRIUMF, Vancouver, Canada V6T 2A3
\newline
$^{  b}$ and Royal Society University Research Fellow
\newline
$^{  c}$ and Institute of Nuclear Research, Debrecen, Hungary
\newline
$^{  d}$ and Department of Experimental Physics, Lajos Kossuth
University, Debrecen, Hungary
\newline
$^{  e}$ on leave of absence from the University of Freiburg
\newline
\vfill\eject

\section{Introduction}\label{sec:introduction}
The \Pgt\ lepton is the only lepton heavy enough to decay into
hadrons. A comparison of the inclusive hadronic decay rate of the
\Pgt\ with QCD predictions can give fundamental parameters of the
theory.  The energy regime governed by $m_\Pgt = 1.777\,{\rm GeV}$ is
regarded as a compromise region between the low and high energy
regimes where non-perturbative and perturbative QCD dominate,
respectively. In fact, \Pgt\ decay is probably the lowest-energy
process from which the coupling constant $\alphas$ can be
cleanly extracted
\cite{art:Braaten88,art:Braaten89,art:Narison88,art:Braaten,art:Pich92}
without large complications from non-perturbative effects, while the
perturbative expansion still converges well.

In this analysis the most important quantity to measure is the strong
coupling constant $\alphas(m_\Pgt^2)$. The \lq running\rq\ of
$\alphas$, for energy scales smaller than $m_\Pgt$, can be tested with
the integrated differential decay rate into hadrons $\rd R_\Pgt/\rd
s$, where $\sqrt{s}$ denotes the mass of the final-state hadronic
system and $R_\Pgt = \Gamma(\Pgt \rightarrow \Ph \Pgngt)/\Gamma(\Pgt
\rightarrow \Pe \Pgne \Pgngt)$ is the hadronic decay width of the tau
normalized to the decay width of the tau going into electron and
neutrinos.  This is possible as the hadronic decay rate $R_\Pgt(s_0)$
depends on the strong coupling constant $\alphas(s_0)$ at the scale
$s_0$ only, where $s_0$ denotes the upper integration limit for the
integral over $\rd R_\Pgt/\rd s$.

The measured $\alphas(m_\Pgt^2)$ can be transformed into a value for
$\alphas(\mzsq)$ through the renormalization group equation
($\beta$-function). In doing that, the relative error of $\alphas(s)$
decreases like the decrease of $\alphas(s)$ itself.  After the
evolution to the \PZz\ mass the strong coupling is reduced to
$\alphas(\mzsq) \simeq (1/3) \alphas(m_\Pgt^2)$ and its error is
reduced to $\Delta \alphas(\mzsq) \simeq (1/9) \Delta
\alphas(m_\Pgt^2)$.  Hence, the significance of this 
measurement compares favorably with other $\alphas(\mzsq)$
determination methods~\cite{art:Burrows}.
 
Inclusive observables like the hadronic decay rate \Rtaus\ have been
calculated in perturbative QCD to $\Order(\alphas^3)$.  Some remaining
theoretical uncertainties due to corrections in powers of $1/m_\Pgt^2$
can be avoided if the differential decay rate $\rd R_\Pgt/\rd s$ is
measured and compared to the theory by means of its spectral moments
which are weighted integrals over $\rd R_\Pgt/\rd s$. As a result, the
power corrections and $\alphas$ can be simultaneously
determined from a fit.  While $R_\Pgt(m_\Pgt^2)$ can be precisely
determined from the leptonic branching ratios and the \Pgt\ lifetime,
$\rd R_\Pgt/\rd s$ involves a measurement of the invariant mass of the
hadronic system. Thus, an exclusive reconstruction of all hadronic
final states in \Pgt\ decays is necessary.

In this paper an analysis is presented using data taken with the OPAL
detector at LEP at energies within $\pm 3\,{\rm GeV}$ of the \PZz\ 
peak.  The analysis includes measurements of the differential decay
rates $\rd R_{\Pgt,{\rm V/A}}/\rd s$ for vector (V) and axial-vector
(A) decays and their respective spectral moments.  Using these
moments, fits of QCD predictions are made extracting the strong
coupling constant $\alphas(m_\Pgt^2)$ and parameters of the
non-perturbative expansion, most notably the contributions of
dimension 6 and 8 operators.  The measurement is based on a set of
spectral moments defined by the same weighting functions used by
ALEPH~\cite{art:Aleph93,art:Andreas98} and CLEO~\cite{art:CLEO}.

The differential decay rates themselves can be re-expressed in terms
of spectral functions of the vector and axial-vector current, $v(s)$
and $a(s)$.  This measurement serves for saturation tests of QCD sum
rules at the \Pgt-mass scale by comparing the experimental values of
the sum rules with chiral QCD predictions.  Furthermore, by evaluating
the moment integrals between zero and $s_0$, where $\sqrt{s_0}$ is an
energy smaller than $m_\Pgt$, the \lq running\rq\ of $\alphas$ is
tested in a single experiment.

The theoretical framework for inclusive observables from hadronic
\Pgt\ decays is described in section~\ref{sec:QCD}.  After a short
description of the OPAL detector in section~\ref{sec:OPAL} the
selection of hadronic \Pgt\ decays is described in
section~\ref{sec:selection}. In section~\ref{sec:unfolding} the
unfolding procedure is described. The measured and unfolded spectra
are discussed in section~\ref{sec:discussion} followed by a
description of the systematic uncertainties in section~\ref{sec:syst}.
Section~\ref{sec:results} contains the results for the moments of
$R_\Pgt$ and for the spectral functions.  The extraction of the strong
coupling constant and of the power corrections, from fits to the
moments of \Rtaus, is discussed in sections~\ref{sec:alphas}
and~\ref{sec:power}.  Section~\ref{sec:running} describes the test of
the \lq running\rq\ of \alphas.  The application of QCD sum rules to
the spectral functions is discussed in
section~\ref{sec:QCD_sum_rules}.  Finally, the results are summarized
in section~\ref{sec:conclusions}.


\section{Theoretical description of hadronic \btau\ decays}\label{sec:QCD}
QCD predictions of inclusive observables in hadronic \Pgt\ decays have
been calculated including perturbative and non-perturbative
contributions.  These observables can be related to the differential,
non-strange hadronic decay width, normalized to the decay width of
$\Pgtm \rightarrow \Pem \Pagne
\Pgngt$~\cite{art:Braaten88,art:Braaten89,art:Narison88,art:Braaten,art:Pich92}:
\begin{equation}
\label{eq:dRds}
\frac{\rd R_{\Pgt,{\rm V/A}}}{\rd s} = 12 \pi S_{\rm EW} |V_{\rm ud}|^2 
\frac{1}{m_\Pgt^2} 
\left(1-\frac{s}{m_\Pgt^2}\right)^2\left[\left(1+2\frac{s}{m_\Pgt^2}\right) 
{\rm Im} {\mit\Pi}_{\rm V/A}^{(1)}(s) + 
{\rm Im} {\mit\Pi}_{\rm V/A}^{(0)}(s)\right], 
\end{equation}
where $s$ denotes the square of the invariant mass of the hadronic
system and the labels $\rm V$ and $\rm A$ stand for the vector and
axial-vector contributions, respectively\footnote{The notation $\rm
  V/A$ will be used throughout the paper to indicate vector and
  axial-vector contributions, respectively.}.  $S_{\rm EW} = 1.0194$
is an electroweak correction term~\cite{art:Marciano} and $|V_{\rm
  ud}|^2 = 0.9512 \pm 0.0008$ is the squared CKM weak mixing matrix
element~\cite{art:PDG96}.  The functions ${\rm Im} {\mit\Pi}$ are
proportional to the spectral functions for the non-strange currents
with angular momenta $J=1$ and $J=0$ as indicated by the superscripts.
The latter spectral function vanishes for the vector current, since no
scalar particle has been observed in $\Pgt$ decays, while ${\rm
  Im}{\mit\Pi}_{\rm A}^{0}$ is given by the pion pole, assuming that
the pion is the only pseudo-scalar final-state in non-strange \Pgt\ 
decays:
\begin{equation}
  \label{eq:pionpole}
  {\rm Im}{\mit\Pi}_{\rm A}^{0}(s) = 
        \frac{m_\Pgt^2}{12\pi S_{\rm EW} |V_{\rm ud}|^2}
        \left(1 - \frac{s}{m_\Pgt^2}\right)^{-2}
        \frac{B(\Pgt \rightarrow \Pgp \Pgngt)}{
              B(\Pgt \rightarrow \Pe \Pgne \Pgngt)}
        \frac{1}{N_\Pgp}
        \frac{\rd N_\Pgp}{\rd s},
\end{equation}
with $N_\Pgp$ being the number of selected \Pgt\ decays into pions.
The spectral functions for the vector and the axial-vector currents
are defined in equation~(\ref{eq:spectral}).

Within the framework of QCD weighted integrals or moments of
(\ref{eq:dRds}) have been calculated~\cite{art:Diberder}:
\begin{equation}
\label{eq:rkl} 
R_{\Pgt,{\rm V/A}}^{kl}(s_0) = \int\limits_0^{s_0} 
        {
        \rd s
        \left( 1 - \frac{s}{s_0}\right)^k
        \left(\frac{s}{m_\Pgt^2}\right)^l
        \frac{\rd R_{\Pgt,{\rm V/A}}}{\rd s}
        }.      
\end{equation}
The moments are used to compare the experiment with theory. In what
follows, ten moments for $kl = 00,10,11,12,13$ for $\rm V$ and $\rm A$
are used.  The first moments~$R_{\Pgt,{\rm V/A}}^{00}(m_\Pgt^2)$\ are
the total normalized decay rates of the \Pgt\ into vector and
axial-vector mesons given by (\ref{eq:dRds}) integrated over $s$. In
the na{\"\i}ve parton model these two rates are identical and add up to the
number of colors. Since only non-strange currents are considered in
this work the na{\"\i}ve expectation has to be multiplied by $|V_{\rm
  ud}|^2$.  Including the perturbative and non-perturbative
contributions, equation (\ref{eq:rkl}) is usually written as
\cite{art:Diberder}:
\begin{equation}
\label{eq:rklQCD}      
R_{\Pgt,{\rm V/A}}^{kl}(s_0) = 
        \frac{3}{2} S_{\rm EW} |V_{\rm ud}|^2 
        \left(1 + 
              \delta_{\rm EW}^{\prime kl}(s_0) +
              \delta_{\rm pert}^{kl}(s_0) +   
              \sum_{D = 2,4,6,\dots}
                \delta_{\rm V/A}^{D,kl}(s_0)
        \right),        
\end{equation}
where $S_{\rm EW}$ is the same multiplicative correction as in
equation~(\ref{eq:dRds}) and $\delta_{\rm EW}^{\prime kl}$ are
additive electroweak corrections. The latter has been calculated for
$kl = 00$ only~\cite{art:Braaten90} yielding $\delta_{\rm EW}^{\prime
  00}(m_\Pgt^2) = \frac{5}{12} \frac{\alpha(m_\Pgt^2)}{\pi} = 0.0010$.
In the higher moments it is assumed that this term scales with the
integral over the weight functions in equations~(\ref{eq:dRds})
and~(\ref{eq:rkl}) like the $\Order(\alphas)$ correction:
\begin{equation}
\label{eq:delta_EW}
        \delta_{\rm EW}^{\prime kl}(s_0) = 
        \frac{R_{\Pgt,{\rm na\ddot{\im}ve}}^{kl}(s_0)}
        {R_{\Pgt,{\rm na\ddot{\im}ve}}^{00}(m_\Pgt^2)}
        \delta_{\rm EW}^{\prime 00}(m_\Pgt^2).
\end{equation}
Therefore the $\delta_{\rm EW}^{\prime kl}$ contribution to the
moments is small ($\sim 0.1\,\%$) and the uncertainty due to this term
is neglected in the analysis.  The other factors in
equation~(\ref{eq:rklQCD}) are explained in more detail below.


\subsection{Perturbative correction terms 
  {\boldmath$\delta_{\rm pert}^{kl}$}}\label{subsec:deltapert} The
perturbative term $\delta_{\rm pert}^{kl}$ is known to third order in
\alphas~\cite{art:Braaten} and partly known to fourth order in
\alphas~\cite{art:Diberder}. For $kl=00$ and $s_0 = m_\Pgt^2$ it is:
\begin{equation}
\label{eq:FOPT}
\delta_{\rm pert}^{00}(m_\Pgt^2) =    \frac{\alphas(m_\Pgt^2)}{\pi}
                       + 5.2023         \frac{\alphas^2(m_\Pgt^2)}{\pi^2} 
                       + 26.366         \frac{\alphas^3(m_\Pgt^2)}{\pi^3} 
                       + (78.003 + K_4) \frac{\alphas^4(m_\Pgt^2)}{\pi^4} 
                       + \Order\left( \alphas^5(m_\Pgt^2)\right).
\end{equation}
This result which truncates after the fourth power of \alphas\ is
refered to as Fixed Order Perturbation Theory (FOPT).  Different
attempts have been made to obtain a resummation of some of the higher
order terms.  The resummation scheme proposed in~\cite{art:Diberder}
compensates for higher order logarithmic terms in \alphas\ by
expressing the $\delta_{\rm pert}^{kl}(s_0)$ terms by
contour-integrals in the complex $s$-plane along the circle $|s| =
s_0$ and solving numerically for each $\alphas(s)$ along the circle
(Contour-Improved Perturbation Theory, CIPT).  The different
$\alphas(s)$ values on the circle can be calculated from
$\alphas(m_\Pgt^2)$ by solving numerically the $\beta$-function:
\begin{equation}
  \label{eq:beta-fct}
  \frac{\rd a}{\rd{\rm ln} s} = 
  \beta(a) = 
  -\beta_1 a^2
  -\beta_2 a^3
  -\beta_3 a^4
  -\beta_4 a^5
  +\Order(a^6),
\end{equation}
with $a=\alphas(s)/\pi$, $\beta_1 = 9/4$, $\beta_2 = 4$,
$\beta_3^{\overline{\rm MS}} = 10.0599$ and $\beta_4^{\overline{\rm
    MS}} = 47.2306$~\cite{art:Ritbergen} for $3$ quark flavors. The
last two coefficients are renormalization scheme dependent and the
quoted values belong to the $\overline{\rm MS}$-scheme.  The third
method considered in this paper resums the leading term of the
$\beta$-function to all orders in \alphas\ by inserting so-called
Renormalon Chains
(RCPT)~\cite{art:Neubert,art:Maxwell95,art:Maxwell96}\footnote{ The
  fixed-order corrected version (up to the third order in \alphas)
  quoted in the lower portion of table 6 in ref.~\cite{art:Neubert} is
  used in the fit.}.

One of the leading theoretical uncertainties for FOPT and CIPT comes
from the unknown $\Order\left( \alphas^4 \right)$ correction~$K_4$.
Expanding the perturbative corrections in terms of CIPT gives:
\begin{equation}
\label{eq:KnAn}
1+\delta_{\rm pert}^{kl}(s_0) = \sum_{n\ge0} K_n A_{n}^{kl}(s_0),
\end{equation}
where the functions $A_n^{kl}$ are the weighted contour integrals.
For $kl=00$ the function is:
\begin{equation}
A_n^{00}(s_0) = \frac{1}{2\pi i} \oint_{|s|=s_0} \frac{\rd s}{s}
               \left(\frac{\alphas(-s)}{\pi}\right)^n\left(
       2\frac{s_0}{m_\Pgt^2} - 2\frac{s_0^3}{m_\Pgt^6}
                             +  \frac{s_0^4}{m_\Pgt^8} 
  -    2\frac{s  }{m_\Pgt^2} + 2\frac{s  ^3}{m_\Pgt^6}  
                             -  \frac{s  ^4}{m_\Pgt^8}
       \right). \label{eq:An}
\end{equation}
In the $\overline{\rm MS}$-scheme and for three flavors the first four
terms are: $K_0 = K_1 = 1$, $K_2 = 1.63982$, $K_3 =
6.37101$~\cite{art:Chetyrkin,art:Dine,art:Celmaster,art:Gorishnii,%
art:Surguladze}. A bold guess for $K_4$ gives $K_4 \approx K_3(K_3/K_2) 
\approx 25$~\cite{art:Diberder}.  Similar estimates are given in
\cite{art:Pich97,art:Kataev}.  A central value of $K_4 = 25$ is used,
with an uncertainty of $\Delta K_4 = \pm 50$ in the perturbative
expansions for CIPT and FOPT.

Another major theoretical uncertainty is the choice of renormalization
scale $\mu$ in the \alphas\ dependence of $\delta_{\rm
  pert}^{kl}(s_0)$.  The scale ratio $\zeta = \mu^2/s_0$ is varied
from $0.4$ to $2.0$ in all three models described
above as suggested in~\cite{art:Pich92}.

The choice of the renormalization scheme (RS) can also alter the
result.  Following the prescription in~\cite{art:Pich92} the third
coefficient of the $\beta$-function $\beta_3^{\rm RS}$ is varied
between $0.0$ and $2.0\,\beta_3^{\overline{\rm MS}}$ in order to
obtain the uncertainty due to different renormalization schemes.

\subsection{Power correction terms 
  {\boldmath$\delta_{\rm V/A}^{D,kl}$}}\label{subsec:deltanonpert} In
the framework of the Operator Product Expansion (OPE)
\cite{art:Wilson} the non-perturbative contributions are expressed as
a power series in terms of $1/m_\Pgt^2$ absorbing the long-distance
dynamics into vacuum matrix elements $\langle
\Op(\tilde{\mu})\rangle$~\cite{art:SVZ1,art:SVZ2,art:SVZ3,art:Braaten}.
Thus, they can be written as sums over power corrections of different
dimensions, $D$:
\begin{equation}
  \label{eq:nonpert}
  \delta_{\rm non\md pert, V/A}^{kl}(s_0) =
  \sum_{D = 2,4,6,\dots}
  \delta_{\rm V/A}^{D,kl}(s_0).
\end{equation}
In contrast to the perturbative part described in the previous section
the power corrections differ for the vector and the axial-vector
currents.

In equation~(\ref{eq:nonpert}) the correction of dimension $D=2$ is a
mass correction term and therefore belongs to the perturbative part.
The $D=4$ term is the first term with major non-perturbative
contributions, namely the quark condensates for the three light
flavors $\langle\overline{\psi}\psi\rangle_{\rm u,d,s}$ and the gluon
condensate $\langle\frac{\alphas}{\pi}\,GG\rangle$.  If one neglects
the small $s$-dependence of the power corrections, the $\delta_{\rm
  V/A}^{D,kl}$ terms can be expressed for all $kl$ values by a product
of the same (vector/axial-vector) operator of dimension $D$ (or the
power correction for $kl = 00$) and a simple integral over the
$kl$-dependent weight-functions~\cite{art:Diberder}:
\begin{eqnarray}
  \label{eq:OPE}
  \delta_{\rm V/A}^{D,kl}(s_0) = & 8\pi^2  &
  \begin{array}{cccccc}
     \Sy D=2 & \Sy D=4 & \Sy D=6 & \Sy D=8 & \Sy D=10 &\Sy kl \\ 
    \left(
      \begin{array}{c} 
       1\\ 1\\ 0\\ 0\\ 0 
      \end{array}
    \right. &
    \begin{array}{c} 
     0\\ \frac{m_\Pgt^2}{s_0}\\ -1\\ 0\\ 0 
    \end{array} &
    \begin{array}{c} 
     -3\\ -3\\ -\frac{m_\Pgt^2}{s_0}\\ 1\\ 0
    \end{array} &
    \begin{array}{c} 
     -2\\ -2-3\frac{m_\Pgt^2}{s_0}\\ 3\\ \frac{m_\Pgt^2}{s_0}\\ -1 
    \end{array} &
    \left.
      \begin{array}{c} 
       0\\ -2\frac{m_\Pgt^2}{s_0}\\ 2+3\frac{m_\Pgt^2}{s_0}\\ -3\\ 
        -\frac{m_\Pgt^2}{s_0} 
      \end{array}
    \right) &
      \begin{array}{c} 
        \Sy 00\\ \Sy 10\\ \Sy 11\\ \Sy 12\\ \Sy 13
      \end{array}
    \end{array} \\ \nonumber
  & \times & \sum_{{\rm dim}\,\Op\, =\, D} 
  \frac{\Ce_{\rm V/A}(\tilde{\mu})\,
    \langle\Op(\tilde{\mu})\rangle}{m_\Pgt^D},
\end{eqnarray}
where each entry in the matrix belongs to a particular dimension~$D$
and a particular moment $kl$, as denoted by the first row and the last
column.  The parameter $\tilde{\mu}$ is an arbitrary factorization
scale which separates the long-distance non-perturbative effects,
which are absorbed in the vacuum matrix elements
$\langle\Op(\tilde{\mu})\rangle$, from short-distance perturbative
effects which are incorporated in the Wilson coefficients $\Ce_{\rm
  V/A}(\tilde{\mu})$~\cite{art:Diberder}.

This approach is used for the dimension $D=6$ and $D=8$ terms, taking
$\delta_{\rm V/A}^{6/8,00}$ as free parameters.  For the dimension
$D=2$ and $D=4$ terms the full $s$-dependence is taken into account
for the theoretical description of the moments \cite{art:Diberder}.
The least precisely known $D=4$ parameter, the gluon condensate, which
is known only to $50\,\%$ \cite{art:Braaten}, is also taken as a free
parameter in the fit, while the $D=2$ term is calculated from the
quark masses and the strong coupling.

Terms with dimensions higher then $8$ are neglected in this
analysis as they do not contribute to $R_{\Pgt,{\rm V/A}}^{00}$ as
can be seen from equation~(\ref{eq:OPE}).


\section{OPAL detector}\label{sec:OPAL}
A detailed description of the OPAL detector can be found
in~\cite{art:OPAL}.  A brief description of the features relevant for
this analysis follows.

A high-precision silicon microvertex detector surrounds the beam pipe.
It covers the angular region of $| \cos{\theta} | \leq 0.8$ and
provides tracking information in the $r$-$\varphi$ (and $z$ after
1992) directions\footnote{In the OPAL coordinate system the $x$-axis
  is horizontal and points to the center of LEP.  The $y$-axis is
  vertical and the $z$-axis is in the $\Pem$ beam direction. The angle
  $\theta$ is defined relative to the
  $z$-axis.}~\cite{art:OPAL_SI1,art:OPAL_SI2}.  Charged particles are
tracked in a central detector enclosed inside a solenoid that provides
a uniform axial magnetic field of $0.435\,{\rm T}$.  The central
detector consists of three drift chambers: a high-resolution vertex
detector, the large-volume jet chamber and the $z$-chambers.  The jet
chamber records the momentum and energy loss of charged particles over
$98\,\%$ of the solid angle and the $z$-chambers are used to improve
the track position measurement in the $z$
direction~\cite{art:OPAL_CJ}.

Outside the solenoid coil are scintillation counters which measure the
time-of-flight from the interaction region and aid in the rejection of
cosmic events.  Next is the electromagnetic calorimeter (ECAL) that is
divided into a barrel ($|\cos{\theta}|<0.82$) and two endcap ($ 0.81 <
|\cos{\theta}| < 0.98 $) sections.  The barrel section is composed of
$9440$ lead-glass blocks pointing to the interaction region. Each
block subtends approximately $10 \times 10 \,{\rm cm}^2$ with a depth
of $24.6$ radiation lengths.  The two endcap sections consist of
dome-shaped arrays, each having $1132$ lead-glass blocks, mounted
coaxial with the beam, where each block covers $9.2 \times 9.2\,{\rm
  cm}^2$ with a typical depth of $22$ radiation lengths.  The hadron
calorimeter (HCAL) is beyond the electromagnetic calorimeter and
instrumented with layers of limited streamer tubes in the iron of the
solenoid magnet return yoke.  In the region $|\cos{\theta}|<0.81$ this
detector typically has a depth of $8$ interaction lengths.  The hadron
calorimeter is covered by the muon chamber system, composed of four
layers of drift chambers in the barrel region and four layers of
limited streamer tubes in the endcap region.


\section{Event selection and reconstruction of \btau\ decays}
\label{sec:selection}
OPAL data collected from 1990 to 1995 is used in this analysis.  The
data were taken within $\pm 3\,{\rm GeV}$ of the \PZz resonance.  The
Monte Carlo samples used in this analysis consist of $600\,000$
\Pgt-pair events generated at $\sqrt{s} = \mz$ with {\sc Koralz
  4.0}~\cite{art:koralz}.  Their decays were modelled with {\sc Tauola
  2.4}~\cite{art:tauola} and then processed through the {\sc
  Geant}~\cite{art:geant} OPAL detector simulation~\cite{art:gopal}.
The non-\Pgt\ background Monte Carlo samples consist of $1\,000\,000$
\Pq\Paq\ events generated with {\sc Jetset 7.4}~\cite{art:jetset},
$800\,000$ Bhabha events generated with {\sc Radbab
  2.0}~\cite{art:Babamc1,art:Babamc2}, $600\,000$ \Pgm-pair events
generated with {\sc Koralz 4.0}~\cite{art:koralz} and $800\,000$
events from two-photon processes generated with {\sc Vermaseren
  1.01}~\cite{art:vermaseren,art:vermaseren2}.


\subsection{Selection of \btau-lepton candidates}\label{subsec:preselection}
The standard \Pgt\ selection procedure as described
in~\cite{art:TPsel} begins with the rejection of cosmic rays,
multi-hadronic events and events from two-photon processes.  Cosmic
rays are rejected by the time-of-flight information of the tracks.
Multihadrons are removed from the sample by requiring two narrow jets
(cones with a half opening angle of $35\,^\circ$) and up to six tracks
in the event.  The events from two-photon processes are eliminated by
allowing an acollinearity angle of up to $15\,^\circ$ between the two
jets.

The remaining event sample contains tau pairs, Bhabha events, and muon
pairs.  Events with an energy deposit of more than $0.8 \times 2
E_{\rm beam}$ are identified as Bhabhas. An event is classified as a
muon pair if two tracks carry energy of more than~$0.6 \times 2 E_{\rm
  beam}$ and if both tracks have at least two hits in the muon
chambers and almost no energy deposit in the ECAL.  The remaining
events are classified as \Pgt\ pairs if the polar angle of the total
cone momentum calculated from track momenta and ECAL clusters
satisfies $|\cos\theta| < 0.95$ for both cones.

After this selection both cones in each event are treated
independently. The non-tau background is further reduced by requiring
one or three tracks in each cone with a total charge of plus or minus
one.  A total of $297\,988$ \Pgt\ candidates survive these selection
criteria with an estimated non-\Pgt\ background fraction of $3.9\,\%$.


\subsection{Identification of \btau-decay modes}\label{subsec:likelihood}
A Maximum Likelihood selection as used in previous publications (see
e.g.~\cite{art:CP}) is applied to the data and the Monte Carlo samples
to distinguish between the following decay modes: $\Pgtm \rightarrow
\Pgngt {\rm X}^-$, where~${\rm X}^-$ is one of $\Pem \Pagne$, $\Pgmm
\Pagngm$, $\Pgpm$, $\Pgpm \Pgpz$, $\Pgpm 2\Pgpz$, $\Pgpm 3\Pgpz$,
$2\Pgpm \Pgpp$, $2\Pgpm \Pgpp \Pgpz$, $2\Pgpm \Pgpp 2\Pgpz$. The
charge and parity conjugated modes are implicitly assumed for $\Pgtp
\rightarrow \Pagngt {\rm X}^+$ decays. Fourteen reference
distributions are used to distinguish between the different one-prong
channels and five reference distributions are used in the three-prong
case. Decays with charged kaons instead of pions are suppressed by a
cut on the specific energy loss $\rd E/\rd x$ in the drift chamber.
Decays into electrons are distinguished from the other modes by the
ratio $E/p$ of the ECAL energy associated with the cone over the track
momentum, and the $\rd E/\rd x$ information. Muons are identified by
the number of hits in the muon chambers and the outermost HCAL layers.
The different hadronic decay modes with zero, one, two or three
neutral pions are separated by using the number of reconstructed
photons in the ECAL (see section~\ref{subsec:pi0recon}).

The decay channels used in this analysis are the three non-strange
one-prong modes with at least one neutral pion: $\Pgp \Pgpz$, $\Pgp
2\Pgpz$ and $\Pgp 3\Pgpz$, and the three non-strange three-prong
modes: $3\Pgp$, $3\Pgp \Pgpz$, $3\Pgp 2\Pgpz$.  A total of $\NDAtotal$
\Pgt\ candidates are selected\footnote{For the decay mode $\Pgp \Pgpz$
  a cut on $|\cos\theta| < 0.9$ is used in order to reduce the
  background from Bhabha events.} in these channels with an estimated
background fraction of $\NBUtotal\,\%$ including misidentified \Pgt\ 
decays and the remaining non-\Pgt\ background fraction of $0.8\,\%$.
Details about the treatment of the cross-talk between the signal
channels due to misidentified \Pgt\ decays are subject to
sections~\ref{sec:unfolding} and~\ref{sec:discussion}.

The most important observable for the discrimination between vector
and axial-vector channels is the number of neutral pions in a cone. A
new method to reconstruct neutral pions in \Pgt\ decays has been
employed, which is described in the following section.


\subsection{Reconstruction of neutral pions}\label{subsec:pi0recon}
Neutral pions are identified by their decay into two photons. Since
photons are only detected in the ECAL, an iterative fit of photon
energies and directions to the observed energy deposits in the ECAL
blocks is performed.

The energy deposition in an ECAL block can be expressed as a function
of the photon energy and the photon direction.  This is done by
parameterizing the integrated energy density of an electromagnetic
shower for each ECAL block.  In the barrel region of the ECAL where
the blocks have a quasi-pointing geometry only lateral shower
profiles need to be parameterized.  They can be approximated by the
sum of two exponential distributions representing core and halo
components~\cite{art:Akopdjanov}.  For the endcap region, where the
blocks are oriented parallel to the beam, lateral and longitudinal
profiles are important. The longitudinal profile is reasonably
described by the gamma distribution~\cite{art:PDG96}.

The mean energy deposit of a minimum ionizing particle is subtracted
from all ECAL blocks hit by a charged particle.  The fit then finds
the smallest number of photons needed to explain the measured energies
and provides their corresponding three-vectors.

Energy depositions from hadronic interactions of charged pions in the
ECAL are accounted for by assigning photon candidates which are close
to track intersections with the ECAL to the track.  The maximum angle
allowed between a photon candidate and a track to which the photon
candidate can be assigned depends on the polar angle of the track and
varies between $1.2\,^\circ$ and $1.7\,^\circ$ in the barrel region and
between $2.0\,^\circ$ and $3.4\,^\circ$ in the endcap region.  A
photon candidate close to a track is still classified as a photon if
the total energy of photon candidates assigned to this track exceeds
the measured track momentum.

\begin{figure}[htb]
\centering
\resizebox{0.49\textwidth}{!}{%
\includegraphics{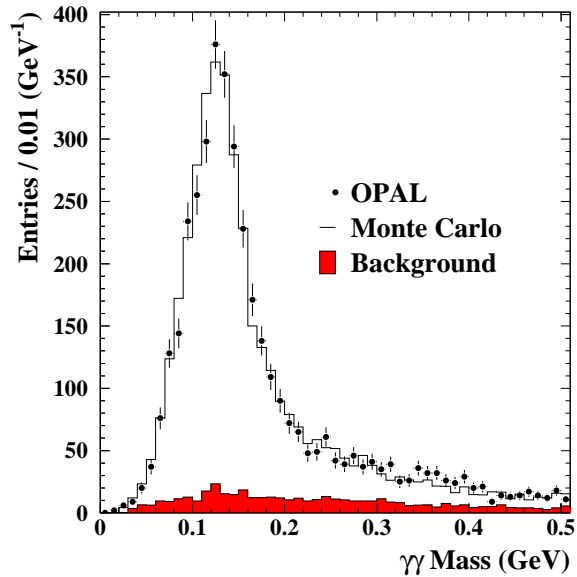}}
\caption{\em The \Pgg\Pgg-mass in the \Pgp\Pgpz\ channel for
  decays with two reconstructed photons with a minimal energy of
  $\mathit{0.5\,\mathit{GeV}}$.  OPAL data is shown as data points;
  the total Monte Carlo prediction is given by the open histogram and
  the shaded histogram denotes the $\mathit{\tau}$ and
  non-$\mathit{\tau}$ background.}\label{fig:pi0}
\end{figure}

All possible two-photon combinations are then used to find \Pgpz\ 
candidates.  The combination resulting in the largest number of \Pgpz\ 
candidates with an average invariant mass deviation from the
\Pgpz\ mass less than $1.5\,\sigma$ is selected.  The error on the
invariant two-photon mass, $\sigma$, is calculated from the error
matrices of the above photon fit.  The \Pgpz\ four-momenta are then
calculated from the energies and directions of the photon pairs of the
selected combination after a constrained fit to the \Pgpz\ mass.
Figure~\ref{fig:pi0} shows the photon-pair mass for selected
\Pgp\Pgpz\ candidates before the \Pgpz-mass constraint.

For all the one-prong modes a minimum energy of $0.7\,{\rm GeV}$ for
each reconstructed \Pgpz\ is required while $E_\Pgpz > 2.0\,{\rm GeV}$
is required in the three-prong modes to suppress fake \Pgpz's
introduced by the energy deposition of charged pions in the ECAL.

The granularity of the ECAL allows the reconstruction of both photons
from a \Pgpz\ only if its energy is below $12\,{\rm GeV}$. The \Pgpz's
with larger energies have photons which are merged in the ECAL.
Therefore all the photon candidates with energies above this value are
considered to be \Pgpz's and their momentum is calculated from the
reconstructed energy corrected by the mass of the \Pgpz.


\section{Unfolding procedure}\label{sec:unfolding}
The Monte Carlo predictions for the measured spectra and their
background contributions are corrected with the most recent
constrained branching ratios of the \Pgt\ given in~\cite{art:PDG96}.
Effects due to limited detector resolution and efficiency are
accounted for by a regularized unfolding technique~\cite{art:Blobel}.

To unfold measured distributions in $s$ (the squared mass of the
hadronic final state) the detector simulation is used to create
response matrices which map the generated distribution in $x \equiv
s_{\rm true}$ to a $y \equiv s_{\rm meas}$ distribution one would
measure including all detector effects.  The following convolution
integral describes the general relation between a true distribution
$f(x)$ and a measured distribution $g(y)$:
\begin{equation}
  \label{eq:unf1}
  g(y) = \int\limits_{x_{\rm min}}^{x_{\rm max}} \rd x\,
  A(y,x)\,\epsilon(x)\,f(x) + b(y),
\end{equation}
where $A(y,x)$ is the detector response function, $b(y)$ denotes the
background distribution, and $\epsilon(x)$ is the selection efficiency.
Equation~(\ref{eq:unf1}) can be simplified to a matrix equation of the
form:
\begin{equation}
  \label{eq:unf2}
  {\bf g} = {\bf A} \cdot {\bf f} + {\bf b}.
\end{equation}
This is performed in two steps. First, the true distribution $f(x)$ is
parameterized with a set of $m$ parameters $f_j$ and $m$ basis
functions $p_j(x)$ which are defined below in
equation~(\ref{eq:bsplines}):
\begin{equation}
  \label{eq:discret1}
  f(x) =  \sum_{j=1}^{m} f_j\,f_{\rm MC}(x)\,p_j(x),
\end{equation}
with $f_{\rm MC}(x)$ being the generated Monte Carlo distribution.  By
defining:
\begin{equation}
  \label{eq:ajy}
  A_j(y) = \int\limits_{x_{\rm min}}^{x_{\rm max}} \rd x\,
  \epsilon(x)\,f_{\rm MC}(x)\,A(y,x)\,p_j(x),
\end{equation}
equation~(\ref{eq:unf1}) takes the form:
\begin{equation}
  \label{eq:unf3}
  g(y) = \sum_{j=1}^m f_j\,A_j(y) + b(y).
\end{equation}
In the second step, $g(y)$, $b(y)$ and $A_j(y)$ are represented by $n$
bins:
\begin{equation}
  \label{eq:discret2}
  g_i    = \int\limits_{y_{i-1}}^{y_i} \rd y\,g(y); \qquad
  b_i    = \int\limits_{y_{i-1}}^{y_i} \rd y\,b(y); \qquad
  A_{ij} = \int\limits_{y_{i-1}}^{y_i} \rd y\,A_j(y).
\end{equation}
The basis functions $p_j(x)$ used in equations~(\ref{eq:discret1})
and~(\ref{eq:ajy}) are chosen as cubic B-splines and thus have the
following form:
\begin{equation}
  \label{eq:bsplines}
  p_j(x) = \frac{1}{6}\times\left\{ \begin{array}{lll}
  z^3                  & z=(x-t_{j\p{+1}})/d & \quad t_{j\p{+1}}\le x<t_{j+1}\\
  (1+3(1+z(1-z))z)     & z=(x-t_{j+1})/d     & \quad t_{j+1}    \le x<t_{j+2}\\
  (1+3(1+z(1-z))(1-z)) & z=(x-t_{j+2})/d     & \quad t_{j+2}    \le x<t_{j+3}\\
  (1-z)^3              & z=(x-t_{j+3})/d     & \quad t_{j+3}    \le x<t_{j+4}\\
  0                    &                     & \quad {\rm otherwise}
  \end{array}\right.,
\end{equation}
where $d=(x_{\rm max}-x_{\rm min})/(m-3)$ is the distance between
adjacent knots $t_k = x_{\rm min} + (k-4)\,d$ for $k=1, \dots, m-1$
knots and $m$ splines.

The coefficient vector ${\bf f} = \{f_j\}$ is now observed in a fit to
the data bins $g_i$, and the unfolded result can be obtained with
equation~(\ref{eq:discret1}).  This particular choice of basis
functions and normalization leads to the simple prediction $f_j = 1$
for all $j$ if the Monte Carlo generated distribution and the unfolded
result are identical.
 
In certain cases unfolding produces results with unphysical behavior.
Statistically insignificant components of the fitted coefficient
vector ${\bf f}$ can lead to large oscillations of the unfolded
distribution.  Therefore the unfolding needs to be modified by a
regularization step which suppresses the statistically insignificant
parts of the solution.  This is achieved by applying a smooth damping
function to the unfolded result.  The magnitude of the fluctuations is
measured by the total curvature $r({\bf f})$ of the function
$f(x)/f_{\rm MC}(x)$:
\begin{equation}
  \label{eq:curvature}
  r({\bf f}) = \int\limits_{x_{\rm min}}^{x_{\rm max}}\rd x\,{\left[\left(
      \frac{\rd^2}{\rd x^2}\frac{f(x)}{f_{\rm MC}(x)}\right)\right]^2} = 
  \int\limits_{x_{\rm min}}^{x_{\rm max}}\rd x\,
  {\left[\sum_{j=1}^m f_j\, 
      \frac{\rd^2}{\rd x^2} p_j(x)\right]^2} =
  {\bf f}^{\rm T} \cdot {\bf C} \cdot {\bf f},
\end{equation}
where ${\bf C}$ is a constant, symmetric, positive semidefinite matrix
obtained from the second derivatives of the basis functions $p_j$.
The regularized result is now obtained by adding the total curvature
$r({\bf f})$ weighted with a regularization parameter~$\rho$ to the
$\chi^2$ in the fit and minimizing the sum:
\begin{equation}
  \label{eq:chisq}
  \chi^2_{\rm reg}({\bf f}) = \chi^2({\bf f}) 
  + \frac{1}{2}\,\rho\,r({\bf f}). 
\end{equation}
The final unfolded distribution in $s$ is given by weighting the Monte
Carlo distribution in $s_{\rm true}$ with the regularized coefficient
vector ${\bf f}$ obtained from the fit~(\ref{eq:discret1}).

This method is bias-free as long as the detector simulation is correct
for all $s$ and is independent of the used Monte Carlo distributions
provided that only the statistically insignificant components of the
fitted contributions are damped.  Possible biases due to the detector
simulation are accounted for in the systematic errors as described in
section~\ref{sec:syst}. The correct choice of the regularization
parameter~$\rho$ can be tested in the following way:
\begin{itemize}
\item[a)] The coefficients $f_j$, which are correlated and have in
  general different errors, can be transformed into a set of
  independent parameters $a_j$ which have unit variance and are sorted
  with the regularization measure $\rho\,r({\bf f})$ in order of
  decreasing significance~\cite{art:Blobel}.  All $a_j$ with $j
  > n(\rho)$, where $n(\rho)$ is the number of effective coefficients
  remaining after damping with the parameter $\rho$, have to be
  consistent with zero.
\item[b)]Furthermore the $\chi^2$-probability of the fit without the
  regularization term $r({\bf f})$ should increase for the regularized
  coefficient vector ${\bf f}$.
\end{itemize}
The regularization parameter is chosen according to these criteria.

The background, in a particular one-prong (three-prong) channel,
consists mainly of misidentified other one-prong (three-prong) \Pgt\ 
decays, introducing correlations between the spectra.  In order to
provide a proper treatment of the correlations, the three one-prong
(three-prong) channels are unfolded simultaneously.  In addition to
the three detector response matrices for the three one-prong
(three-prong) signal modes, six more detector response matrices are
used, mapping the Monte Carlo generated distribution in $s_{\rm true}$
of a background channel to the background part introduced by this
channel in the $s_{\rm meas}$ distribution of a signal channel as
correlated background.  Non-\Pgt\ background and other misidentified
\Pgt\ decays are treated as uncorrelated background $b_i$ as described
above.


\section{Discussion of the measured spectra}
\label{sec:discussion}
\begin{figure}[htbp]
\centering
\resizebox{0.49\textwidth}{!}{%
\includegraphics{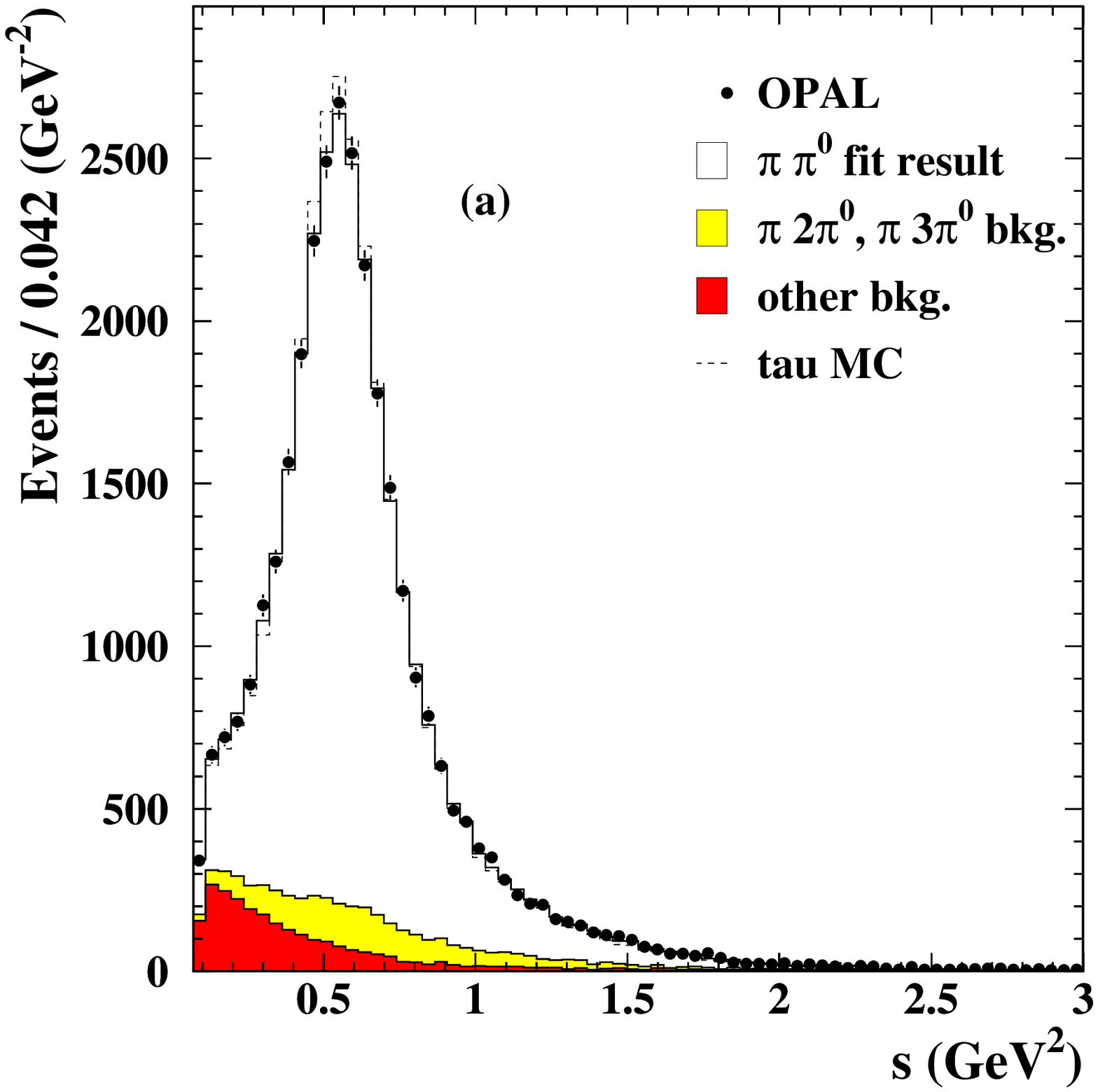}}
\resizebox{0.49\textwidth}{!}{%
\includegraphics{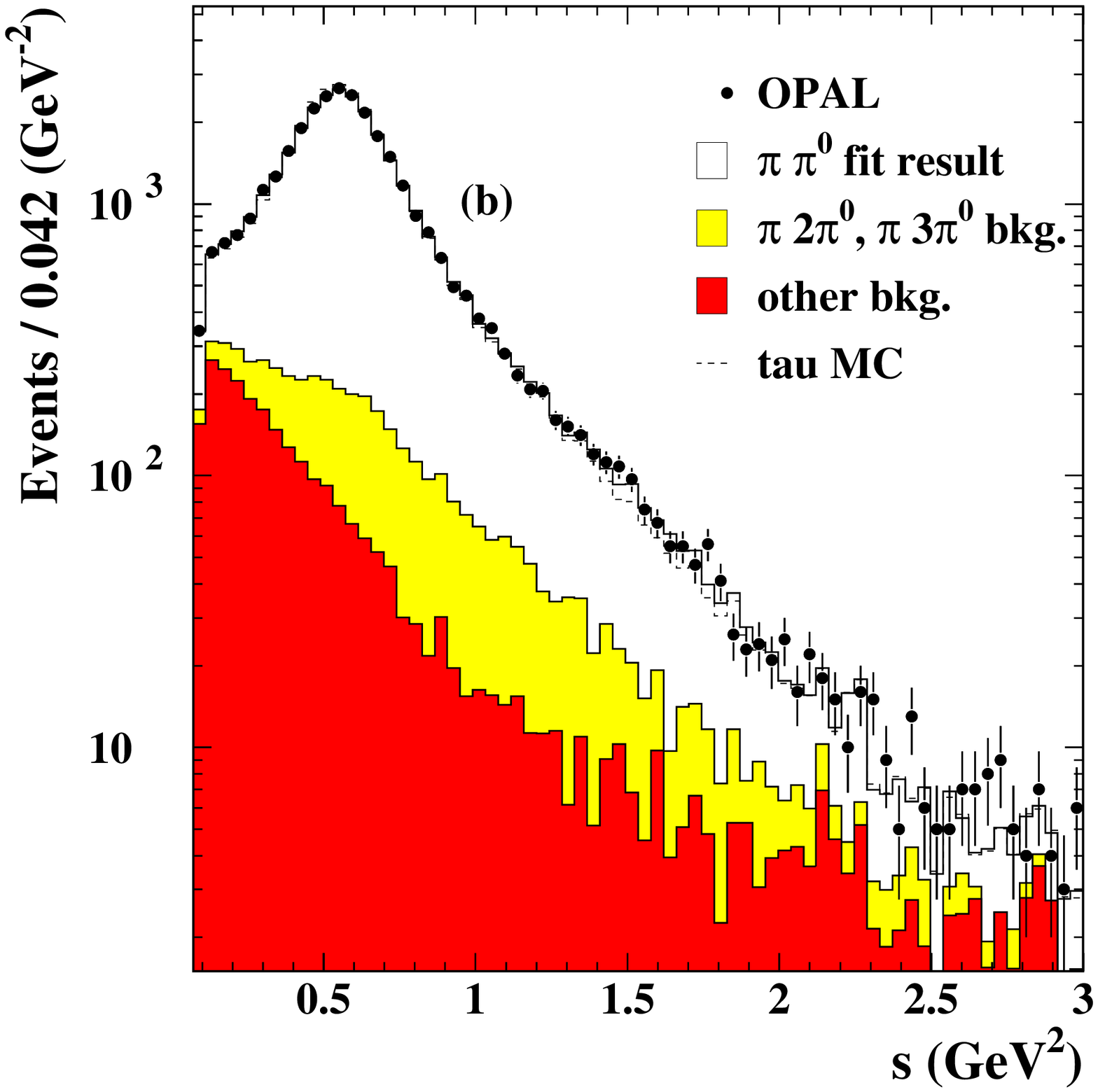}}
\resizebox{0.49\textwidth}{!}{%
\includegraphics{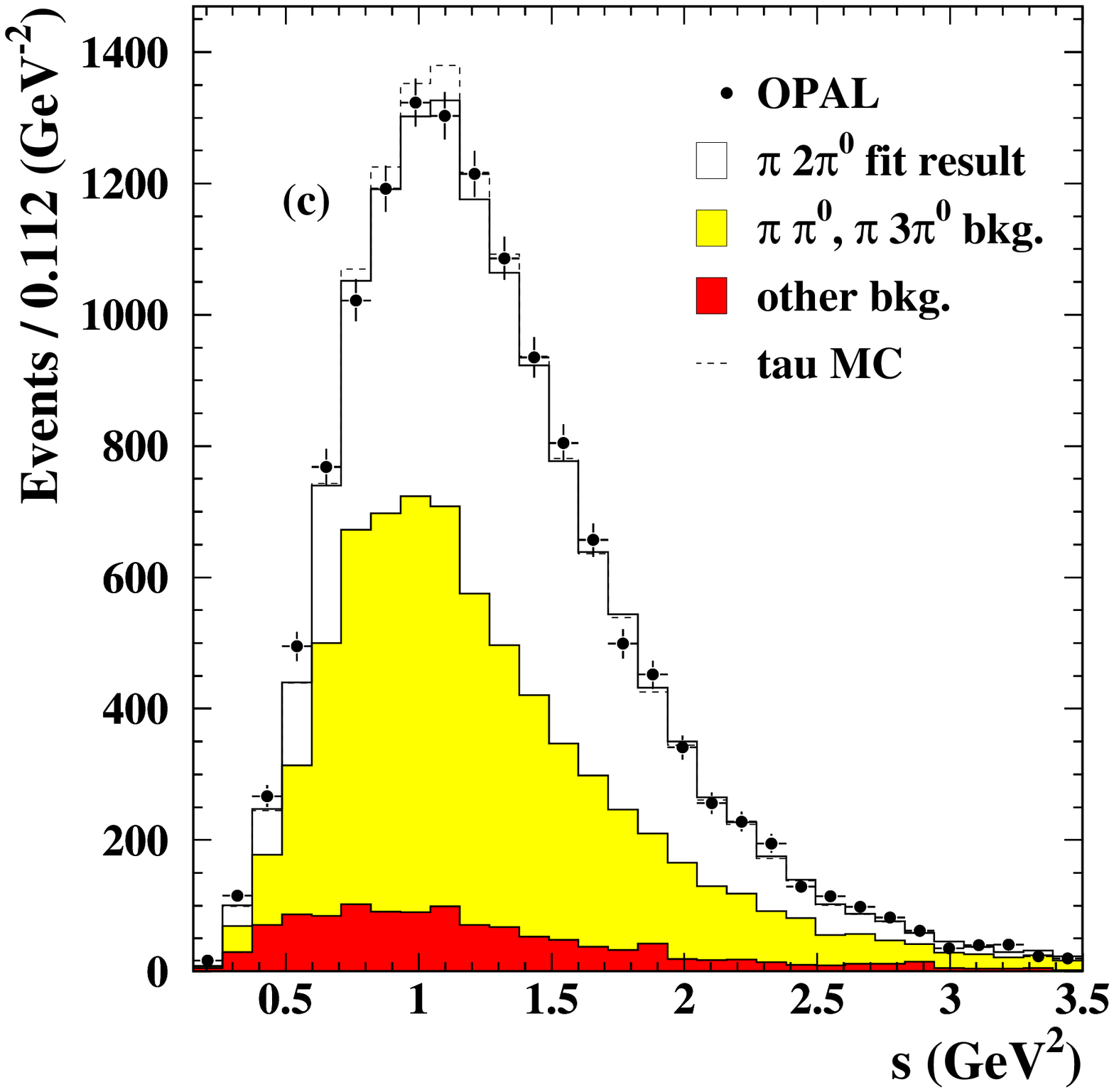}}
\resizebox{0.49\textwidth}{!}{%
\includegraphics{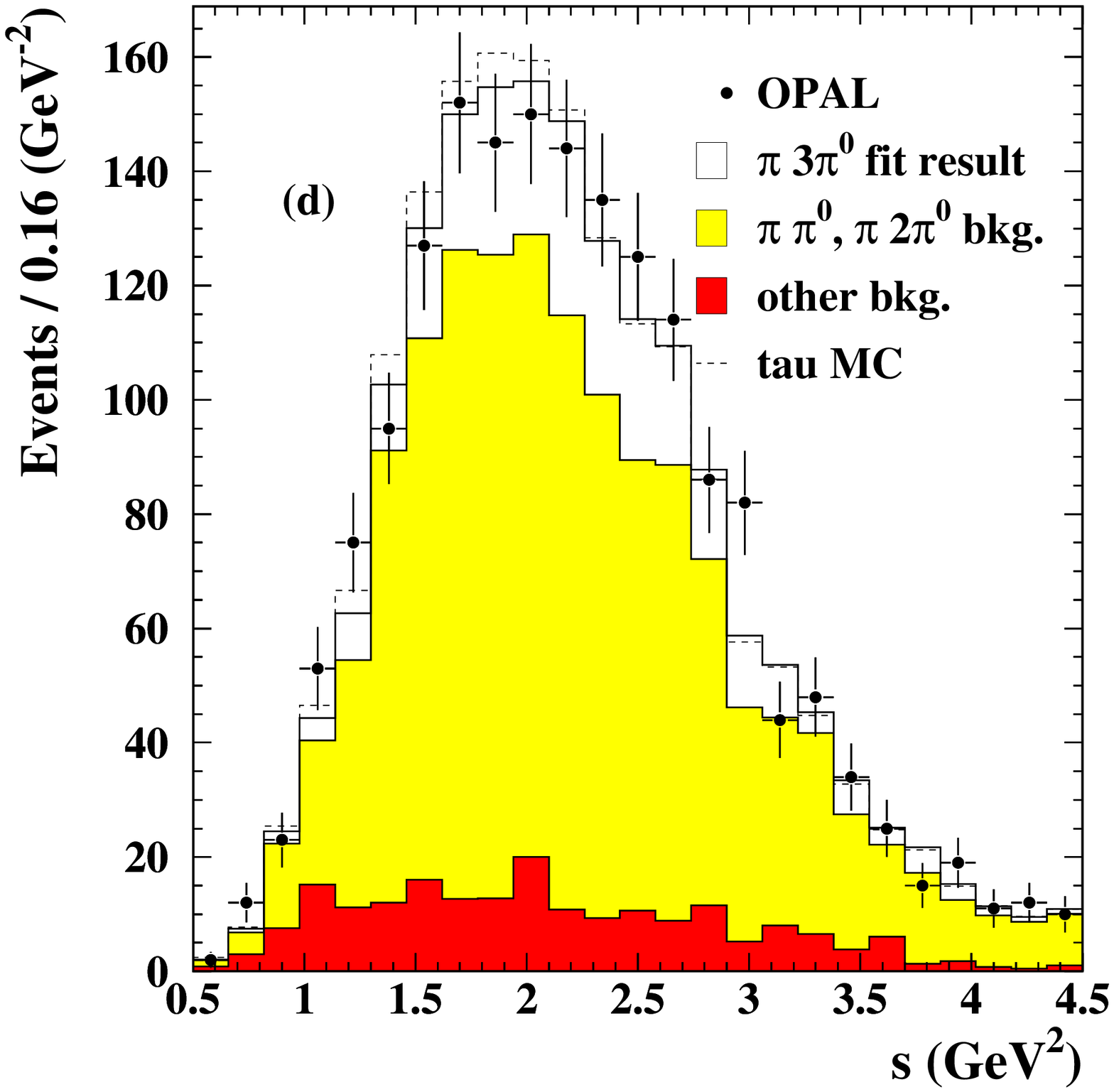}}
\caption{\em The measured $\mathit{s_\mathit{meas}}$ spectra for 
  1-prong decays.  Plots (a) and (b) are the $\mathit{\Pgp \Pgpz}$
  channel, (c) and (d) are the $\mathit{\Pgp 2\Pgpz}$ and
  $\mathit{\Pgp 3\Pgpz}$ modes, respectively.  The points denote OPAL
  data (statistical errors only).  The open histograms show the fitted
  spectra after the regularized unfolding, refolded into detector
  space.  The background contributions from simultaneously unfolded
  channels (correlated background) are shown as light grey areas while
  the background from other sources (uncorrelated background) is
  represented in dark grey.}
\label{fig:dataQ2-1pr}
\end{figure}
\begin{figure}[htbp]
\centering
\resizebox{0.49\textwidth}{!}{%
\includegraphics{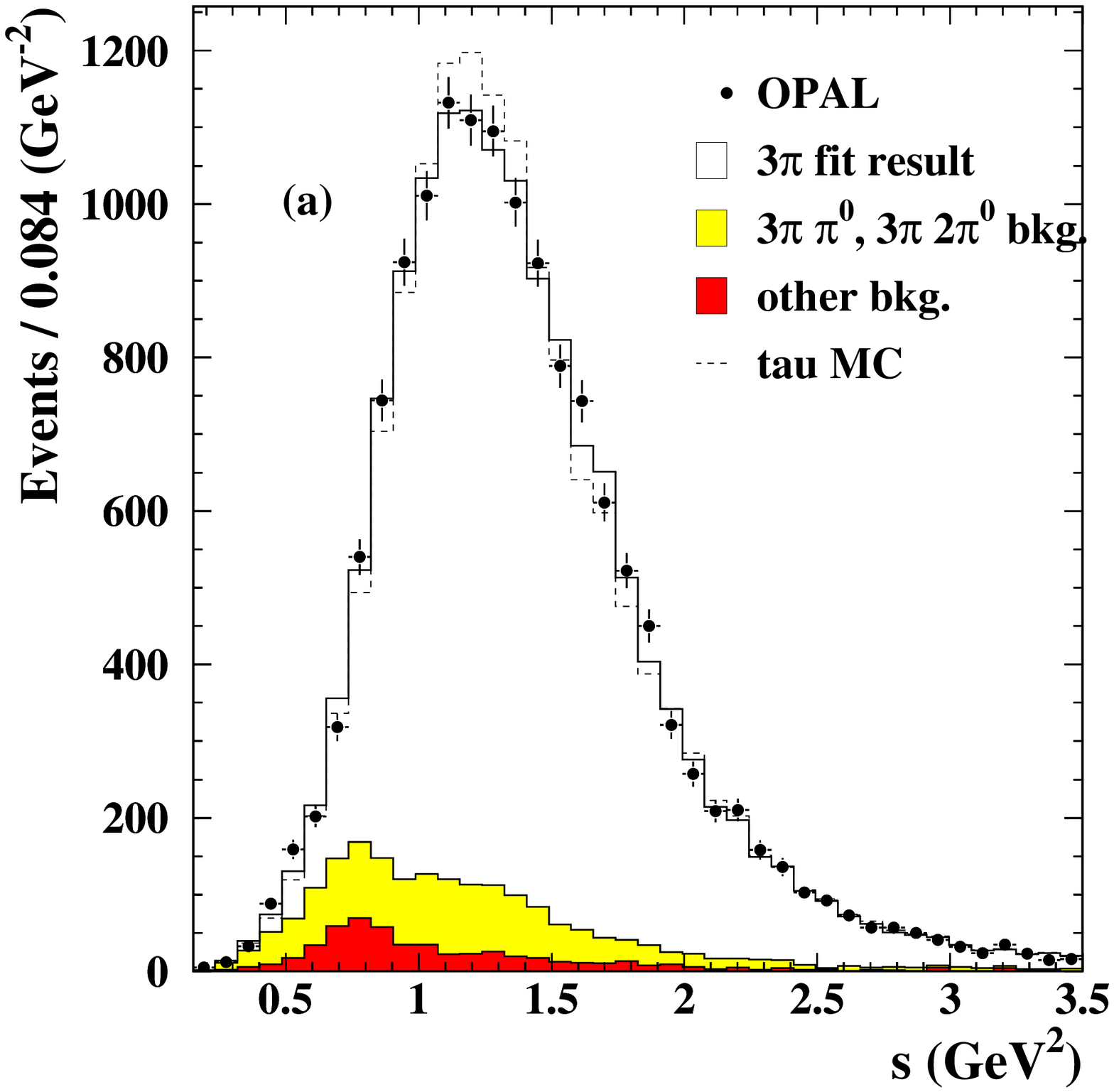}}
\resizebox{0.49\textwidth}{!}{%
\includegraphics{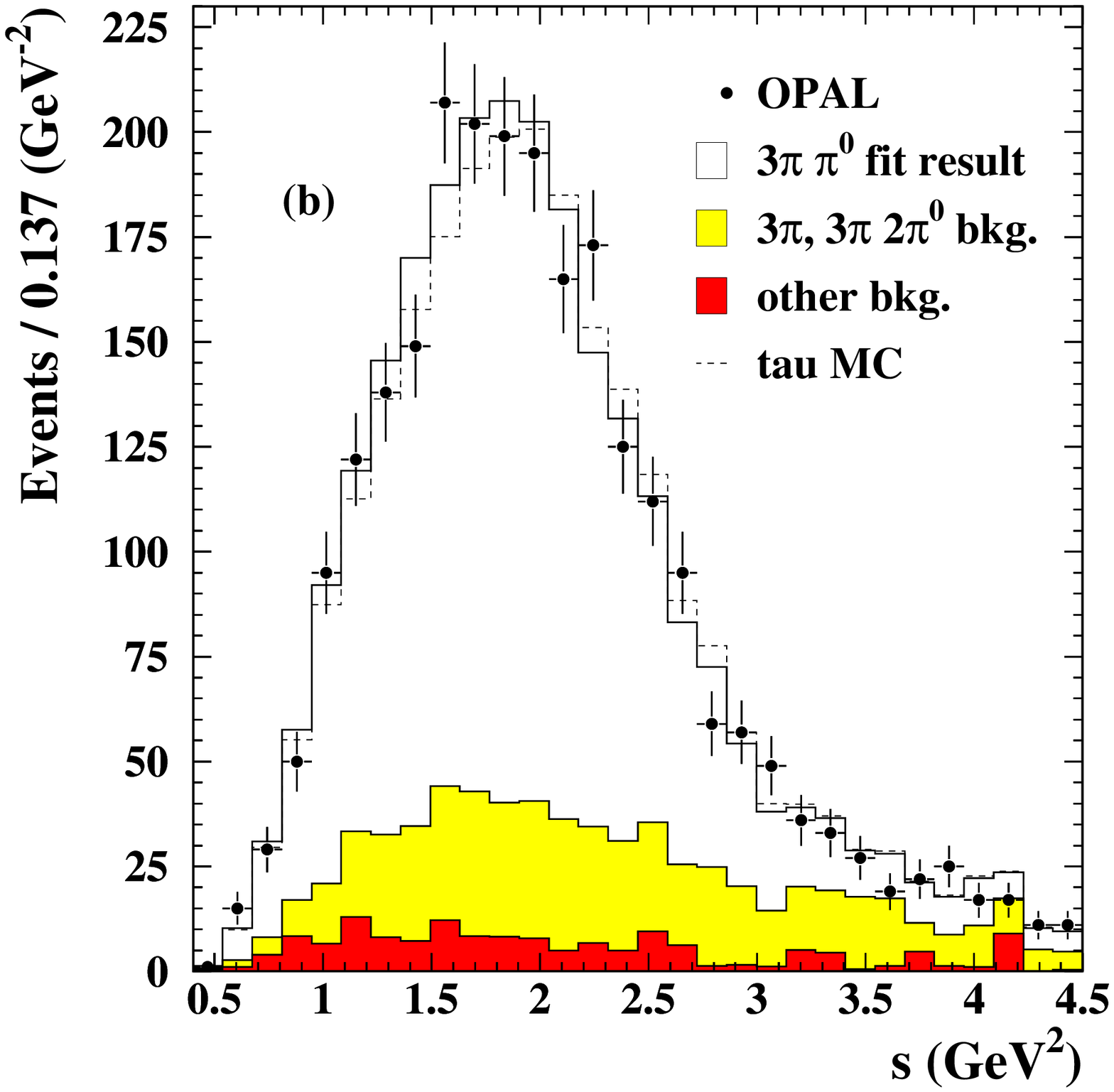}}
\resizebox{0.49\textwidth}{!}{%
\includegraphics{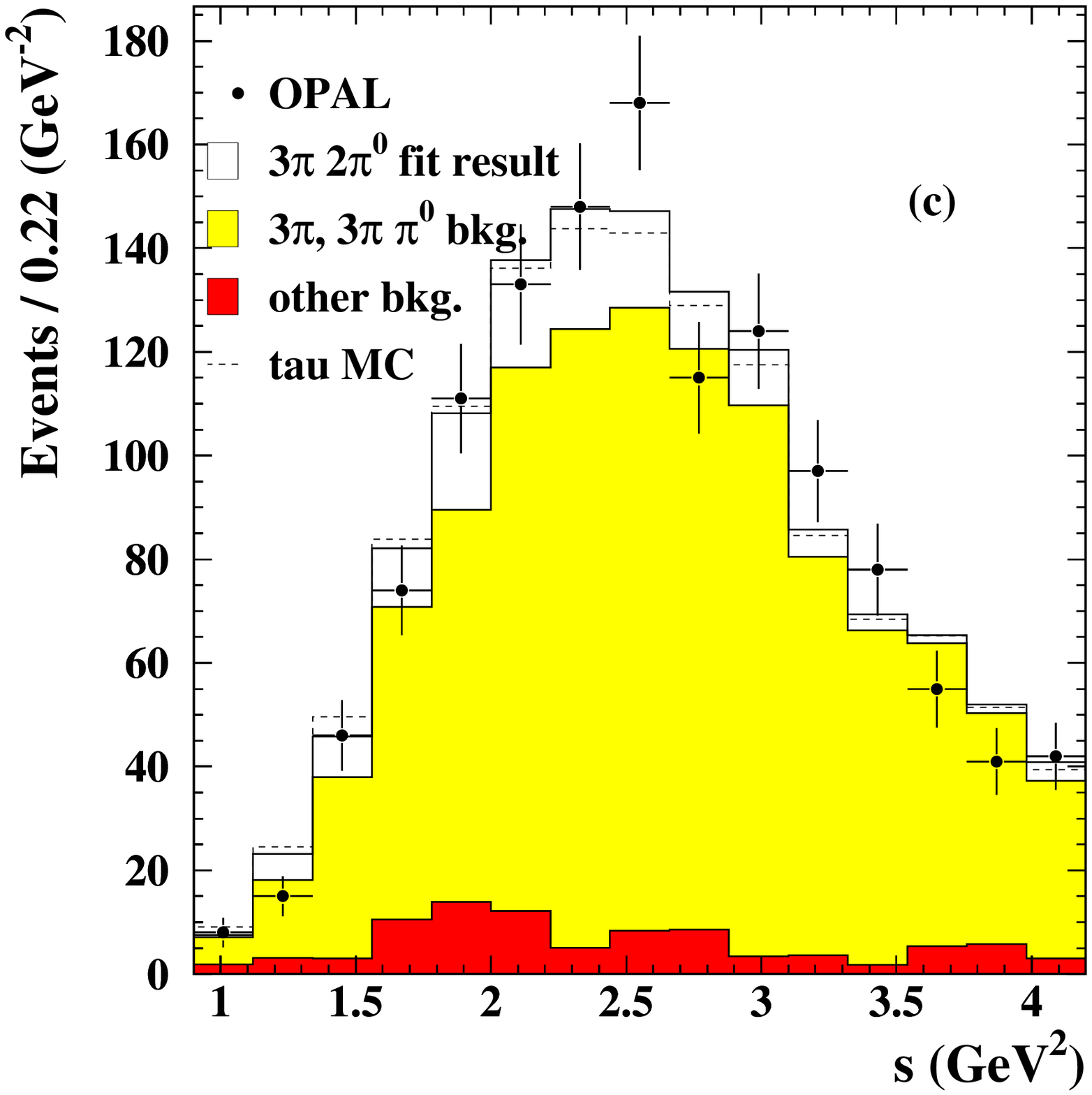}}
\caption{\em The measured $\mathit{s_\mathit{meas}}$ spectra for 
  3-prong decays.  Plot (a) is the $\mathit{3\Pgp}$ channel, (b) and
  (c) are the $\mathit{3\Pgp \Pgpz}$ and $\mathit{3\Pgp 2\Pgpz}$
  modes, respectively.  The points denote OPAL data (statistical
  errors only).  The open histograms show the fitted spectra after the
  regularized unfolding, refolded into detector space.  The background
  contributions from simultaneously unfolded channels (correlated
  background) are shown as light grey areas while the background from
  other sources (uncorrelated background) is represented in dark
  grey.}
\label{fig:dataQ2-3pr}
\end{figure}
The selection efficiencies and background fractions from
simultaneously unfolded channels (correlated background) and other
\Pgt- and non-\Pgt-background sources (uncorrelated background) are
listed in table~\ref{tab:eff}.
\begin{table}[htbp]
  \begin{center}
    \begin{tabular}{|c|rrrr|}
      \hline
       \rule[-11pt]{0pt}{30pt} channel &\multicolumn{1}{c}{efficiency}&
       \multicolumn{1}{c}{\parbox{2.2cm}{correlated\\ background}}& 
       \multicolumn{1}{c}{\parbox{2.4cm}{uncorrelated\\ background}}&
       \multicolumn{1}{c|}{\parbox{1.5cm}{selected\\ decays}}\\
      \hline
      $\Pgp \Pgpz$                     &
      $(\EFVapapz\pm\EFEapapz)\,\%$    &
      $(\NBCapapz\pm\EBCapapz)\,\%$    &
      $(\NBUapapz\pm\EBUapapz)\,\%$    &
      $\NDAapapz$ \\
      $\Pgp 2\Pgpz$                    &
      $(\EFVapbpz\pm\EFEapbpz)\,\%$    &
      $(\NBCapbpz\pm\EBCapbpz)\,\%$    &
      $(\NBUapbpz\pm\EBUapbpz)\,\%$    &
      $\NDAapbpz$ \\
      $\Pgp 3\Pgpz$                    &
      $(\EFVapcpz\pm\EFEapcpz)\,\%$    &
      $(\NBCapcpz\pm\EBCapcpz)\,\%$    &
      $(\NBUapcpz\pm\EBUapcpz)\,\%$    &
      $\NDAapcpz$ \\
      \hline
      $3\Pgp $                         &
      $(\EFVcpopz\pm\EFEcpopz)\,\%$    &
      $(\NBCcpopz\pm\EBCcpopz)\,\%$    &
      $(\NBUcpopz\pm\EBUcpopz)\,\%$    &
      $\NDAcpopz$ \\
      $3\Pgp \Pgpz$                    &
      $(\EFVcpapz\pm\EFEcpapz)\,\%$    &
      $(\NBCcpapz\pm\EBCcpapz)\,\%$    &
      $(\NBUcpapz\pm\EBUcpapz)\,\%$    &
      $\NDAcpapz$ \\
      $3\Pgp 2\Pgpz$                   &
      $(\EFVcpbpz\pm\EFEcpbpz)\,\%$    &
      $(\NBCcpbpz\pm\EBCcpbpz)\,\%$    &
      $(\NBUcpbpz\pm\EBUcpbpz)\,\%$    &
      $\NDAcpbpz$ \\
      \hline
    \end{tabular}
    \caption{\em Efficiencies, background fractions and total number of 
                 selected \Pgt\ decays.}
    \label{tab:eff}
  \end{center}
\end{table}
Figures~\ref{fig:dataQ2-1pr} and \ref{fig:dataQ2-3pr} show the
measured $s_{\rm meas}$ distributions of the six channels used in this
analysis in comparison to the fitted signal after the regularized
unfolding, and the Monte Carlo predictions.

The $3\Pgp$ spectrum shows a significant deviation from the shape
predicted by the Monte Carlo (the dashed histogram) as has been
observed in previous analyses of the $3\Pgp$ decay
current~\cite{art:Ute}.  There is also a slight deviation on the left
side of the peak in the $\Pgp \Pgpz$ channel and in the upper tail
region. The other modes are statistically consistent with their Monte
Carlo predictions.
 
The $\chi^2$ values for the one-prong and three-prong fits after the
regularization step are $\chi^2_{\rm 1-pr.}/{\rm d.o.f.} = 94.0/109$
and $\chi^2_{\rm 3-pr.}/{\rm d.o.f} = 71.4/69$ leading to the
$\chi^2$-probabilities $0.85$ and $0.40$, respectively.

\begin{figure}[htbp]
\centering
\resizebox{0.414\textwidth}{!}{%
\includegraphics{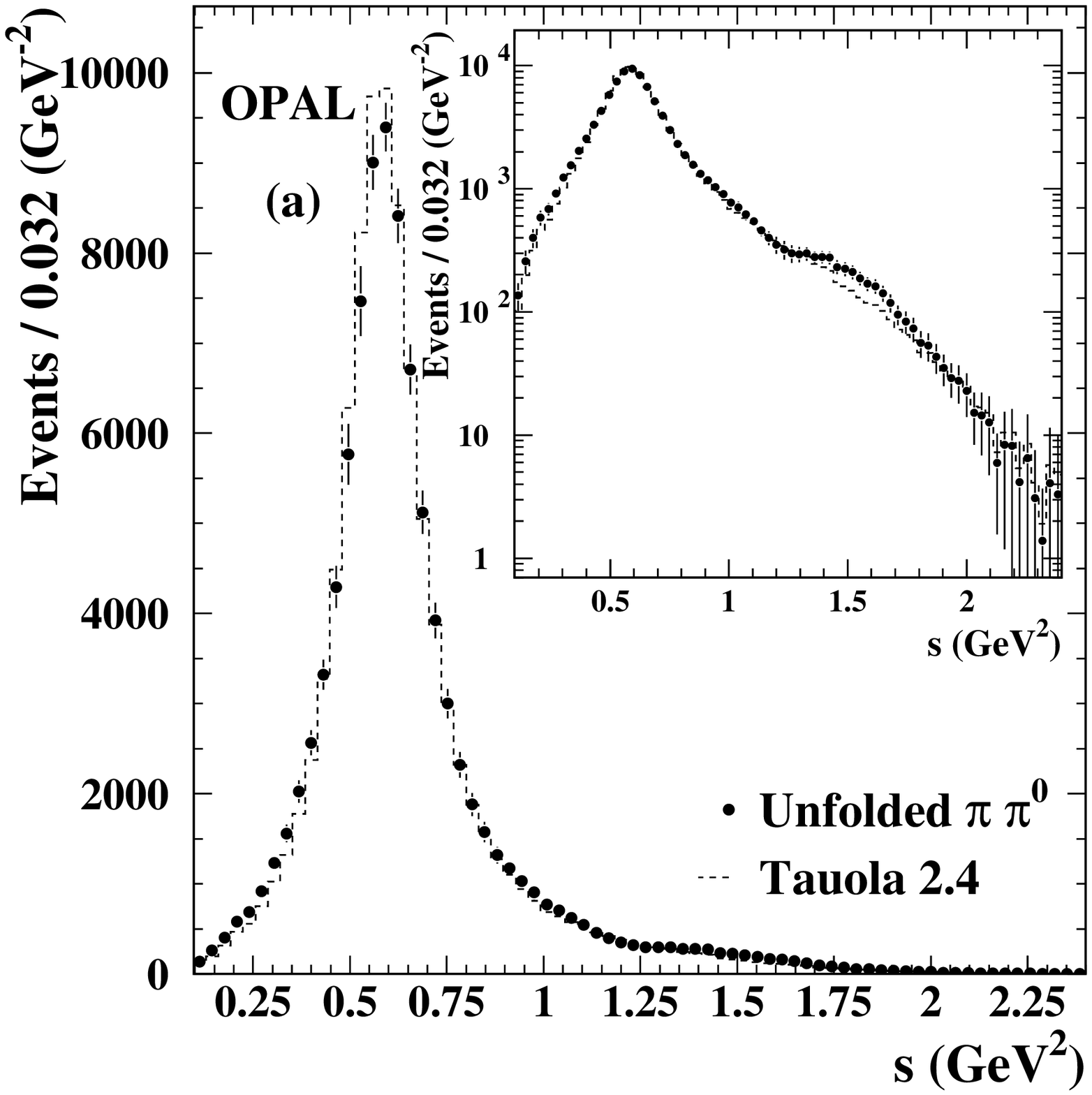}}
\resizebox{0.414\textwidth}{!}{%
\includegraphics{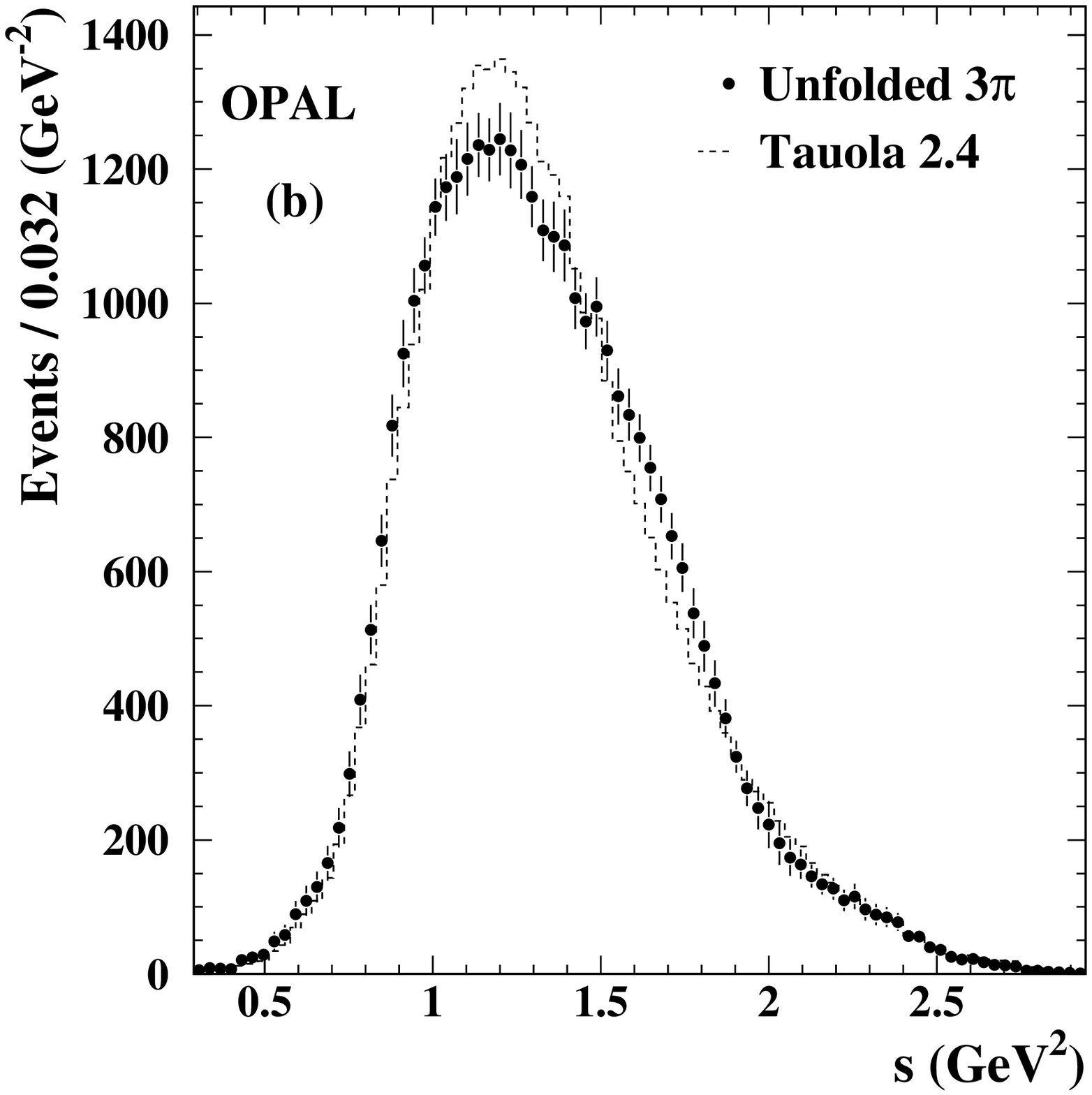}}
\resizebox{0.414\textwidth}{!}{%
\includegraphics{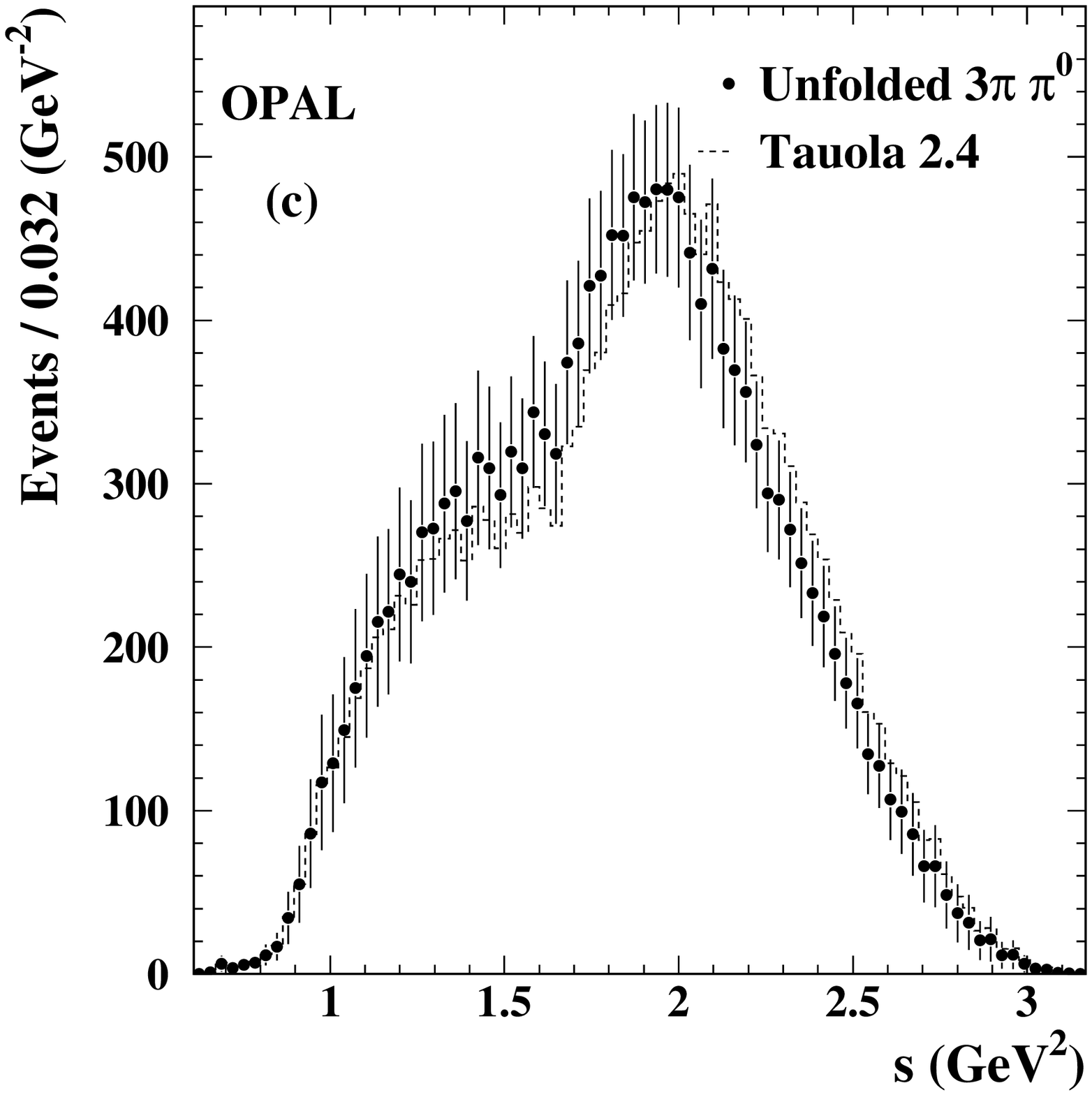}}
\resizebox{0.414\textwidth}{!}{%
\includegraphics{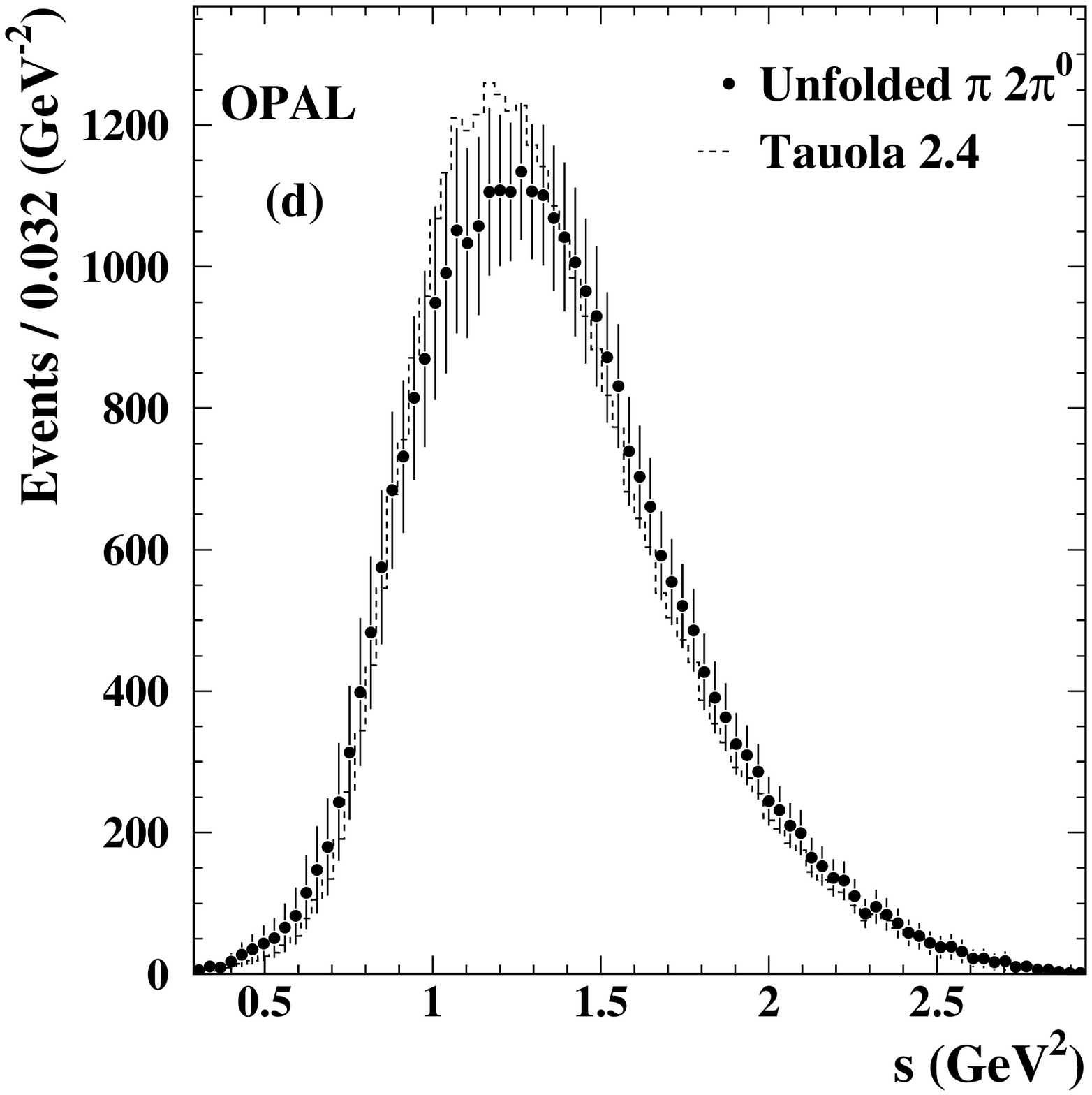}}
\resizebox{0.414\textwidth}{!}{%
\includegraphics{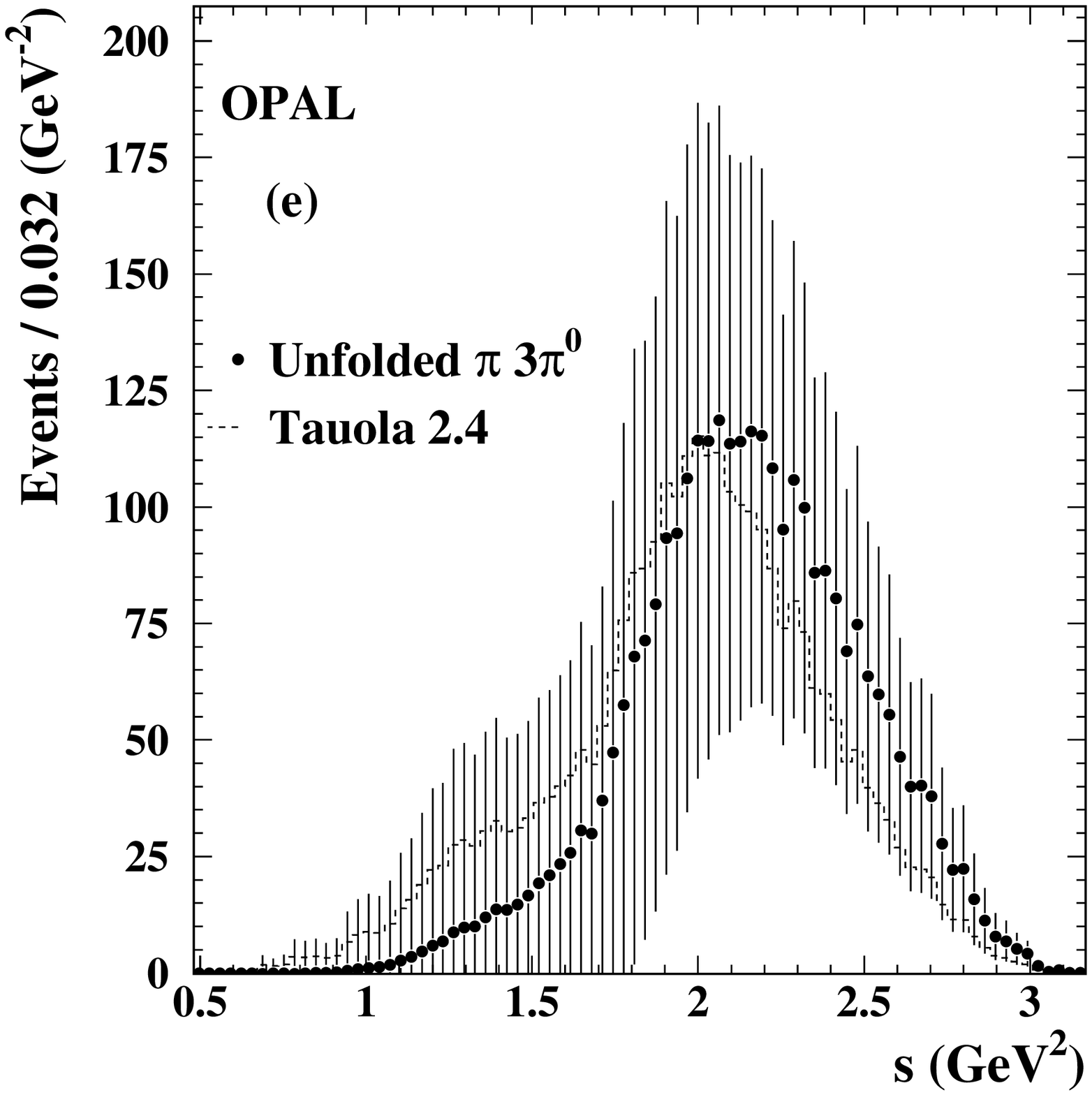}}
\resizebox{0.414\textwidth}{!}{%
\includegraphics{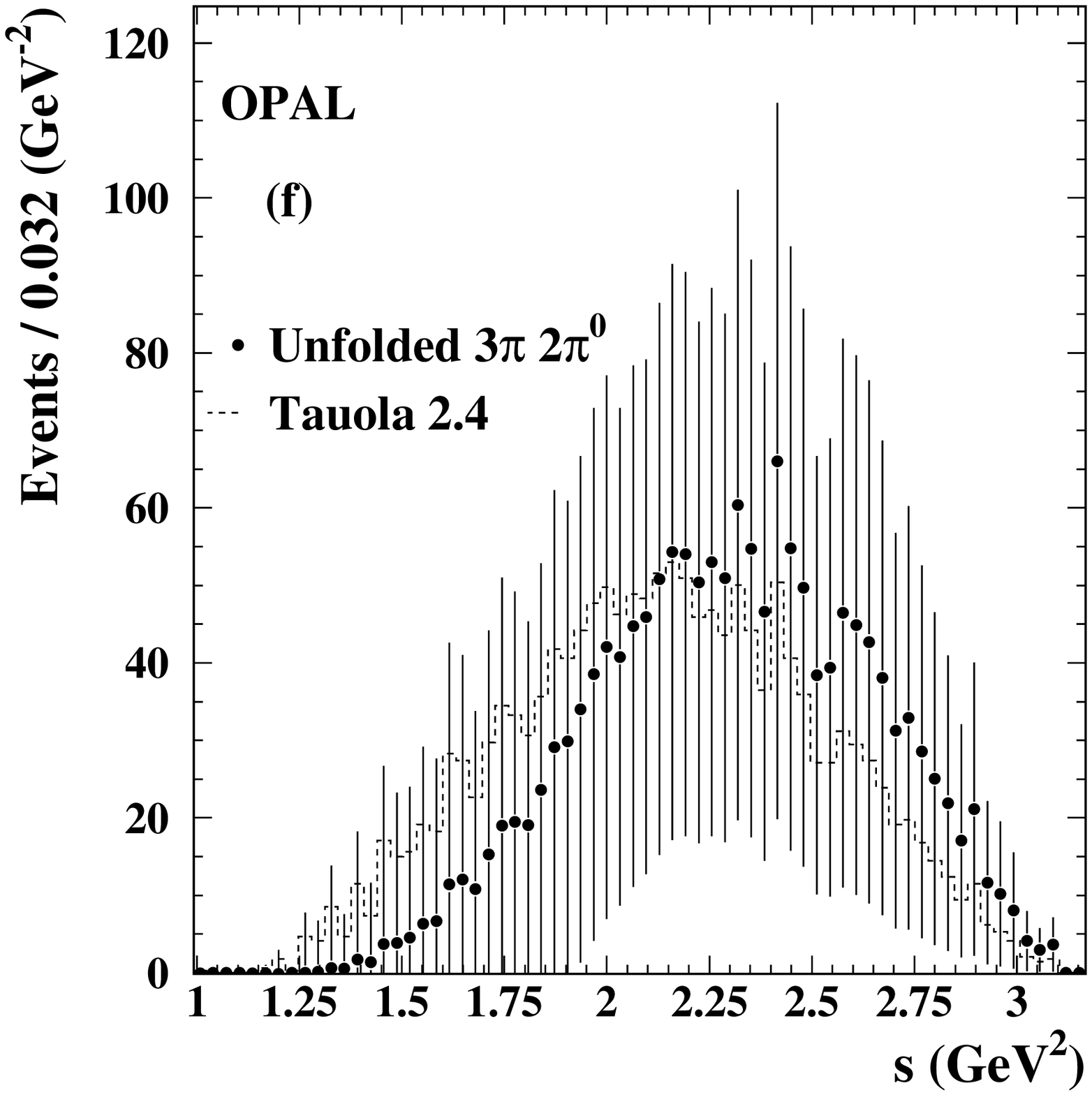}}
\caption{\em The unfolded $\mathit{s_\mathit{true}}$ spectra. Shown are 
  the three vector channels (left) and the three axial-vector channels
  (right) together with the Monte Carlo prediction. There are strong
  correlations between the data points due to the unfolding.  The
  plots (a),(d),(e) are the unfolded spectra of plots (a),(c),(d) in
  figure~\ref{fig:dataQ2-1pr} and the plots (b),(c),(f) are the
  unfolded spectra of the plots (a),(b),(c) in
  figure~\ref{fig:dataQ2-3pr}. The error bars include statistical and
  systematic uncertainties.}
\label{fig:unfQ2}
\end{figure}

The unfolded distributions of the measured spectra are shown in 
figure~\ref{fig:unfQ2}.
The plotted data points are strongly correlated due to the unfolding
procedure. The deviations from the Monte Carlo prediction seen in
figures~\ref{fig:dataQ2-1pr} and~\ref{fig:dataQ2-3pr} are still
present after the unfolding, most prominently in the $\Pgp \Pgpz$ and
the $3\Pgp$ channel. The enhancement in the upper tail (see
figures~\ref{fig:dataQ2-1pr} (b) and~\ref{fig:unfQ2} (a)) of the $\Pgp
\Pgpz$ distribution can be explained within the K{\"u}hn--Santamaria
model~\cite{art:KS} by enlarging the fraction of $\Pgra$'s and
$\Pgrb$'s in the $\Pgr$ decay amplitude.  A similar correction to the
three-pion current, modelled as a Breit--Wigner decay chain $\Pai
\rightarrow \Pgr \Pgp \rightarrow 3\Pgp$~\cite{art:KS} in the Monte
Carlo, does not account for the observed discrepancy.

 
\section{Systematic uncertainties}\label{sec:syst}
Possible origins for systematic effects on the reconstructed value for
the squared hadronic mass, $s_{\rm meas}$, come from the uncertainty
in the energy scale for reconstructed photons and the uncertainty in
the momentum scale for tracks, while the wrong choice of the
regularization parameter~$\rho$ in the unfolding can distort the
unfolded distributions.

The energy resolution can be tested by measuring the invariant mass of
the two photons from \Pgpz\ decays. A systematic shift in the observed
mass in the data compared to the detector simulation can be translated
into a scale factor for the reconstructed photon energies.  Deviations
of $(0.5 \pm 0.9)\,{\rm MeV}$ for $m_\Pgpz$ have been observed between
data and Monte Carlo~(figure \ref{fig:pi0}). This corresponds to an
energy scale factor of $1.004 \pm 0.007$. The energies of the
reconstructed photons in the Monte Carlo samples are varied by $\pm
0.7\,\%$ in order to estimate the systematic error due to this
effect\footnote{Since the invariant two-photon mass depends also on
  the angle between the two photons, this energy scale factor accounts
  for systematic uncertainties in the energy resolution and the
  angular resolution of the ECAL.}.

The uncertainty in the momentum of the tracks have been studied using
\Pgm\ pairs.  The Monte Carlo is corrected for observed deviations
between data and Monte Carlo in the mean and the width of the momentum
distribution.  The momenta and the momentum resolution of all tracks
in the Monte Carlo samples are scaled due to the uncertainties in
these corrections, thus leading to the quoted systematic errors.

The damping parameter $\rho$ in the regularization step of the
unfolding procedure is calculated from the number of effectively
remaining spline coefficients after the regularization.  This number
is chosen so that the test conditions a) and b) given in
section~\ref{sec:unfolding} are satisfied.  The default value ($16$
effective splines from $48$ total splines for the 1-prong fit and $16$
effective splines from $36$ total splines for the 3-prong fit) is
varied by $\pm 4$ for both fits, where the range is derived from Monte
Carlo tests of the unfolding procedure:
the tests consist of unfolding fake data samples in the
$\Pgr\rightarrow\Pgp\Pgpz$ channel.
The mass and the width of the $\Pgr$ in the fake data samples at the
generator level are different from the values used in the standard Monte
Carlo which is used to create the response matrix. 
The allowed range of the damping parameter is then determined by
comparing the unfolded fake data samples with their generator level 
distributions for different choices of the damping parameter, for which 
the test conditions a) and b) are satisfied.
Within this range the unfolded distributions reproduce the 
mass spectrum of the modified $\Pgr$ without biases towards the 
generator distribution of the standard Monte Carlo. 
The uncertainty due to the variation of the damping parameter is
added as a systematic error on the unfolded results.

Uncertainties of statistical nature from the errors on the branching
ratios (see table~\ref{tab:branching}), the limited statistics of
signal and background Monte Carlo samples, and on the efficiencies are
incorporated in the unfolding procedure by adding them in quadrature
to the statistical errors on the data.

Systematic effects related to photon and $\Pgpz$ detection efficiency
are largely covered by the systematic errors.
\begin{table}[htb]
  \begin{center}
    \begin{tabular}{|c|c|c|c|c|}
      \hline
       $\Pgt \rightarrow \Pgngt {\rm X}$ & 
       $B [\%]$ & $w_{\rm V}$ & $w_{\rm A}$ & comment\\
      \hline
      $\Pe \Pgne$       & $17.83 \pm 0.08$ & --    & --      & \\
      $\Pgm \Pgngm$     & $17.35 \pm 0.10$ & --    & --      & \\
      $\Pgp \Pgpz$      & $25.24 \pm 0.16$ & $1.0$ & $0.0$   & \\
      $3\Pgp \Pgpz$     & $\p{0}4.26 \pm 0.09$ & $1.0$ & $0.0$   & 
      including $\Pgo \Pgp$ and $\Pgo \Pgp \Pgpz$\\
      $\Pgp 3\Pgpz$     & $\p{0}1.14 \pm 0.14$ & $1.0$ & $0.0$   & \\
      $\Pgp$            & $11.31 \pm 0.15$ & $0.0$ & $1.0$   & \\
      $3\Pgp$           & $\p{0}9.26  \pm 0.12$ & $0.0$ & $1.0$   & 
      $3\Ph - 2\PK \Pgp - \PK 2\Pgp$ including $\Pgo \Pgp$\\
      $\Pgp 2\Pgpz$     & $\p{0}9.27  \pm 0.14$ & $0.0$ & $1.0$   & \\  
      $3\Pgp 2\Pgpz$    & $\p{0}0.50  \pm 0.05$ & $0.0$ & $1.0$   & 
      including $\Pgo \Pgp \Pgpz$ and $\Pgh \Pgp \Pgpz$\\ 
      $5\Pgp$           & $\p{0}0.075 \pm 0.007$& $0.0$ & $1.0$   & MC \\
      $\Pgp 4\Pgpz$     & $\p{0}0.12  \pm 0.06$ & $0.0$ & $1.0$   & MC \\
      $3\Pgp 3\Pgpz$    & $\p{0}0.11  \pm 0.06$ & $1.0$ & $0.0$   & MC \\
      $5\Pgp \Pgpz$     & $\p{0}0.022 \pm 0.005$& $1.0$ & $0.0$   & MC \\
      $\PK \PKz$        & $\p{0}0.16  \pm 0.03$ & $1.0$ & $0.0$   & MC \\
      $2\PK \Pgp$       & $\p{0}0.10  \pm 0.03$ & $0.5 \pm 0.5$   & 
      $0.5 \pm 0.5$     & MC \\ 
      $2\PKz \Pgp$      & $\p{0}0.10  \pm 0.02$ & $0.5 \pm 0.5$   & 
      $0.5 \pm 0.5$     & MC \\ 
      $\PK \PKz \Pgpz$  & $\p{0}0.14  \pm 0.03$ & $0.5 \pm 0.5$   & 
      $0.5 \pm 0.5$     & MC \\ 
      $\Pgo \Pgp$       & $\p{0}0.21  \pm 0.01$ & $1.0$ & $-0.2$  & 
      MC excluding $3\Pgp \Pgpz$ \\ 
      $\Pgo \Pgp \Pgpz$ & $\p{0}0.046  \pm 0.007$ & $-0.25$ & $1.0$ & 
      excluding $3\Pgp 2\Pgpz$ \\ 
      $\Pgh \Pgp \Pgpz$ & $\p{0}0.17  \pm 0.03$ & $1.0$ & $-0.24$ & MC \\
      ${\rm X}_{\rm strange}$ & $\p{0}2.67 \pm 0.14$ & --    & --      & \\
      \hline
    \end{tabular}
    \caption{\em Branching ratios for the hadron modes and lepton
      channels. Shown are the fitted values from the Particle Data
      Group~\cite{art:PDG96} and the contributing weights for the
      vector and axial-vector current. Channels marked with MC are \lq
      generator-level\rq\ Monte Carlo channels included in the
      spectra. Negative weights are used to subtract inclusively
      measured contributions from the wrong current.}
    \label{tab:branching}
  \end{center}
\end{table}


\section{Results}\label{sec:results}
\subsection{Moments of \boldmath $R_\tau$}
\label{subsec:Moments}
The unfolded spectra of the hadronic modes shown in
figure~\ref{fig:unfQ2} are normalized to their branching fractions and
summed up to the vector and axial-vector spectra with their
appropriate weights:
\begin{equation}
\label{eq:rkldata}      
R_{\Pgt,{\rm V/A}}^{kl}(s_0) = 
        \int\limits_0^{s_0} \rd s
        \left(1 - \frac{s}{s_0}\right)^k
        \left(\frac{s}{m_\Pgt^2}\right)^l \sum_{\Ph_{\rm V/A}}
        \frac{B(\Pgt \rightarrow \Ph_{\rm V/A} \Pgngt)}{
              B(\Pgt \rightarrow \Pe \Pgne \Pgngt)}
        \frac{w_{\rm V/A}}{N_{\rm V/A}}\frac{\rd N_{\rm V/A}}{\rd s},
\end{equation}
where $N_{\rm V/A}$ is the number of taus that decay into the hadron
$\Ph_{\rm V/A}$ plus neutrino, and $w_{\rm V/A}$ denotes the
appropriate weight of the hadronic mode to the vector or axial-vector
current.  The branching ratios of the hadronic modes (and the lepton
channels), together with their contributing weights for the vector and
axial-vector spectra, are summarized in table~\ref{tab:branching}.

The hadronic modes $\Pgo\Pgp$, $\Pgo\Pgp\Pgpz$ and $\Pgh\Pgp\Pgpz$
involve decays of \Pgo's and \Pgh's, and do not conserve isospin
symmetry, since their decay can occur via the electromagnetic
interaction. Therefore, the unfolded distributions in the $3\Pgp$
mode, which is considered to belong to the axial-vector current, and
in the $3\Pgp\Pgpz$ mode, which belongs to the vector current, are
contaminated by decays not belonging to the assigned currents (e.g.
$\Pgo\Pgp \rightarrow 3\Pgp$), and thus need to be corrected.
Since~$\sim 71\,\%$ of the $3\Pgp 2\Pgpz$ mode consist of $\Pgo \Pgp
\Pgpz$ decays, this channel is used for the $\Pgo \Pgp \Pgpz$
corrections.  Corrections for the other \Pgo\ and \Pgh\ modes are made
with the Monte Carlo.  Decay modes which are not reconstructed from
the data have also to be included in the total vector and axial-vector
spectra.  Their distributions are taken from the Monte Carlo. The
$2\PK \Pgp$ modes contribute to both classes to an unknown amount.  A
weight of $(50 \pm 50)\,\%$ is used for both currents and a
correlation of~$-100\,\%$ between the vector and the axial-vector
weights is assumed.  The errors assigned to Monte Carlo spectra are
taken to be $\pm 100\,\%$, in order to take a possible mismodelling of
the Monte Carlo into account.
\begin{table}[htb]
  \begin{center}
    \begin{tabular}{|c|ll|ll|}
      \hline
      moment $kl$ & $R_{\rm V}^{kl}$ & total error &
                    $R_{\rm A}^{kl}$ & total error \\
      \hline
      $00$ & $\RTVaa$ & $\RTVEXaa$  & 
             $\RTAaa$ & $\RTAEXaa$ \\
      $10$ & $\RTVba$ & $\RTVEXba$  & 
             $\RTAba$ & $\RTAEXba$ \\
      $11$ & $\RTVbb$ & $\RTVEXbb$  & 
             $\RTAbb$ & $\RTAEXbb$ \\
      $12$ & $\RTVbc$ & $\RTVEXbc$  & 
             $\RTAbc$ & $\RTAEXbc$ \\
      $13$ & $\RTVbd$ & $\RTVEXbd$  & 
             $\RTAbd$ & $\RTAEXbd$ \\
      \hline            
    \end{tabular}
    \caption{\em The measured moments $\mathit{R_{V/A}^{kl}}$, for 
      $\mathit{kl = 00, 10, 11, 12, 13}$. The errors shown represent
      statistical and systematic uncertainties.}
    \label{tab:rkl}
  \end{center}
\end{table}
\begin{table}[htb]
  \begin{center}
    \begin{tabular}{|cc|ll|llll|}
      \hline
    &      & & & 
             \multicolumn{4}{c|}{systematic errors}\\
    & $kl$ & data stat.& branching ratios & MC~stat.    & 
                         $E$ scale   & $p$ scale   & 
                         regularization \\  
      \hline
        &  $00$ &\hfil--\hfil& $\RTVBRaa$ &\hfil--\hfil& 
                 \hfil--\hfil&\hfil--\hfil&\hfil--\hfil\\ 
        &  $10$ & $\RTVDAba$ & $\RTVBRba$ & $\RTVMCba$ & 
                  $\RTVESba$ & $\RTVPSba$ & $\RTVRGba$ \\
$\rm V$ &  $11$ & $\RTVDAbb$ & $\RTVBRbb$ & $\RTVMCbb$ &  
                  $\RTVESbb$ & $\RTVPSbb$ & $\RTVRGbb$ \\
        &  $12$ & $\RTVDAbc$ & $\RTVBRbc$ & $\RTVMCbc$ &  
                  $\RTVESbc$ & $\RTVPSbc$ & $\RTVRGbc$ \\
        &  $13$ & $\RTVDAbd$ & $\RTVBRbd$ & $\RTVMCbd$ &  
                  $\RTVESbd$ & $\RTVPSbd$ & $\RTVRGbd$ \\
      \hline            
        &  $00$ &\hfil--\hfil& $\RTABRaa$ &\hfil--\hfil& 
                 \hfil--\hfil&\hfil--\hfil&\hfil--\hfil\\ 
        &  $10$ & $\RTADAba$ & $\RTABRba$ & $\RTAMCba$ & 
                  $\RTAESba$ & $\RTAPSba$ & $\RTARGba$ \\
$\rm A$ &  $11$ & $\RTADAbb$ & $\RTABRbb$ & $\RTAMCbb$ &  
                  $\RTAESbb$ & $\RTAPSbb$ & $\RTARGbb$ \\
        &  $12$ & $\RTADAbc$ & $\RTABRbc$ & $\RTAMCbc$ &  
                  $\RTAESbc$ & $\RTAPSbc$ & $\RTARGbc$ \\
        &  $13$ & $\RTADAbd$ & $\RTABRbd$ & $\RTAMCbd$ &  
                  $\RTAESbd$ & $\RTAPSbd$ & $\RTARGbd$ \\
      \hline            
    \end{tabular}
    \caption{\em Statistical and systematic uncertainties of the measured
      moments. The upper (lower) portion of the table contains the
      result for the vector (axial-vector) current.}
    \label{tab:rklerr}
  \end{center}
\end{table}
\begin{table}[htb]
  \begin{center}
    \begin{tabular}[t]{|c|cccc|}
      \hline
      $\rm V$  & $00$       & $10$       & $11$       & $12$          \\
      \hline                           
      $10$     & $\CVaaVba$ &            &            &               \\
      $11$     & $\CVaaVbb$ & $\CVbaVbb$ &            &               \\
      $12$     & $\CVaaVbc$ & $\CVbaVbc$ & $\CVbbVbc$ &               \\
      $13$     & $\CVaaVbd$ & $\CVbaVbd$ & $\CVbbVbd$ & $\CVbcVbd$    \\
      \hline            
    \end{tabular}\hfil
    \begin{tabular}[t]{|c|rrrrr|}
      \hline
      ${\rm A} \backslash {\rm V}$
           & $ 00$      & $ 10$      & $ 11$      & $ 12$      & $ 13$     \\
      \hline                                      
      $00$ & $\CVaaAaa$ & $\CVbaAaa$ & $\CVbbAaa$ & $\CVbcAaa$ & $\CVbdAaa$\\
      $10$ & $\CVaaAba$ & $\CVbaAba$ & $\CVbbAba$ & $\CVbcAba$ & $\CVbdAba$\\
      $11$ & $\CVaaAbb$ & $\CVbaAbb$ & $\CVbbAbb$ & $\CVbcAbb$ & $\CVbdAbb$\\
      $12$ & $\CVaaAbc$ & $\CVbaAbc$ & $\CVbbAbc$ & $\CVbcAbc$ & $\CVbdAbc$\\
      $13$ & $\CVaaAbd$ & $\CVbaAbd$ & $\CVbbAbd$ & $\CVbcAbd$ & $\CVbdAbd$\\
      \hline            
    \end{tabular}\hfil
    \begin{tabular}[t]{|c|cccc|}
      \hline
      $\rm A$  & $00$       & $10$       & $11$       & $12$          \\
      \hline
      $10$     & $\CAaaAba$ &            &            &               \\
      $11$     & $\CAaaAbb$ & $\CAbaAbb$ &            &               \\
      $12$     & $\CAaaAbc$ & $\CAbaAbc$ & $\CAbbAbc$ &               \\
      $13$     & $\CAaaAbd$ & $\CAbaAbd$ & $\CAbbAbd$ & $\CAbcAbd$    \\
      \hline            
    \end{tabular}
    \caption{\em Correlations between the measured moments 
      $\mathit{R_{V/A}^{kl}}$ in percent. The left (right) table gives
      the correlations between the moments of the vector (axial-vetor)
      current; the table in the middle shows the correlations between
      the moments of different currents.}
    \label{tab:rklcorr}
  \end{center}
\end{table}

The moments $R_{\rm V/A}^{kl}$ are given in table~\ref{tab:rkl}.  The
errors on the moments are subdivided into statistical uncertainties
due to the data statistics, the uncertainties comming from the
branching ratio errors, and systematic uncertainties induced by the
limited Monte Carlo statistics and the variations of the energy scale,
the momentum scale, and the regularization parameter
(table~\ref{tab:rklerr}).  Correlations between the moments are given
in table~\ref{tab:rklcorr}.


\subsection{Spectral functions}
\label{subsec:spectral}
The vector and axial-vector spectral functions are given by inverting
equation~(\ref{eq:dRds}):
\begin{eqnarray}
        v/a (s) & = & 2\pi\,{\rm Im} {\mit\Pi}_{\rm V/A}^{(1)}(s) 
       \nonumber \\ 
       & = & 
        m_\Pgt^2 \left[ 6 S_{\rm EW} |V_{\rm ud}|^2 
        \left(1 - \frac{s}{m_\Pgt^2}\right)^2
        \left(1 + 2\frac{s}{m_\Pgt^2}\right)
        \right]^{-1} \nonumber \\
        & & \times \sum_{\Ph_{\rm V/A}} 
        \frac{B(\Pgt \rightarrow \Ph_{\rm V/A} \Pgngt)}{
              B(\Pgt \rightarrow \Pe \Pgne \Pgngt)}
        \frac{w_{\rm V/A}}{N_{\rm V/A}}
        \frac{\rd N_{\rm V/A}}{\rd s},
       \label{eq:spectral} 
\end{eqnarray}
where the sum is performed over hadronic final states $h_{\rm V/A}$
with angular momentum $J=1$.

\begin{figure}[htbp]
\centering
\resizebox{0.49\textwidth}{!}{%
\includegraphics{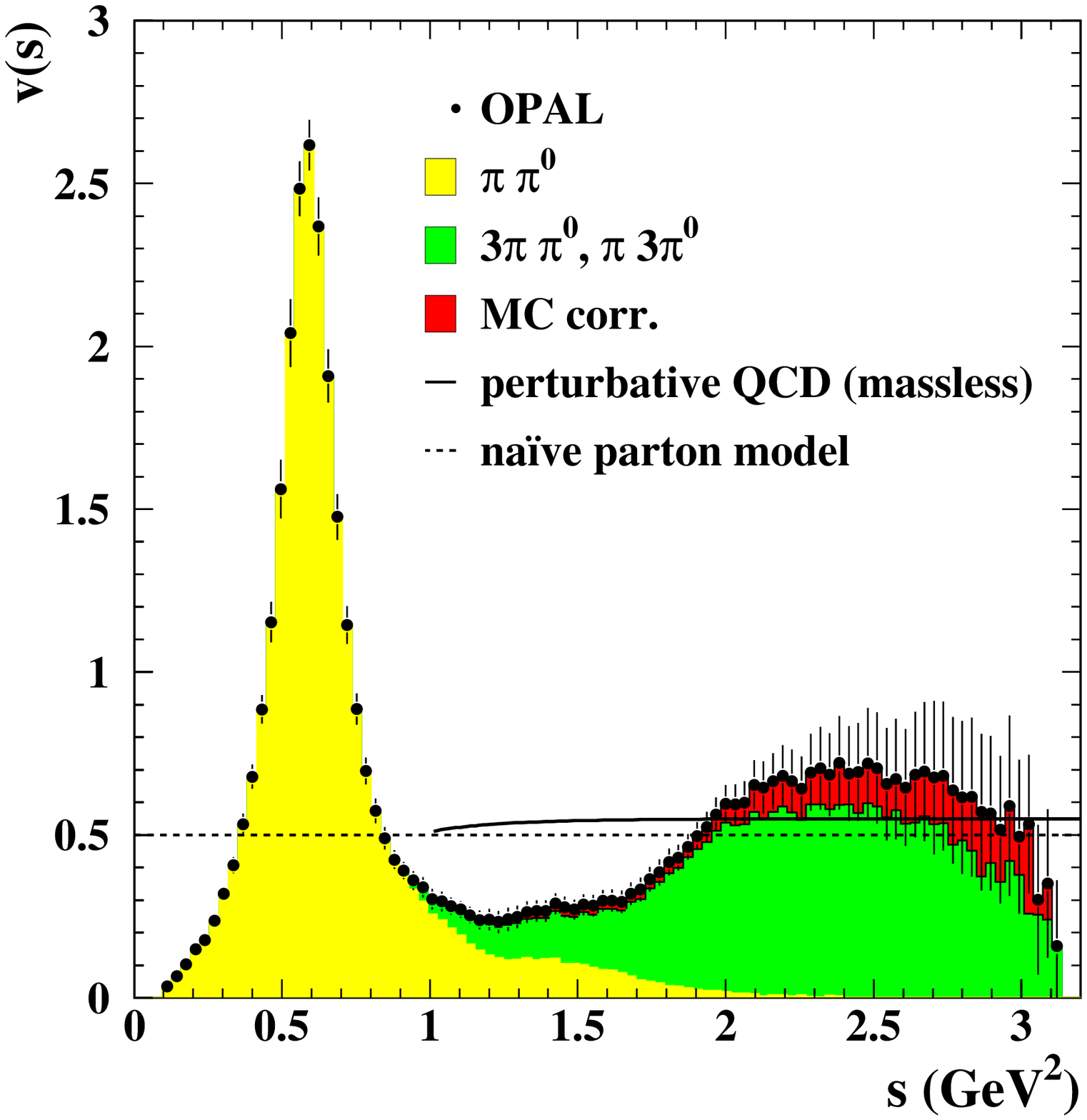}}
\resizebox{0.49\textwidth}{!}{%
\includegraphics{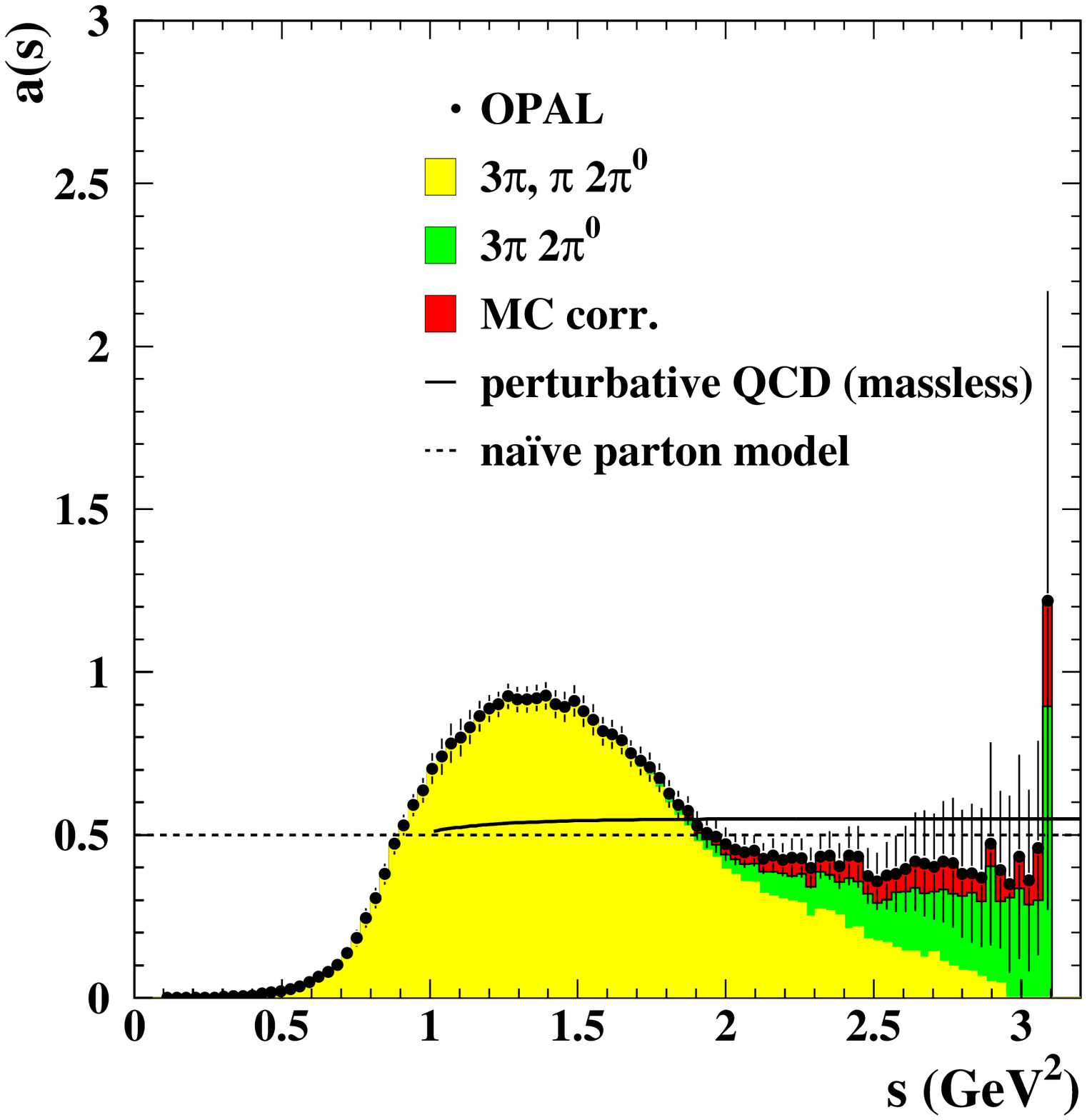}}
\resizebox{0.666\textwidth}{!}{%
\includegraphics{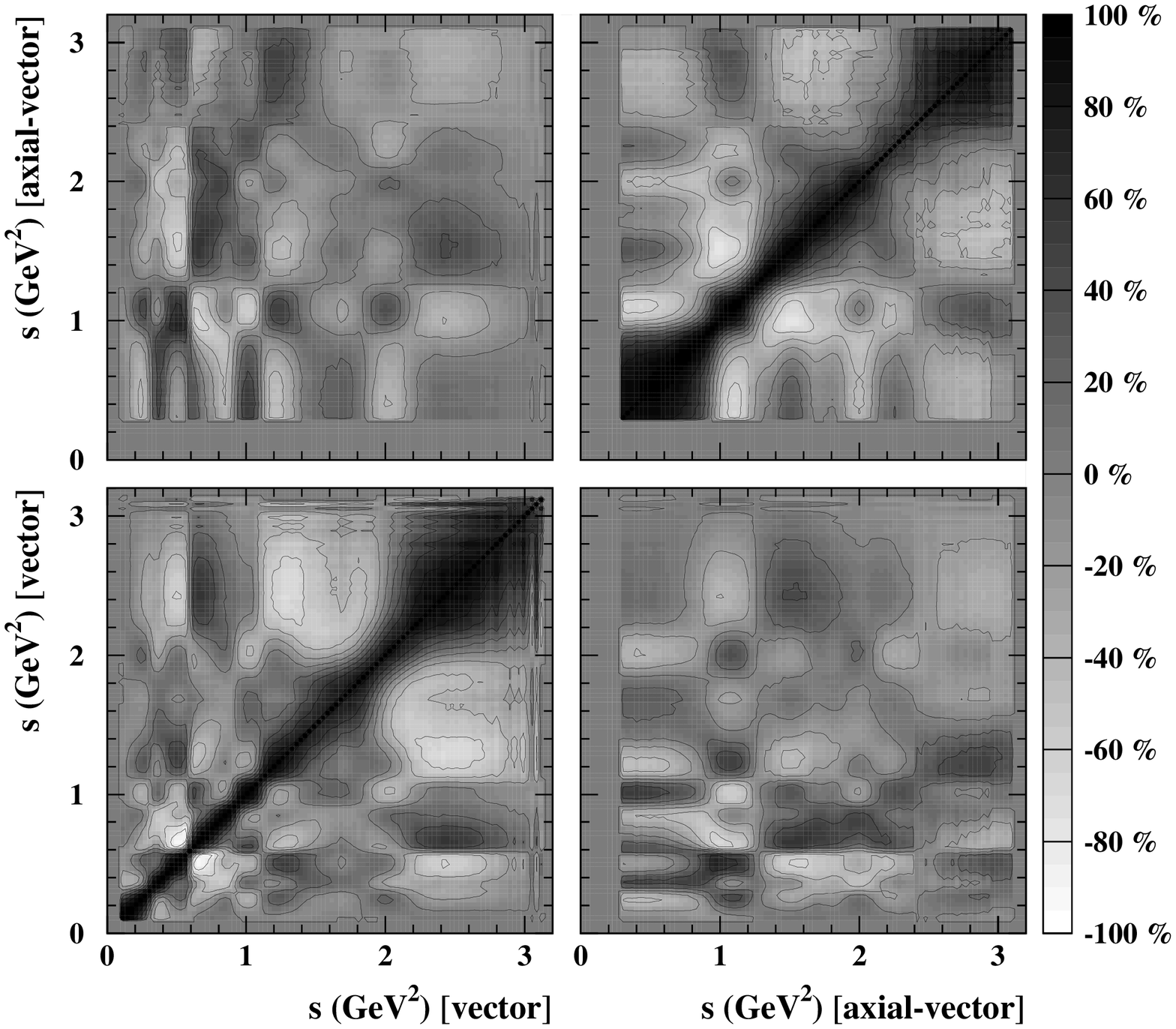}}
\caption{\em 
  The vector and axial-vector spectral functions. Shown are the sums
  of all contributing channels as data points (upper two plots).  Some
  exclusive contributions are shown as shaded areas. The na{\"\i}ve parton
  model prediction is shown as dashed line, while the solid line
  depicts the perturbative, massless QCD prediction for
  $\mathit{\alpha_s(\mzsq) = 0.122}$.  The error bars include
  statistical and systematic uncertainties. The pion pole is
  subtracted from the axial-vector spectrum. The lower plot shows the
  correlations of the two spectral functions in continuous gray-levels
  from white to black which correspond to the correlations in percent
  from $\mathit{-100\,\%}$ to $\mathit{+100\,\%}$. The contour lines
  are drawn in equidistant steps of $\mathit{20\,\%}$.}
\label{fig:spectral}
\end{figure}

\begin{figure}[htbp]
\centering
\resizebox{0.49\textwidth}{!}{%
\includegraphics{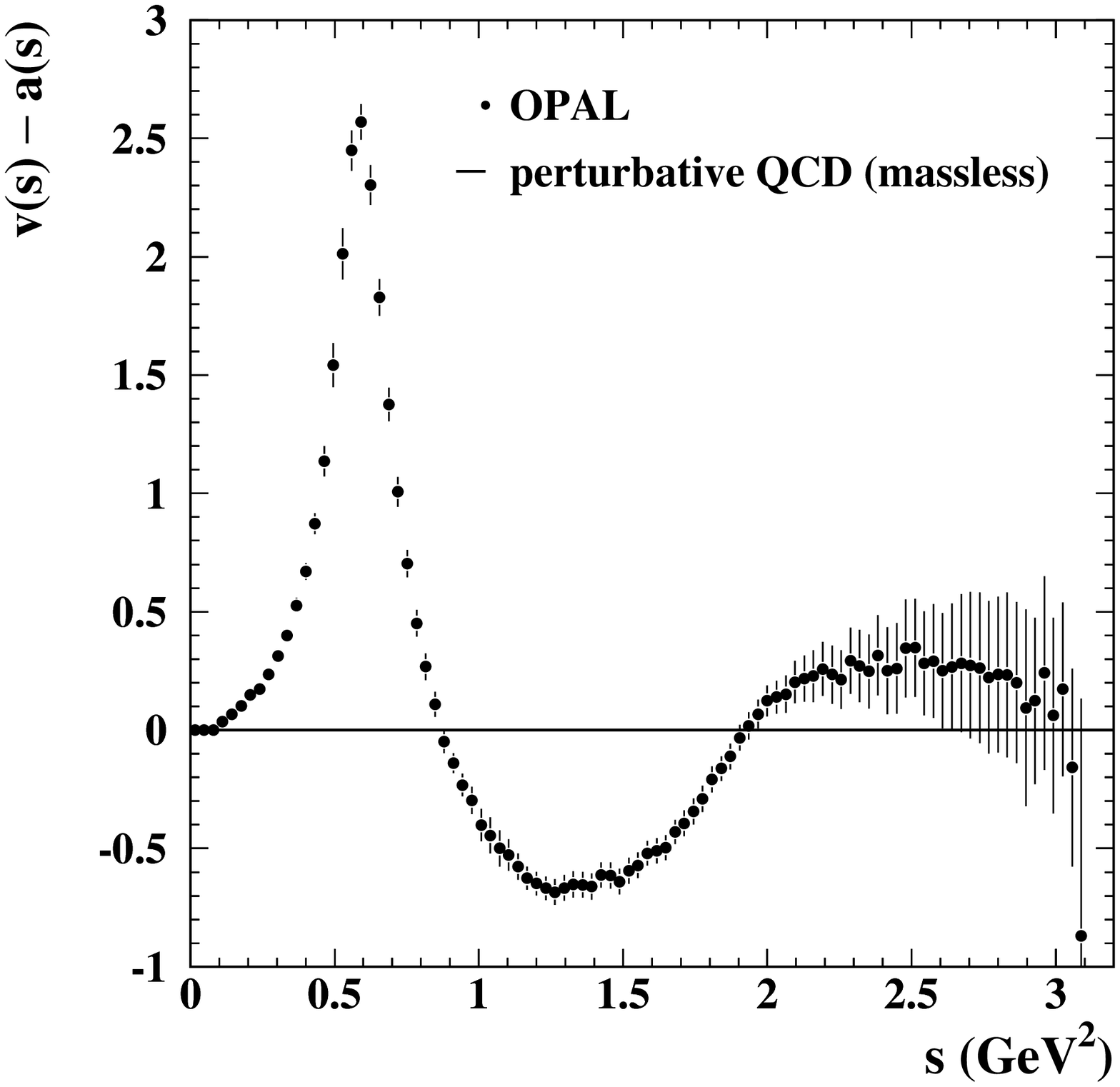}}
\resizebox{0.49\textwidth}{!}{%
\includegraphics{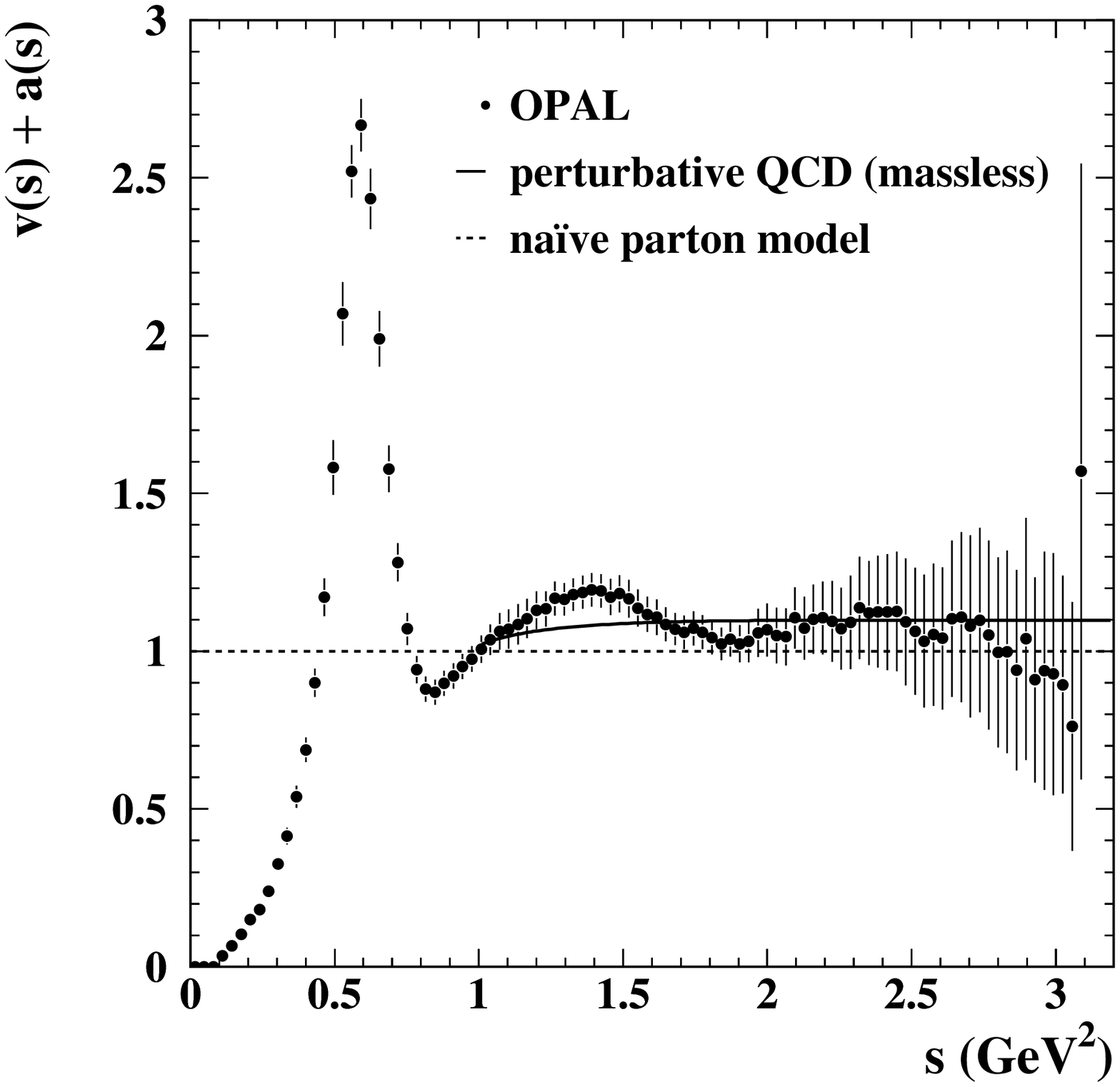}}
\caption{\em The difference (sum) of the spectral functions 
  $\mathit{v(s)-a(s)}$ ($\mathit{v(s)+a(s)}$). The dashed line is the
  na{\"\i}ve parton model expectation and the solid lines depict the
  prediction of massless, perturbative QCD as in
  figure~\ref{fig:spectral}. For $\mathit{v(s)-a(s)}$ both predictions
  are identically zero.}
\label{fig:v+-a}
\end{figure}

The spectral functions (and their correlations) are shown in
figure~\ref{fig:spectral} together with the flat na{\"\i}ve parton model
prediction $v_{\rm na\ddot{\im}ve}(s) = a_{\rm na\ddot{\im}ve}(s) =
1/2$ and the prediction of perturbative QCD (massless) for
$\alphas(\mzsq) = 0.122$ which increases the na{\"\i}ve prediction by
$\approx 10\,\%$.  As a result of the regularized unfolding, the
bin-to-bin correlations are of the order of $+80\,\%$ ($-50\,\%$) for
bin distances of $0.1\,{\rm GeV^2}$ ($\approx 1\,{\rm GeV}^2$). The
correlation between vector and axial-vector spectral function varies
from $-60\,\%$ to $+60\,\%$.
Figure~\ref{fig:v+-a} shows the difference and the sum of the two
measured spectral functions. The function $v(s)-a(s)$ should vanish in
the limit of perturbative, massless QCD. The deviation from this
prediction, e.g. due to the \Pgr\ and \Pai\ resonances, indicates the
large sensitivity of this distribution to non-perturbative effects.
The QCD prediction for $v(s)+a(s)$ which is $\approx 10\,\%$ above the
na{\"\i}ve expectation $v(s)+a(s) = 1$ as in figure~\ref{fig:spectral}
gives a reasonable description of the region $s > 1\,{\rm GeV}^2$. The
structure due to the narrow resonances in the region below $s \simeq
1\,{\rm GeV}^2$ is however not described by perturbative QCD.


\section{Measurement of the strong coupling \balphas}\label{sec:alphas}
Since the perturbative expansions for vector and axial-vector currents
are identical while the non-perturbative parts have opposite sign but
the same order of magnitude for both currents, two different fits are
used for the extraction of \alphas\ and the power corrections,
respectively.  The sum of vector and axial-vector moments is most
sensitive to perturbative QCD and is used for the measurement of
\alphas\ (fit 1) while the separate moments of both currents are used
to obtain the power corrections (fit 2).  In addition to the moments
listed in table~\ref{tab:rkl} it is possible to include the
measurements of the \Pgt\ lifetime $\tau_\Pgt$ and the branching ratio
$B_\Pgm = B(\Pgt \rightarrow \Pgm \Pgngm \Pgngt)$ in fit 1 since each
of them can be used to predict the total hadronic decay rate of the
\Pgt\ lepton:
\begin{eqnarray}
  R_\Pgt(\tau_\Pgt) & = &\frac{1}{\Gamma_\Pe}\frac{1}{\tau_\Pgt} - 1 - 
           \frac{\Gamma_\Pgm}{\Gamma_\Pe}\label{eq:tautau},\\
  R_\Pgt(B_\Pgm)   & = &\frac{\Gamma_\Pgm}{\Gamma_\Pe}\frac{1}{B_\Pgm} - 1 - 
           \frac{\Gamma_\Pgm}{\Gamma_\Pe}\label{eq:bmu}.
\end{eqnarray}
Both equations assume lepton universality so that the following
equation holds:
\begin{equation}
  \label{eq:universality}
  B_\Pgm = B_\Pe \frac{\Gamma_\Pgm}{\Gamma_\Pe},
\end{equation}
with $\frac{\Gamma_\Pgm}{\Gamma_\Pe} = 0.9726$~\cite{art:Marciano} and
$\Gamma_\Pe = 4.0329 \cdot 10^{-13}\,{\rm
  GeV}$~\cite{art:Tsai,art:Marciano}.  The non-strange decay rate of
the \Pgt\ lepton is then obtained by subtracting $R_{\Pgt,{\rm s}} =
B_{\rm s} / B_\Pe = 0.150 \pm 0.008$~\cite{art:PDG96} from the
weighted average $R_\Pgt(B_\Pgm,\tau_\Pgt)$ of $R_\Pgt(\tau_\Pgt)$ and
$R_\Pgt(B_\Pgm)$\footnote{ $R_{\Pgt,{\rm s}}$ is subtracted from
  $R_\Pgt$, since the induced dependency on the mass of the strange
  quark would lead to a larger uncertainty in the fits if $R_\Pgt$
  would be used instead.}.  In principle the electron branching ratio
$B_\Pe$ could also be used to determine $R_\Pgt$ but this has a
$96\,\%$ correlation with $R_\Pgt$ from the hadronic modes due to the
correlations of the constrained branching ratios in~\cite{art:PDG96}.

Using the world average $\tau_\Pgt = (291.0 \pm 1.5)\,{\rm fs}$ and
the fitted value $B_\Pgm = 0.1735 \pm 0.0010$~\cite{art:PDG96} one
gets:
\begin{equation}
  \label{eq:rvpaPDG96}
  R_\Pgt(B_\Pgm,\tau_\Pgt) - R_{\Pgt,{\rm s}} =  3.485 \pm  0.023.
\end{equation}
From the vector and axial-vector decay rates in table~\ref{tab:rkl}
one gets the following value:
\begin{equation}
  \label{eq:rv+ra}
  R_{\Pgt,{\rm V}} + R_{\Pgt,{\rm A}} =  3.484 \pm  0.024.
\end{equation}

In the first fit, four parameters are used to describe the five
moments, leaving one degree of freedom for the fit: the strong
coupling~$\alphas(m_\Pgt^2)$, the gluon
condensate~$\langle\frac{\alphas}{\pi}\,GG\rangle$ and the dimension 6
and 8 contributions to the $kl=00$ moments $\delta_{\rm V+A}^6$,
$\delta_{\rm V+A}^8$.  The second fit requires six parameters to
predict ten moments (four degrees of freedom): $\alphas(m_\Pgt^2)$,
$\langle\frac{\alphas}{\pi}\,GG\rangle$ and the power corrections
$\delta_{\rm V}^6$, $\delta_{\rm V}^8$, $\delta_{\rm A}^6$ and
$\delta_{\rm A}^8$.  The power corrections from the two fits can be
compared via the following relation:
\begin{equation}
\label{eq:average}
\delta_{\rm V+A}^D = \frac{1}{2} \left( \delta_{\rm V}^D + 
  \delta_{\rm A}^D \right).
\end{equation}
Further inputs for both fits are the quark masses for the three light
quarks~$m_{\rm u,d,s}$
\begin{equation}
  \label{eq:quarkmass} m_{\rm u} = (8.7 \pm 1.5)\,{\rm MeV},\quad m_{\rm d} =
  (15.4 \pm 1.5)\,{\rm MeV},\quad m_{\rm s} = (270 \pm 30)\,{\rm MeV},
\end{equation}
and the quark condensates $\langle\overline{\psi}\psi\rangle_{\rm
  u,d,s} = -\mu_{\rm u,d,s}^3$, with
\begin{equation}
  \label{eq:condensates} \mu_{\rm u} = \mu_{\rm d} = (189 \pm 7)\,{\rm MeV},
  \quad \mu_{\rm s} = (160 \pm 10)\,{\rm MeV}.
\end{equation}
The values are taken from~\cite{art:Braaten}.  The error matrix of the
moments is calculated from the experimental errors on the moments and
their correlations (tables~\ref{tab:rkl}\ and \ref{tab:rklcorr}) and
the theoretical error matrix calculated from the errors on the
quark-masses and quark-condensates.  The results from the fit to the
sum of vector and axial-vector moments is given in
table~\ref{tab:V+A_fit}.  The quoted errors are subdivided into a
statistical error due to the data statistics, the uncertainty induced
by the errors on the branching ratios, an experimental systematic
error from the Monte Carlo statistics and the unfolding procedure, and
a theoretical error including the uncertainties on quark masses, the
variation of the $\Order(\alphas^4)$ coefficient, the renormalization
scheme dependence, and the renormalization scale uncertainty.  The
strong coupling is most sensitive to the $kl=00$ moment, and therefore
the dominant contribution to the experimental uncertainty on $\alphas$
comes from the uncertainties on the branching ratios.
\begin{table}[htb]
  \begin{center}
    \begin{tabular}[t]{|c|c|l|llll|c|}
      \hline
         & & & \multicolumn{4}{c|}{contributing errors} & \\
         theory     & observable & \p{+}value & \p{+}data & \p{+}$B$ &  
         \p{+}syst. & \p{+}theo. & $\chi^2/{\rm d.o.f.}$  \\       
      \hline               
\rule[-6pt]{0pt}{19pt} & $\alphas(m_\Pgt^2)$ & 
                         $\p{+}\CIPTASMT$    & 
                         $\pm \CIPTDAMT$     &
                         $\pm \CIPTBRMT$     &
                         $\pm \CIPTSMMT$     &
                         $\pm \CIPTTMMT$     & \\
\rule[-6pt]{0pt}{19pt} & $\langle\frac{\alphas}{\pi}\,GG\rangle/{\rm GeV^4}$ & 
                         $\CIPTGGVL$        & 
                         $\pm \CIPTGGDA$    &
                         $\pm \CIPTGGBR$    &
                         $\pm \CIPTGGSM$    &
                         $\pm \CIPTGGTM$    & \\
\rule[-6pt]{0pt}{19pt} \raisebox{8.5pt}[0pt][0pt]{CIPT} & $\delta_{\rm V+A}^6$&
                         $\p{+}\CIPTSSVL$   & 
                         $\pm \CIPTSSDA$    &
                         $\pm \CIPTSSBR$    &
                         $\pm \CIPTSSSM$    &
                         $\pm \CIPTSSTM$    & 
                         \raisebox{8.5pt}[0pt][0pt]{$\CIPTCSMT / 1$}    \\
\rule[-6pt]{0pt}{19pt} & $\delta_{\rm V+A}^8$   &
                         $\CIPTSEVL$        & 
                         $\pm \CIPTSEDA$    &
                         $\pm \CIPTSEBR$    &
                         $\pm \CIPTSESM$    &
                         $\pm \CIPTSETM$    & \\
\hline
\rule[-6pt]{0pt}{19pt} & $\alphas(m_\Pgt^2)$ & 
                         $\p{+}\FOPTASMT$    & 
                         $\pm \FOPTDAMT$     &
                         $\pm \FOPTBRMT$     &
                         $\pm \FOPTSMMT$     &
                         $\pm \FOPTTMMT$     & \\
\rule[-6pt]{0pt}{19pt} & $\langle\frac{\alphas}{\pi}\,GG\rangle/{\rm GeV^4}$ & 
                         $\p{+}\FOPTGGVL$   & 
                         $\pm \FOPTGGDA$    &
                         $\pm \FOPTGGBR$    &
                         $\pm \FOPTGGSM$    &
                         $\pm \FOPTGGTM$    & \\
\rule[-6pt]{0pt}{19pt} \raisebox{8.5pt}[0pt][0pt]{FOPT} & $\delta_{\rm V+A}^6$&
                         $\p{+}\FOPTSSVL$        & 
                         $\pm \FOPTSSDA$    &
                         $\pm \FOPTSSBR$    &
                         $\pm \FOPTSSSM$    &
                         $\pm \FOPTSSTM$    & 
                         \raisebox{8.5pt}[0pt][0pt]{$\FOPTCSMT / 1$}    \\
\rule[-6pt]{0pt}{19pt} & $\delta_{\rm V+A}^8$   &
                         $\FOPTSEVL$        & 
                         $\pm \FOPTSEDA$    &
                         $\pm \FOPTSEBR$    &
                         $\pm \FOPTSESM$    &
                         $\pm \FOPTSETM$    & \\
\hline
\rule[-6pt]{0pt}{19pt} & $\alphas(m_\Pgt^2)$ & 
                         $\p{+}\RCPTASMT$   & 
                         $\pm \RCPTDAMT$    &
                         $\pm \RCPTBRMT$    &
                         $\pm \RCPTSMMT$    &
                         $\pm \RCPTTMMT$    & \\
\rule[-6pt]{0pt}{19pt} & $\langle\frac{\alphas}{\pi}\,GG\rangle/{\rm GeV^4}$ & 
                         $\RCPTGGVL$        & 
                         $\pm \RCPTGGDA$    &
                         $\pm \RCPTGGBR$    &
                         $\pm \RCPTGGSM$    &
                         $\pm \RCPTGGTM$    & \\
\rule[-6pt]{0pt}{19pt} \raisebox{8.5pt}[0pt][0pt]{RCPT} & $\delta_{\rm V+A}^6$&
                         $\RCPTSSVL$        & 
                         $\pm \RCPTSSDA$    &
                         $\pm \RCPTSSBR$    &
                         $\pm \RCPTSSSM$    &
                         $\pm \RCPTSSTM$    &
                         \raisebox{8.5pt}[0pt][0pt]{$\RCPTCSMT / 1$}    \\
\rule[-6pt]{0pt}{19pt} & $\delta_{\rm V+A}^8$   &
                         $\RCPTSEVL$        & 
                         $\pm \RCPTSEDA$    &
                         $\pm \RCPTSEBR$    &
                         $\pm \RCPTSESM$    &
                         $\pm \RCPTSETM$    & \\
\hline                       
    \end{tabular}
    \caption{\em The result for $\mathit{\alpha_s(m_\tau^2)}$ and the 
      non-perturbative parameters from the fit to the sum of vector
      and axial-vector moments. Shown are the values for the three
      different descriptions of the perturbative part of the moments
      (see text). The given errors correspond to the data statistics,
      the uncertainty due to the errors on the branching ratios~$B$, a
      systematic error from the Monte Carlo statistics, the energy
      scale, the momentum scale, and the unfolding, and a total
      theoretical uncertainty.}  \label{tab:V+A_fit} \end{center}
\end{table}                                

All three theories lead to similar $\chi^2$ values (see
table~\ref{tab:V+A_fit}) but the spread in the fitted values for
$\alphas(m_\Pgt^2)$ exceeds the total uncertainties by a factor of
two.  A similar spread of the values for $\alpha_s(m_\Pgt^2)$ from the
three models has also been observed
in~\cite{art:Girone,art:Andreas98}, where RCPT has led to the lowest
value and CIPT to the largest value in agreement with our results
(table~\ref{tab:V+A_fit}).

The differences in the statistical and systematic errors on $\alphas$
are induced by the scaling of the relative error with $\alphas$ and
thus are compatible for the three fits. The theoretical uncertainties
should also obey this scaling behavior: here the fits for FOPT and
CIPT only include the uncertainty on the unknown $K_4$ coefficient and
hence cannot be compared to the RCPT result. Furthermore the
uncertainty due to the variation of the renormalization scheme
vanishes for RCPT.  The impact on $\alphas$ from the various
theoretical error sources is listed in table~\ref{tab:theoerr}. The
given errors correspond to the spread of the fitted values of
$\alphas$ in fit 1 due to the unknown $\Order\left( \alphas^4 \right)$
dependence $K_4 = 25\pm50$, the choice of renormalization scale $0.4
\le \mu^2/m_\Pgt^2 \le 2.0$, the variation of the renormalization
scheme parameterized with the third coefficient of the
$\beta$-function $0.0 \le \beta_3^{\rm RS}/\beta_3^{\overline{\rm MS}}
\le 2.0$, and the evolution of $\alphas(m_\Pgt^2)$ to the \PZz-mass
scale.
\begin{table}[htbp]
  \begin{center}
    \begin{tabular}[t]{|c|ccc|ccc|}
      \hline
      \rule[-6pt]{0pt}{19pt}                           &
      \multicolumn{3}{c|}{$\Delta\alphas(m_\Pgt^2)$}   &
      \multicolumn{3}{c|}{$\Delta\alphas(\mzsq)$}      \\
      \rule[-6pt]{0pt}{19pt} error source              & 
      CIPT & FOPT & RCPT & CIPT & FOPT & RCPT          \\
      \hline               
      \rule[-6pt]{0pt}{19pt} $-25 \le K_4 \le 75$      & 
      $\pm\CIPTKFMT$ & $\pm\FOPTKFMT$ & --             & 
      $\pm\CIPTKFMZ$ & $\pm\FOPTKFMZ$ & --             \\ 
      \rule[-6pt]{0pt}{19pt} $0.4 \le \mu^2/m_\Pgt^2 \le 2.0$ & 
      $\pm\CIPTZEMT$ & $\pm\FOPTZEMT$ & $\pm\RCPTZEMT$ & 
      $\pm\CIPTZEMZ$ & $\pm\FOPTZEMZ$ & $\pm\RCPTZEMZ$ \\ 
      \rule[-6pt]{0pt}{19pt} $0.0 \le \beta_3^{\rm RS}/
      \beta_3^{\overline{\rm MS}} \le 2.0$             & 
      $\pm\CIPTRSMT$ & $\pm\FOPTRSMT$ & $\pm\RCPTRSMT$ & 
      $\pm\CIPTRSMZ$ & $\pm\FOPTRSMZ$ & $\pm\RCPTRSMZ$ \\ 
      \rule[-6pt]{0pt}{19pt} evolution                 & 
      --             & --             & --             &
      $\pm 0.0003$   & $\pm 0.0003$   & $\pm 0.0003$   \\
      \hline
    \end{tabular}
    \caption{\em The theoretical uncertainties on the strong coupling 
      constant.  The errors correspond to the full spread of the
      fitted $\mathit{\alpha_s}$ values in fit 1 due to the variation
      of the parameters listed in the first column.}
    \label{tab:theoerr}
  \end{center}                                
\end{table}

\begin{figure}[htbp]
\centering
\resizebox{\textwidth}{!}{%
\includegraphics{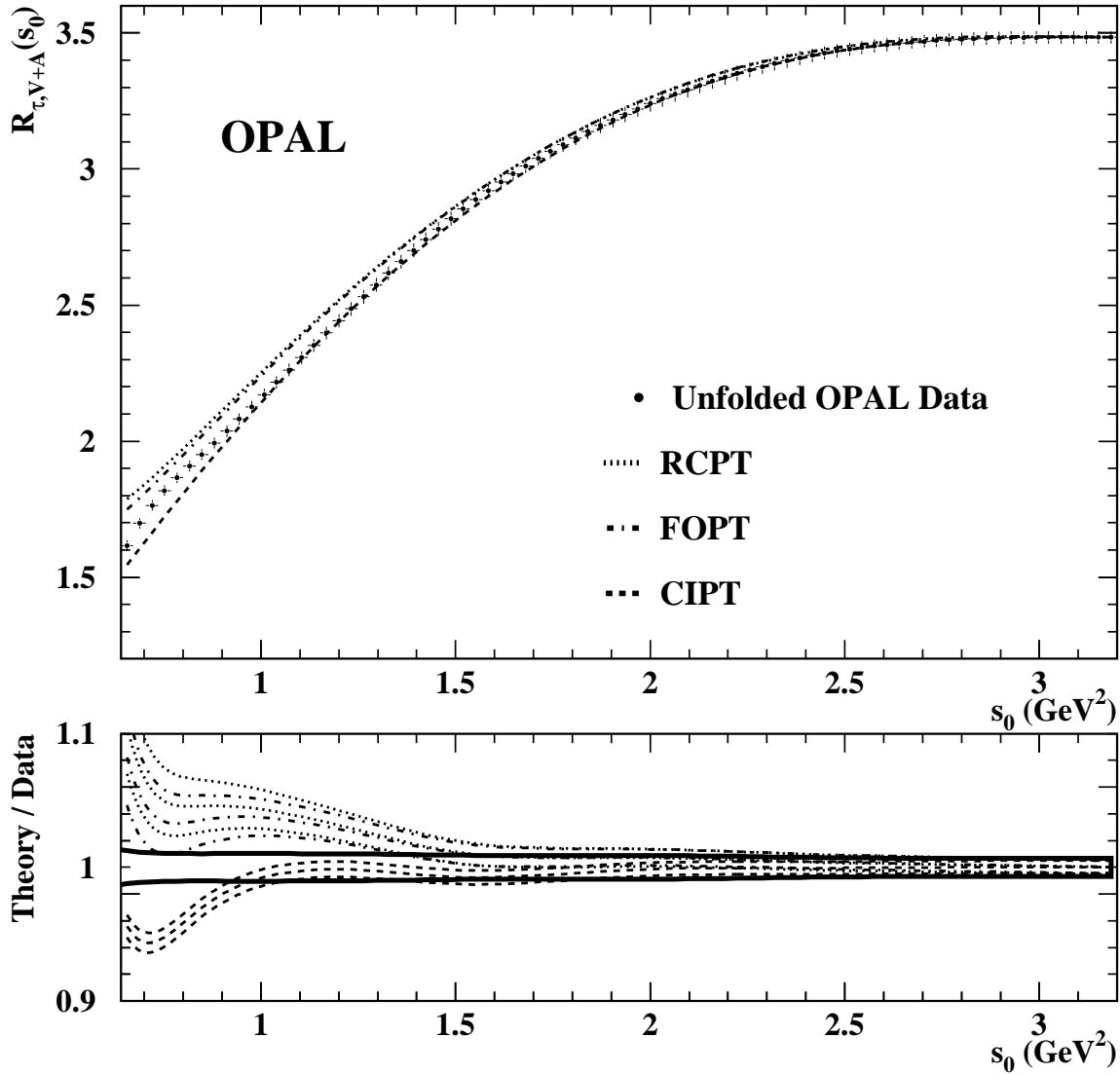}}
\caption{\em The non-strange hadronic decay rate of the $\mathit{\Pgt}$ 
  lepton $\mathit{R_{\Pgt,{V}}(s_0)+R_{\Pgt,{A}}(s_0)}$ versus the
  upper integration limit $s_0$.  The points in the upper plot denote
  OPAL data; the dashed, dashed-dotted and dotted curves represent the
  theoretical predictions of the three theories with the results from
  the fit to the moments at $\mathit{s_0 = m_\Pgt^2}$ used as input.
  The lower plot shows the three theories normalized to the data. The
  three sets of dashed, dashed-dotted and dotted curves indicate
  central values and total experimental errors for each of the three
  theories. The errors on the data are shown as solid curves.}
\label{fig:rvpa00}
\end{figure}

Although the total theoretical uncertainties on \alphas\ are
compatible for all three theories there is a major difference between
FOPT and the two other models: the FOPT fit leads to a significant
larger dependency of the non-perturbative parameters
$\langle\frac{\alphas}{\pi}GG\rangle$ and $\delta_{\rm V+A}^{6/8}$ on
the theoretical uncertainties than CIPT and RCPT.  The dominant effect
comes from the variation of the renormalization scale $\mu^2$.  The
statistical and systematic uncertainties on the power corrections are
very similar for all three theories, agreeing with expectation.
Figure~\ref{fig:rvpa00} shows a comparison of $R_{\Pgt,{\rm
    V}}(s_0)+R_{\Pgt,{\rm A}}(s_0)$ as predicted from the three
theories using the fit results at $s_0 = m_\Pgt^2$ with the data.  The
Contour Improved prediction is consistent with the data from the
\Pgt-mass scale down to $s_0 \approx 1\,{\rm GeV}^2$ while FOPT and
RCPT tend to predict too large values below $s_0 \approx 2\,{\rm
  GeV}^2$.


\subsection{Evolution of \balphas\ from 
  {\boldmath$m_\Pgt$}\ to {\boldmath$\mz$}}\label{subsec:as_mz} The
value of the strong coupling at the mass scale of the \Pgt\ lepton can
be evolved up to the mass scale of the \PZz. This is done by solving
the four-loop $\beta$-function given by equation~(\ref{eq:beta-fct})
numerically in small steps from $m_\Pgt^2$ to $\mzsq$ applying a
three-loop matching condition~\cite{art:Rodrigo} at the flavor
thresholds for $m_{\rm c}(m_{\rm c}) = (1.30 \pm 0.06)\,{\rm GeV}$ and
$m_{\rm b}(m_{\rm b}) = (4.13 \pm 0.06)\,{\rm
  GeV}$~\cite{art:Rodrigo}.  The evolution procedure induces an
additional error of~$\pm 0.0003$~\cite{art:Rodrigo} on the strong
coupling at the \PZz\ mass.  Using the CIPT result for
$\alphas(m_\Pgt^2)$ and $\mz = 91.187\,{\rm GeV}$ the following value
is obtained:
\begin{equation}
  \label{eq:alpha_s_mz}
  \alphas(\mzsq) = \MEANASVLMZ 
  \pm \MEANASXMMZ_{\rm exp} 
  \pm \MEANASTMMZ_{\rm theo}
  \pm 0.0003_{\rm evol}.
\end{equation}
The FOPT fit gives 
\begin{equation}
  \label{eq:alpha_s_mz_FOPT}
  \alphas(\mzsq) = \FOPTASMZ 
  \pm \FOPTEMMZ_{\rm exp} 
  \pm \FOPTTMMZ_{\rm theo}  
  \pm 0.0003_{\rm evol}.
\end{equation}
Finally RCPT gives:
\begin{equation}
  \label{eq:alpha_s_mz_RCPT}
  \alphas(\mzsq) = \RCPTASMZ 
  \pm \RCPTEMMZ_{\rm exp} 
  \pm \RCPTTMMZ_{\rm theo}
  \pm 0.0003_{\rm evol}.
\end{equation}
The different contributions to the theoretical uncertainties are
listed in table~\ref{tab:theoerr}.
The results are in good agreement with the value obtained from fits to  
combined electroweak measurements at LEP and SLD~\cite{art:Lineshape}:
\begin{equation}
  \label{eq:alpha_s_mz_EW}
  \alphas(\mzsq) = 0.120 \pm 0.003.
\end{equation}


\section{Measurement of dimension 6 and 8 operators}
\label{sec:power} 
\begin{table}[htb]
  \begin{center}
    \begin{tabular}[t]{|c|c|l|llll|c|}
      \hline
         & & & \multicolumn{4}{c|}{contributing errors} & \\
         theory & observable & \p{+}value & \p{+}data & \p{+}$B$ &
         \p{+}syst. & \p{+}theo. & $\chi^2/{\rm d.o.f.}$  \\
      \hline               
\rule[-6pt]{0pt}{19pt} & $\alphas(m_\Pgt^2)$& 
                         $\p{+}\CIPTcATVL$  & 
                         $\pm \CIPTcATDA$   &
                         $\pm \CIPTcATBR$   &
                         $\pm \CIPTcATSM$   &
                         $\pm \CIPTcATTM$   & \\
\rule[-6pt]{0pt}{19pt} & $\langle\frac{\alphas}{\pi}\,GG\rangle/{\rm GeV^4}$ & 
                         $\p{+}\CIPTcGGVL$  & 
                         $\pm \CIPTcGGDA$   &
                         $\pm \CIPTcGGBR$   &
                         $\pm \CIPTcGGSM$   &
                         $\pm \CIPTcGGTM$   & \\
\rule[-6pt]{0pt}{19pt} & $\delta_{\rm V}^6$ &  
                         $\p{+}\CIPTcVSVL$  & 
                         $\pm \CIPTcVSDA$   & 
                         $\pm \CIPTcVSBR$   & 
                         $\pm \CIPTcVSSM$   & 
                         $\pm \CIPTcVSTM$   & \\
\rule[-6pt]{0pt}{19pt}   \raisebox{8.5pt}[0pt][0pt]{CIPT} & $\delta_{\rm V}^8$& 
                         $\CIPTcVEVL$       & 
                         $\pm \CIPTcVEDA$   & 
                         $\pm \CIPTcVEBR$   & 
                         $\pm \CIPTcVESM$   & 
                         $\pm \CIPTcVETM$   &
                         \raisebox{8.5pt}[0pt][0pt]{$\CIPTcCHSQ / 4$}  \\
\rule[-6pt]{0pt}{19pt} & $\delta_{\rm A}^6$ &  
                         $\CIPTcASVL$       & 
                         $\pm \CIPTcASDA$   & 
                         $\pm \CIPTcASBR$   & 
                         $\pm \CIPTcASSM$   & 
                         $\pm \CIPTcASTM$   & \\
\rule[-6pt]{0pt}{19pt} & $\delta_{\rm A}^8$ & 
                         $\p{+}\CIPTcAEVL$  & 
                         $\pm \CIPTcAEDA$   &  
                         $\pm \CIPTcAEBR$   &  
                         $\pm \CIPTcAESM$   & 
                         $\pm \CIPTcAETM$   & \\
      \hline                       
\rule[-6pt]{0pt}{19pt} & $\alphas(m_\Pgt^2)$& 
                         $\p{+}\FOPTcATVL$  & 
                         $\pm \FOPTcATDA$   &
                         $\pm \FOPTcATBR$   &
                         $\pm \FOPTcATSM$   &
                         $\pm \FOPTcATTM$   & \\
\rule[-6pt]{0pt}{19pt} & $\langle\frac{\alphas}{\pi}\,GG\rangle/{\rm GeV^4}$ & 
                         $\p{+}\FOPTcGGVL$  & 
                         $\pm \FOPTcGGDA$   &
                         $\pm \FOPTcGGBR$   &
                         $\pm \FOPTcGGSM$   &
                         $\pm \FOPTcGGTM$   & \\
\rule[-6pt]{0pt}{19pt} & $\delta_{\rm V}^6$ &  
                         $\p{+}\FOPTcVSVL$  & 
                         $\pm \FOPTcVSDA$   & 
                         $\pm \FOPTcVSBR$   & 
                         $\pm \FOPTcVSSM$   & 
                         $\pm \FOPTcVSTM$   & \\
\rule[-6pt]{0pt}{19pt}   \raisebox{8.5pt}[0pt][0pt]{FOPT} & $\delta_{\rm V}^8$& 
                         $\FOPTcVEVL$       & 
                         $\pm \FOPTcVEDA$   & 
                         $\pm \FOPTcVEBR$   & 
                         $\pm \FOPTcVESM$   & 
                         $\pm \FOPTcVETM$   &
                         \raisebox{8.5pt}[0pt][0pt]{$\FOPTcCHSQ / 4$}  \\
\rule[-6pt]{0pt}{19pt} & $\delta_{\rm A}^6$ &  
                         $\FOPTcASVL$       & 
                         $\pm \FOPTcASDA$   & 
                         $\pm \FOPTcASBR$   & 
                         $\pm \FOPTcASSM$   & 
                         $\pm \FOPTcASTM$   & \\
\rule[-6pt]{0pt}{19pt} & $\delta_{\rm A}^8$ & 
                         $\p{+}\FOPTcAEVL$  & 
                         $\pm \FOPTcAEDA$   & 
                         $\pm \FOPTcAEBR$   & 
                         $\pm \FOPTcAESM$   & 
                         $\pm \FOPTcAETM$   & \\
    \hline
\rule[-6pt]{0pt}{19pt} & $\alphas(m_\Pgt^2)$& 
                         $\p{+}\RCPTcATVL$  & 
                         $\pm \RCPTcATDA$   &
                         $\pm \RCPTcATBR$   &
                         $\pm \RCPTcATSM$   &
                         $\pm \RCPTcATTM$   & \\
\rule[-6pt]{0pt}{19pt} & $\langle\frac{\alphas}{\pi}\,GG\rangle/{\rm GeV^4}$ & 
                         $\p{+}\RCPTcGGVL$  & 
                         $\pm \RCPTcGGDA$   &
                         $\pm \RCPTcGGBR$   &
                         $\pm \RCPTcGGSM$   &
                         $\pm \RCPTcGGTM$   & \\
\rule[-6pt]{0pt}{19pt} & $\delta_{\rm V}^6$ &  
                         $\p{+}\RCPTcVSVL$  & 
                         $\pm \RCPTcVSDA$   & 
                         $\pm \RCPTcVSBR$   & 
                         $\pm \RCPTcVSSM$   & 
                         $\pm \RCPTcVSTM$   & \\
\rule[-6pt]{0pt}{19pt}   \raisebox{8.5pt}[0pt][0pt]{RCPT} & $\delta_{\rm V}^8$&
                         $\RCPTcVEVL$       & 
                         $\pm \RCPTcVEDA$   & 
                         $\pm \RCPTcVEBR$   & 
                         $\pm \RCPTcVESM$   & 
                         $\pm \RCPTcVETM$   &
                         \raisebox{8.5pt}[0pt][0pt]{$\RCPTcCHSQ / 4$}  \\
\rule[-6pt]{0pt}{19pt} & $\delta_{\rm A}^6$ &  
                         $\RCPTcASVL$       & 
                         $\pm \RCPTcASDA$   & 
                         $\pm \RCPTcASBR$   & 
                         $\pm \RCPTcASSM$   & 
                         $\pm \RCPTcASTM$   & \\
\rule[-6pt]{0pt}{19pt} & $\delta_{\rm A}^8$ & 
                         $\p{+}\RCPTcAEVL$  & 
                         $\pm \RCPTcAEDA$   & 
                         $\pm \RCPTcAEBR$   & 
                         $\pm \RCPTcAESM$   & 
                         $\pm \RCPTcAETM$   & \\
    \hline
    \end{tabular}
    \caption{\em The fit result for $\mathit{\alpha_s}$ and the power 
      corrections from the combined fit to vector and axial-vector
      moments. The given errors correspond to the data statistics, the
      uncertainty due to the errors on the branching ratios~$B$, a
      systematic error from the Monte Carlo statistics, the energy
      scale, the momentum scale, and the unfolding, and a total
      theoretical uncertainty.}
    \label{tab:VaA_fit} \end{center}
\end{table}                                

\begin{table}[htbp]
  \begin{center}
    \begin{tabular}[t]{|c|ccccc|}
      \hline
\rule[-6pt]{0pt}{19pt} & $\alphas(m_\Pgt^2)$               & 
        $\left\langle\frac{\alphas}{\pi} GG\right\rangle$  & 
        $\delta_{\rm V}^6$                                 & 
        $\delta_{\rm A}^6$                                 & 
        $\delta_{\rm V}^8$                                 \\
      \hline 
\rule[-6pt]{0pt}{19pt} $\left\langle\frac{\alphas}{\pi} GG\right\rangle$  &
$\CORRFOATGG$            &                    &                           &
                         &                                                \\
\rule[-6pt]{0pt}{19pt} $\delta_{\rm V}^6$                                 & 
$\CORRFOATVS$            & $\p{+}\CORRFOGGVS$ &                           &
                         &                                                \\
\rule[-6pt]{0pt}{19pt} $\delta_{\rm A}^6$                                 & 
$\CORRFOATAS$            & $\p{+}\CORRFOGGAS$ & $\p{+}\CORRFOVSAS$        &
                         &                                                \\
\rule[-6pt]{0pt}{19pt} $\delta_{\rm V}^8$                                 & 
$\p{+}\CORRFOATVE$ & $\CORRFOGGVE$            & $\CORRFOVSVE$             &
$\CORRFOASVE$            &                                                \\
\rule[-6pt]{0pt}{19pt} $\delta_{\rm A}^8$                                 & 
$\p{+}\CORRFOATAE$ & $\CORRFOGGAE$            & $\CORRFOVSAE$             &
$\CORRFOASAE$            & $\p{+}\CORRFOVEAE$                             \\
      \hline            
    \end{tabular}
    \caption{\em Correlations between the QCD parameters from the fit to the
      moments of the vector and axial-vector current in percent. The
      given numbers are taken from the CIPT fit result.}
    \label{tab:VaA_corr} 
  \end{center}
\end{table}                                
\begin{figure}[htb]
\centering
\resizebox{0.5\textwidth}{!}{%
\includegraphics{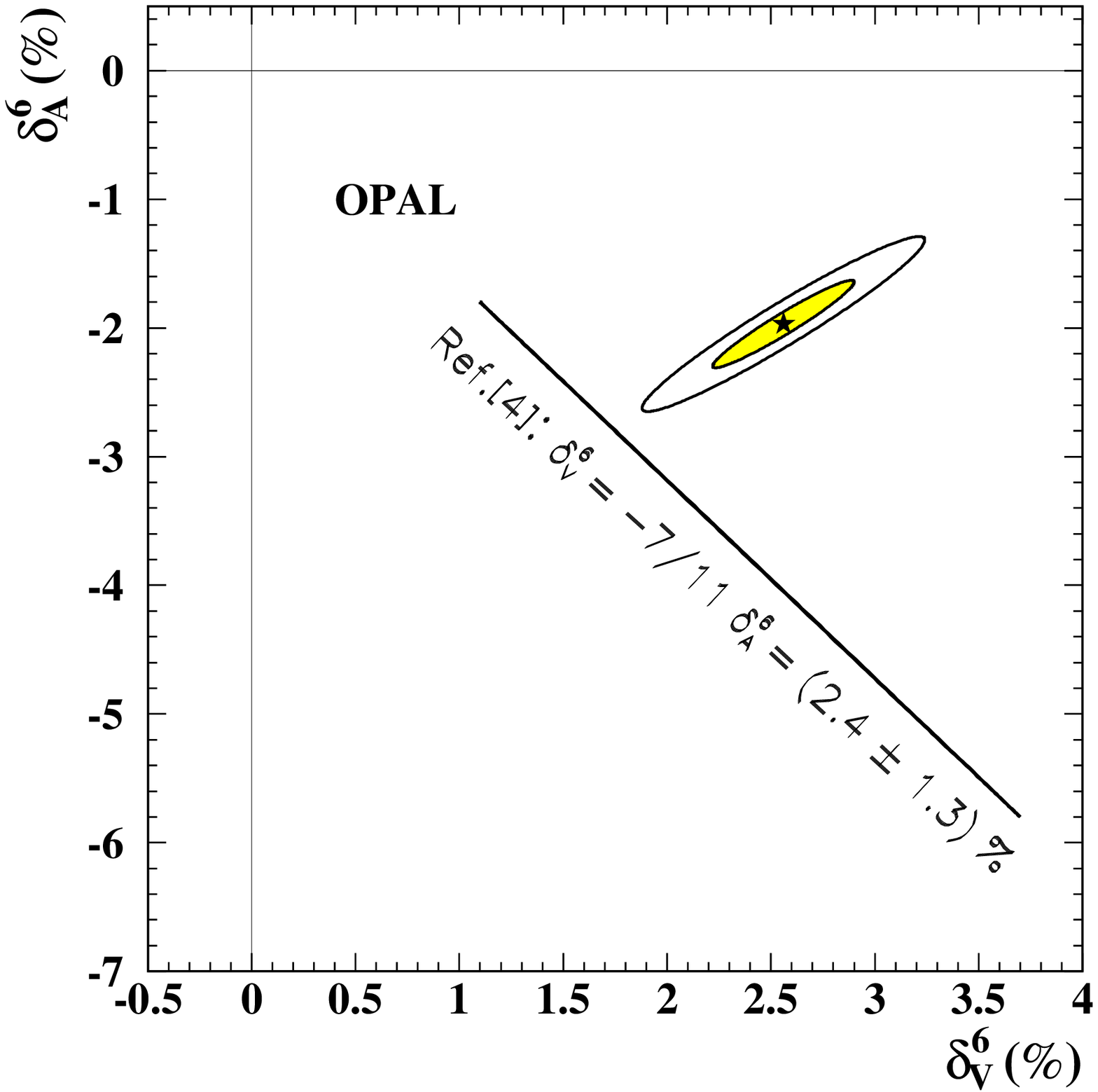}}
\caption{\em The power corrections of dimension 6 to 
  $\mathit{R_{\Pgt,{V/A}}}$.  Shown are the one and two standard
  deviation contours of the correlated result for the vector and
  axial-vector current (CIPT) including experimental and theoretical
  uncertainties. The solid line indicates the theoretical
  prediction given in~\cite{art:Braaten}.}
\label{fig:delta6}
\end{figure}
The results from fit 2 where the separate moments of the vector
current and axial-vector current are used are given in
table~\ref{tab:VaA_fit}.  In contrast to $\alphas$ where the error is
dominated by the theoretical uncertainties, the power corrections are
almost independent of the theoretical uncertainties for CIPT and RCPT.
As mentioned in section~\ref{sec:alphas}, this is not the case for the
FOPT fit which leads to theoretical errors of the order of (or even
larger than) the experimental errors.  Due to the correlated unfolding
of vector and axial-vector spectra a strong positive correlation
between the power corrections of the vector and axial vector current
of the same dimension is observed.  The power corrections of different
dimension but for the same current are anti-correlated. All
correlations of the fit parameters for CIPT are summarized in
table~\ref{tab:VaA_corr}.  The fitted values of the strong coupling
constant in both fits are in excellent agreement for all three models.
The experimental error on $\alphas$ from this fit is larger than in
fit 1 as the additional information from the \Pgt\ lifetime and the
branching ratio $B(\Pgt \rightarrow \Pgm \Pgngm \Pgngt)$ is omitted.
Using equation~(\ref{eq:average}), the separate and total power
corrections are also in good agreement for all three models.  As in
fit 1 all three theories give similar $\chi^2$ values in the fit to
the exclusive moments.  The theoretical uncertainties behave similarly
in fit 1 and fit 2.  The sum of all power corrections $\delta_{\rm
  non\md pert, V/A}$ and $\delta_{\rm non\md pert, V+A}$ to
$R_{\Pgt,{\rm V/A}}$ and $R_{\Pgt,{\rm V}}+R_{\Pgt,{\rm A}}$ including
the dimension 2 quark-mass correction and the dimension 4 correction
obtained from the fitted gluon condensate are:

\begin{eqnarray}
  &  & \p{+}0.0172 \pm 0.0026 \quad 
  {\rm CIPT} \nonumber  \\
  \delta_{\rm non\md pert, V} & = & \p{+}0.0187 \pm 0.0054 \quad 
  {\rm FOPT} \label{eq:d_V}  \\
   &  & \p{+}0.0124 \pm 0.0033 \quad 
  {\rm RCPT} \nonumber, 
\end{eqnarray}
\begin{eqnarray}
   &  & -0.0219 \pm 0.0026 \quad 
  {\rm CIPT} \nonumber \\
  \delta_{\rm non\md pert, A} & = & -0.0204 \pm 0.0050 \quad 
  {\rm FOPT} \label{eq:d_A} \\
   &  & -0.0266 \pm 0.0032 \quad 
  {\rm RCPT} \nonumber, 
\end{eqnarray}
\begin{eqnarray}
   &  & -0.0024 \pm 0.0025 \quad 
  {\rm CIPT} \nonumber \\
  \delta_{\rm non\md pert, V+A} & = & -0.0009 \pm 0.0051 \quad 
  {\rm FOPT} \label{eq:d_V+A} \\
   &  & -0.0071 \pm 0.0031 \quad 
  {\rm RCPT} \nonumber, 
\end{eqnarray}
where the errors include experimental and theoretical uncertainties.
Thus all three theories lead to non-perturbative corrections to
$R_{\Pgt,{\rm V}}$ ($R_{\Pgt,{\rm A}}$) of the order $1.6\,\%$
($-2.3\,\%$), while a large cancellation of both contributions leads
to a total non-perturbative correction to $R_{\Pgt,{\rm V}} +
R_{\Pgt,{\rm A}}$ which is compatible with zero and therefore allows a
precise measurement of the strong coupling constant in fit 1.  The
numbers in table~\ref{tab:VaA_fit} can be compared to the estimates
given in~\cite{art:Braaten}:

\begin{eqnarray}
  \label{eq:non-pert-Braaten}
  \langle\frac{\alphas}{\pi}\,GG\rangle/{\rm GeV^4} & = & 
                   \p{+}0.02  \pm 0.01  \nonumber,\\
  \delta_{\rm V}^6 & = & \p{+}0.024 \pm 0.013, \\
  \delta_{\rm A}^6 & = &           -0.038 \pm 0.020 \nonumber,\\
  \delta_{\rm V/A}^8   & \simeq &  -0.0001 \nonumber.
\end{eqnarray}
Only the power corrections of dimension 8 seem to be underestimated,
while the other estimates are in good agreement with the measured
values.  Figure~\ref{fig:delta6} shows the two power corrections of
dimension 6 (CIPT) together with the theoretical prediction given
in~\cite{art:Braaten}.


\section{Test of the \lq running\rq\ of \balphas}
\label{sec:running}
The fit to the sum of vector and axial-vector moments (fit 1) can be
extended to lower values of $s_0$, thus giving a correlated
measurement of the strong coupling at different scales.  Four
equidistant values for $s_0$ between $1.3\,{\rm GeV}^2$ and $m_\Pgt^2$
are used.  In addition to the five moments at $s_0 = m_\Pgt^2$ the
integrated differential decay rate $R_{\Pgt,{\rm
    V}}^{00}(s_0)+R_{\Pgt,{\rm A}}^{00}(s_0)$ for each additional
$s_0$ value is included in the fit (see figure~\ref{fig:rvpa00}).

\begin{figure}[htb]
\centering
\resizebox{0.6\textwidth}{!}{%
\includegraphics{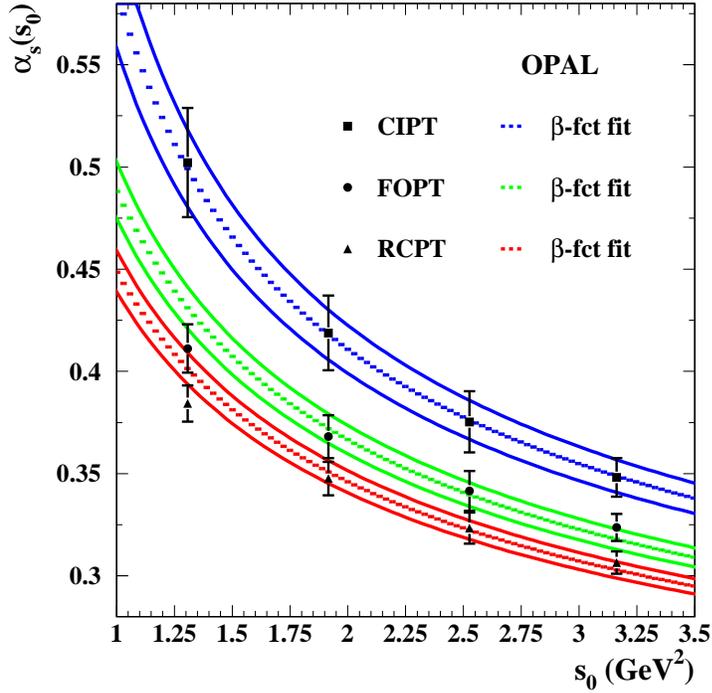}}
\caption{\em The \lq running\rq\  of the strong coupling. The three 
  sets of $\mathit{\alpha_s}$ values are shown as data points. The
  error bars include statistical and systematic uncertainties. The
  dashed curves represent the predictions of the 4-loop
  $\mathit{\beta}$-function obtained from fits to the three sets of
  $\mathit{\alpha_s}$ values not including the values
  $\mathit{\alpha_s(1.3\,{GeV}^2)}$.  The solid lines depict the
  errors from the fits.}
\label{fig:running}
\end{figure}

For the extraction of the \lq running\rq\ of \alphas\ the number of
fit parameters is increased to include the strong coupling
$\alphas(s_0)$ for each $s_0$ value below $m_\Pgt^2$.  The result can
be examined with the four-loop $\beta$-function. This is shown in
figure~\ref{fig:running}, where the $\beta$-function has been refitted
for all three sets of \alphas\ values. The values at $s_0 = 1.3\,{\rm
  GeV^2}$ were not included in the fit. A comparison of these values
with the predicted \lq running\rq\ shows good agreement in case of
CIPT, while a weaker \lq running\rq\ as predicted by the
$\beta$-function is preferred by the FOPT and the RCPT values.

\begin{figure}[htbp]
\centering
\resizebox{\textwidth}{!}{%
\includegraphics{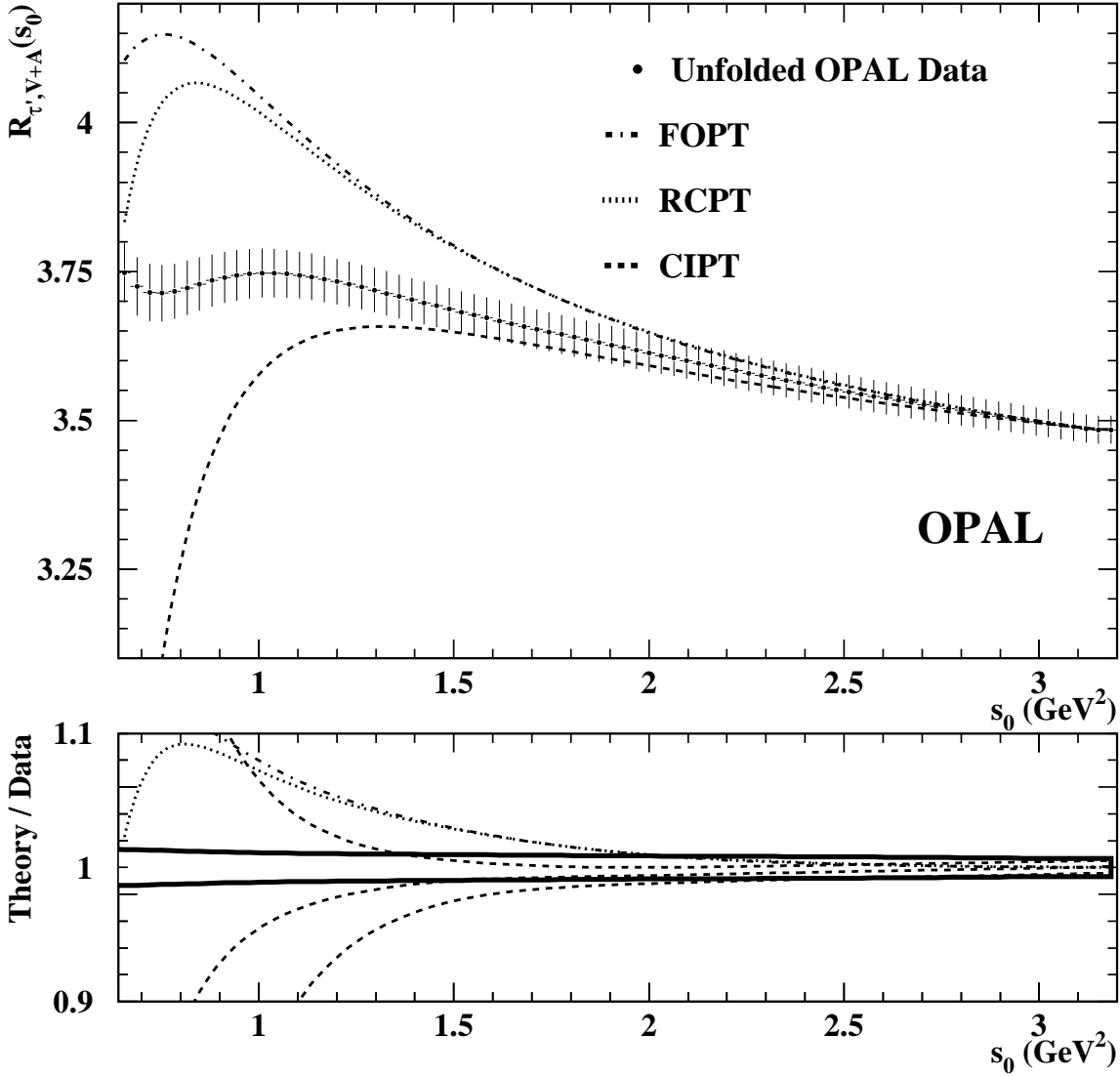}}
\caption{\em The non-strange hadronic decay rate of a hypothetical 
  $\mathit{\Pgt'}$ lepton with $\mathit{m_{\Pgt'}^2 = s_0}$ versus the
  upper integration limit $\mathit{s_0}$. The points in the upper plot
  denote OPAL data; the dashed, dashed-dotted and dotted curves
  represent the theoretical predictions of the three theories with the
  results from the fit to the moments at $\mathit{s_0 = m_\Pgt^2}$
  used as input.  The lower plot shows the three theories normalized
  to the data. The three dashed curves indicate central values and
  total experimental errors for CIPT.  The dashed-dotted and dotted
  curves show central values for FOPT and RCPT. The errors for FOPT
  and RCPT are similar to the CIPT errors and omitted from the plot.
  The errors on the data are shown as solid lines.}
\label{fig:rtvpa00}
\end{figure}

Figures~\ref{fig:rvpa00} and~\ref{fig:running} can be regarded as
tests of the validity of the OPE for $s_0$ values below $m_\Pgt^2$.
It has been questioned if the definition of $R_\Pgt(s_0)$ is still
valid in this region~\cite{art:Braaten}, since the endpoint $s = s_0$
is no longer suppressed by the $(1-s/m_\Pgt^2)^2$ term in front of the
spectral function (see equation~(\ref{eq:dRds})).  By defining the
hadronic decay rate for a hypothetical $\Pgt'$ with a mass of
$m_{\Pgt'} = \sqrt{s_0}$ and inserting $m_{\Pgt'}$ for $m_\Pgt$
in equation~(\ref{eq:dRds}) one gets~\cite{art:Andreas98}:
\begin{equation}
  \label{eq:dRtds}
  R_{\Pgt',{\rm V/A}}(s_0) = 12 \pi S_{\rm EW} |V_{\rm ud}|^2 
  \int\limits_0^{s_0}{ 
    \frac{\rd s}{s_0} 
    \left(1-\frac{s}{s_0}\right)^2\left[\left(1+2\frac{s}{s_0}\right) 
      {\rm Im} {\mit\Pi}_{\rm V/A}^{(1)}(s) + 
      {\rm Im} {\mit\Pi}_{\rm V/A}^{(0)}(s)\right]}, 
\end{equation}
obeying the same quadratic suppression of the endpoint on the real
$s$-axis as $R_{\Pgt,{\rm V/A}}(m_\Pgt^2)$. Figure~\ref{fig:rtvpa00}
shows the sum $R_{\Pgt',{\rm V}}(s_0) + R_{\Pgt',{\rm A}}(s_0)$ versus
the upper integration limit $s_0$.  The error band for CIPT in the
lower plot shows that the uncertainties increase below $s_0 \simeq
1.5\,{\rm GeV}^2$ compared to the error in the lower plot of
figure~\ref{fig:rvpa00}.  While the error on $R_\Pgt(s_0)$ is
dominated by the uncertainty of the perturbative expansion, the error
on $R_{\Pgt'}(s_0)$ originates mainly from its dependency on the
non-perturbative parts.  In contrast to $R_\Pgt$ where these power
corrections stay constant for all $s_0$ (see equation~(\ref{eq:OPE}))
they increase with powers of $1/s_0$ as $s_0$ decreases in the case of
$R_{\Pgt'}$.  As the errors are large for small values of $s_0$ little
can be said about this region.

\section{QCD sum rules}
\label{sec:QCD_sum_rules}
\begin{figure}[htb]
\centering
\resizebox{0.43\textwidth}{!}{%
\includegraphics{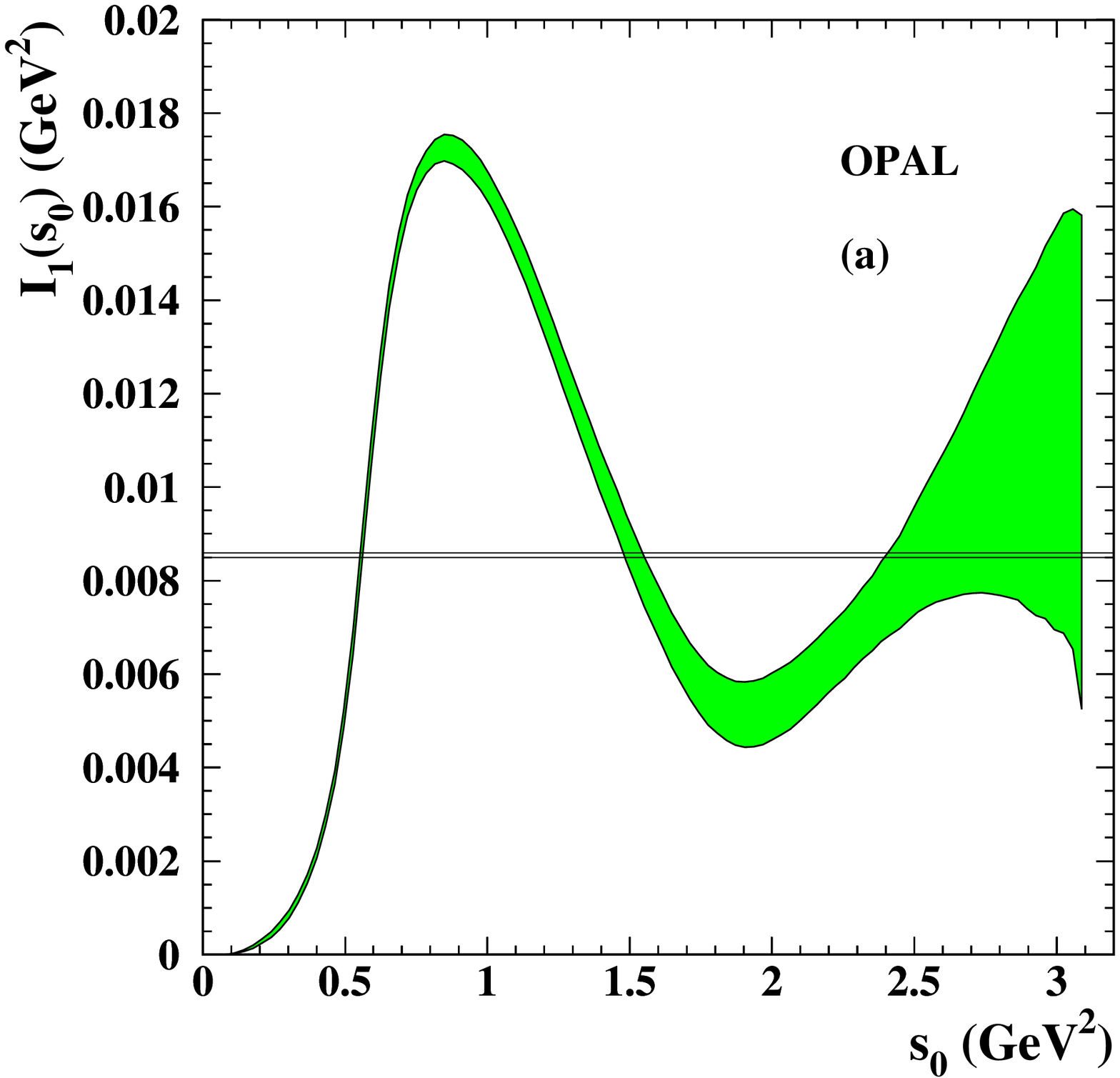}}
\resizebox{0.43\textwidth}{!}{%
\includegraphics{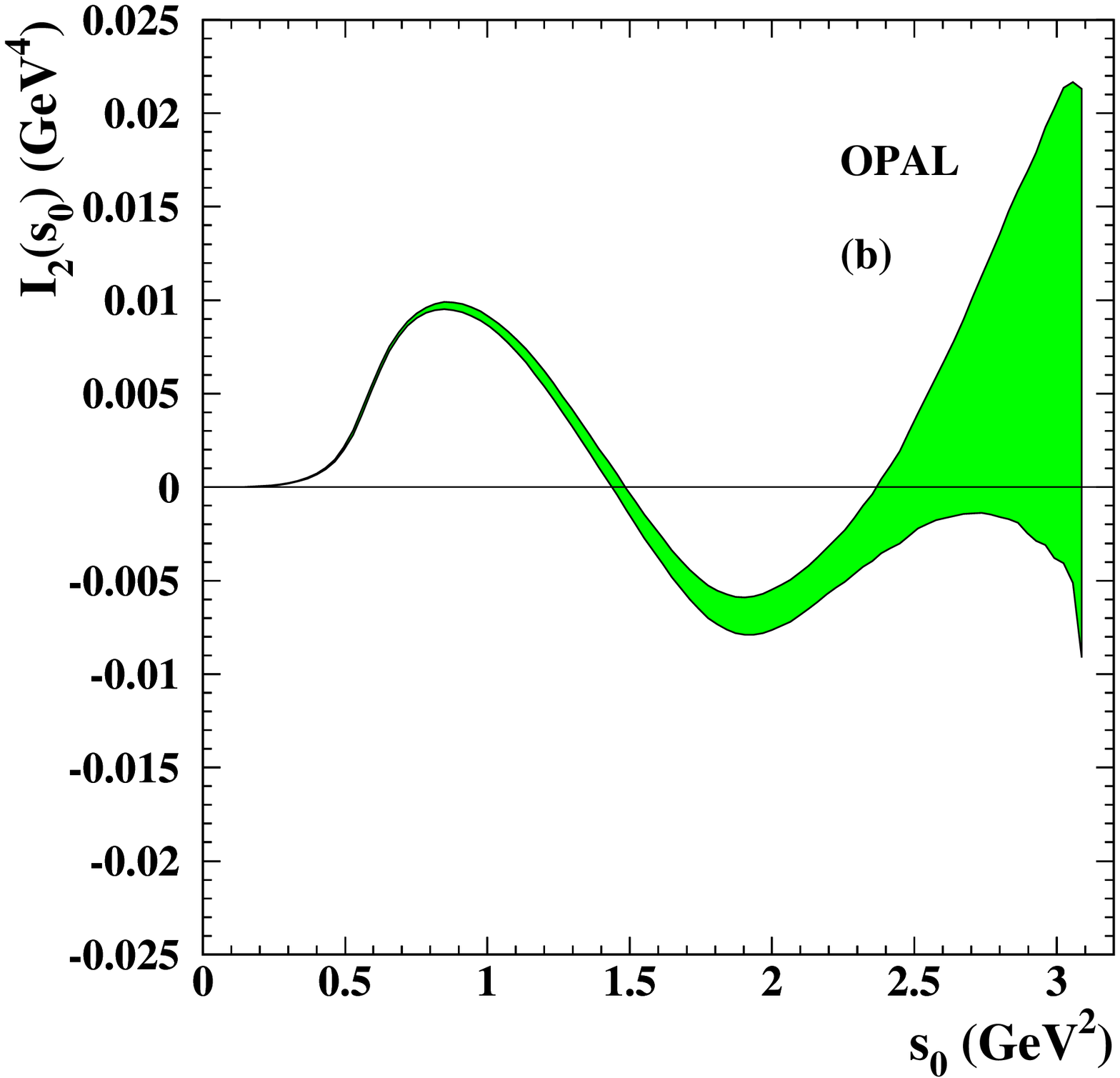}}
\resizebox{0.43\textwidth}{!}{%
\includegraphics{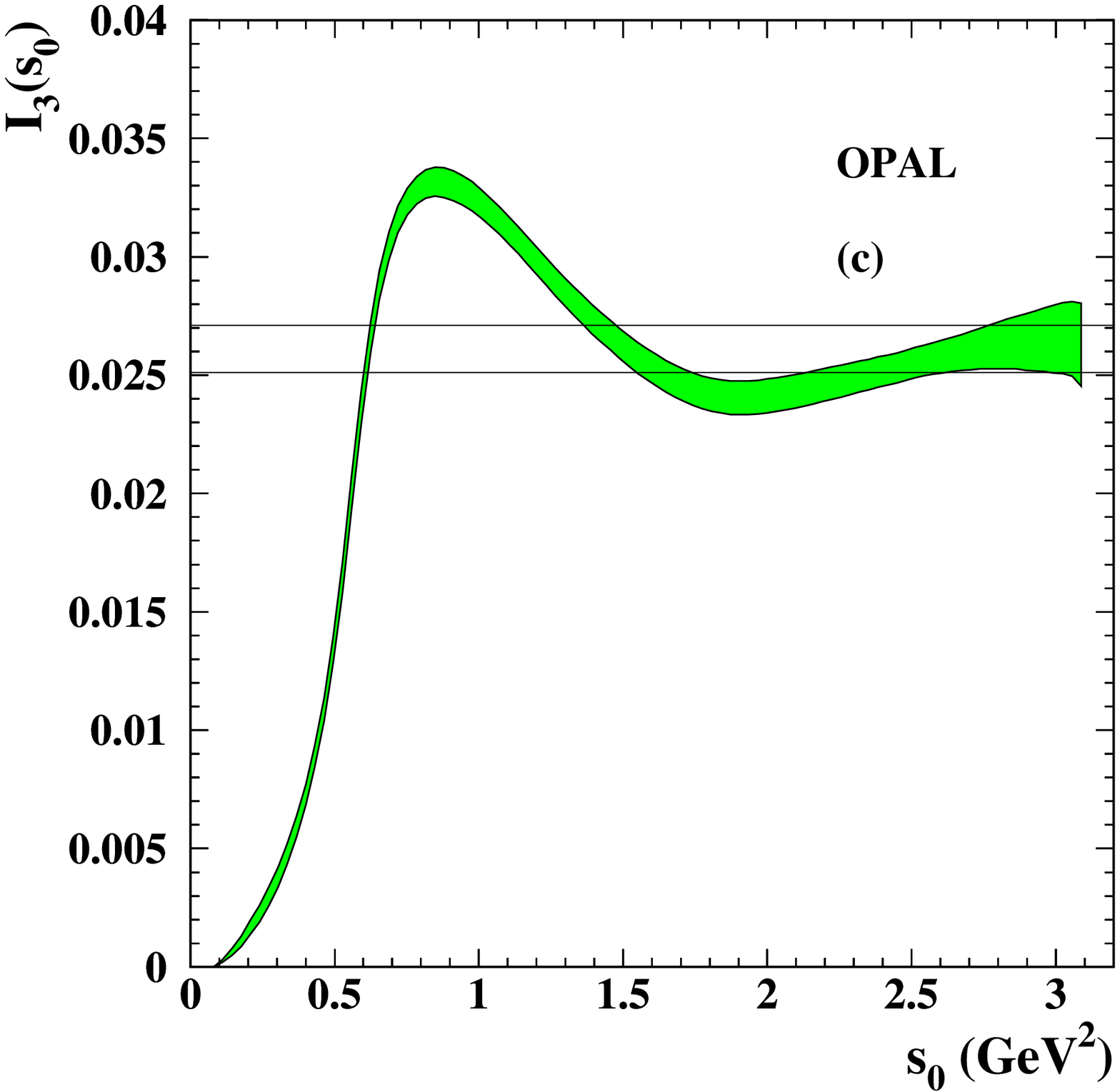}}
\resizebox{0.43\textwidth}{!}{%
\includegraphics{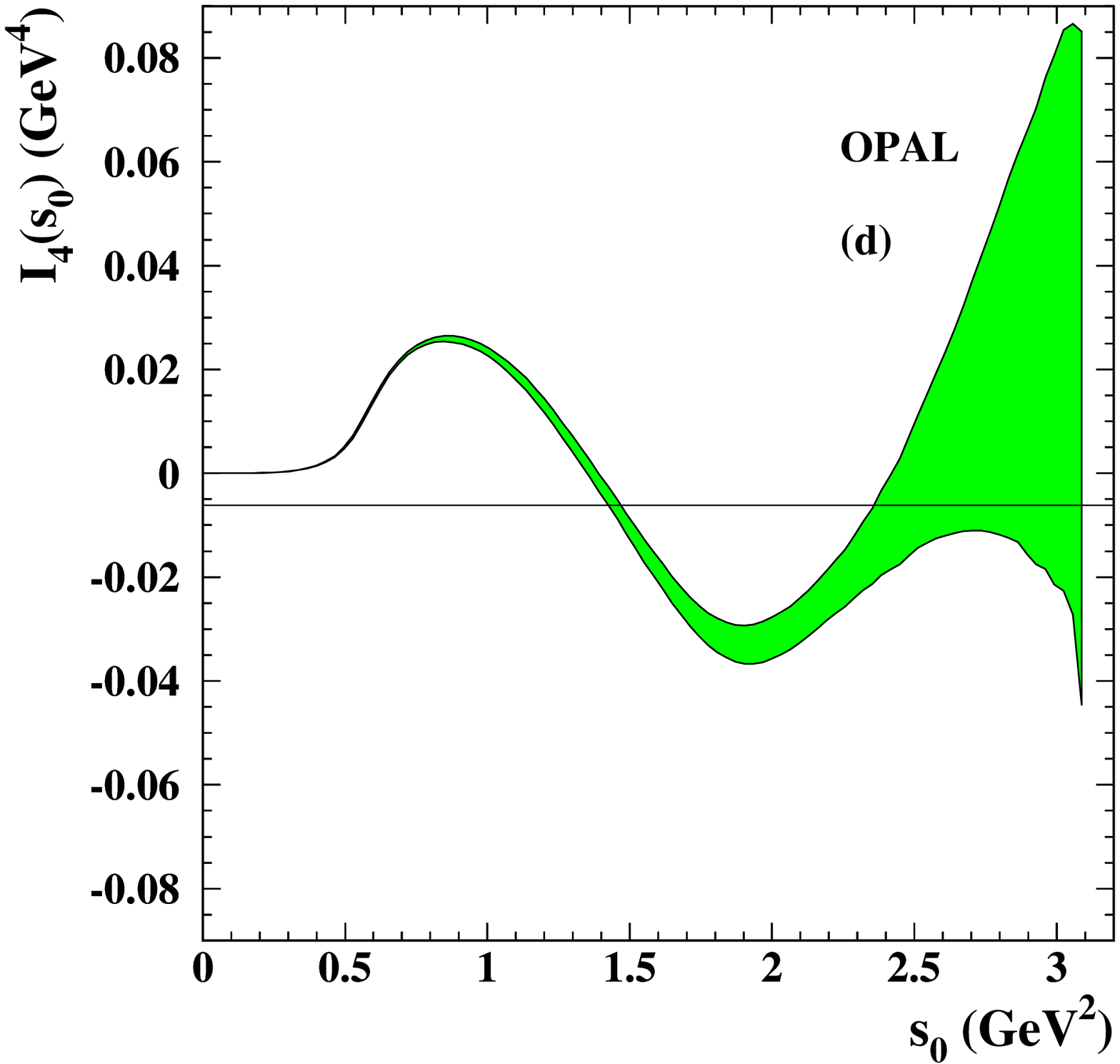}}
\caption{\em QCD sum rules. Equations (\ref{eq:i1})--(\ref{eq:i4}) are shown
  in the plots (a)--(d). Shown are the integrals versus the upper
  integration limit as shaded bands. The chiral prediction is given by
  the lines ($\mathit{\pm 1\,\sigma}$ when two lines are present).}
\label{fig:sum_rules}
\end{figure}
Weighted integrals over the difference of the two measured spectral
functions shown in figure~\ref{fig:v+-a} can be compared to the chiral
predictions of QCD sum rules:
\begin{eqnarray}
  I_1(s_0) & = & \frac{1}{4\pi^2}\int\limits_0^{s_0} \rd s\,\left(
    v(s)-a(s)\right) 
             = f_\pi^2  \label{eq:i1},\\
  I_2(s_0) & = & \frac{1}{4\pi^2}\int\limits_0^{s_0} \rd s\,
  s\,\left(v(s)-a(s)\right)
             = 0 \label{eq:i2},\\
  I_3(s_0) & = & \frac{1}{4\pi^2}\int\limits_0^{s_0} \frac{\rd s}{s}\,
                 \left(v(s)-a(s)\right)
             = f_\pi^2\,\frac{\langle r_\pi^2\rangle}{3}-F_{\rm A} 
             \label{eq:i3},\\
  I_4(s_0) & = & \frac{1}{4\pi^2}\int\limits_0^{s_0} \rd s\,
  s\,{\rm ln}\frac{s}{\lambda^2}\,
                 \left(v(s)-a(s)\right)
             = -\frac{4\pi f_\pi^2}{3\,\alpha} 
                \left(m_{\pi^\pm}^2-m_{\pi^0}^2\right). \label{eq:i4}
\end{eqnarray}
Here the right hand side of each equation is understood to be the
chiral prediction in the limit $s_0 \rightarrow \infty$.
Equation~(\ref{eq:i1}) is known as the first Weinberg sum
rule~\cite{art:Weinberg}, assuming that the only scalar contribution
is given by the pion pole which is related to the pion decay constant
$f_\Pgp = (92.4 \pm 0.26)\,{\rm MeV}$~\cite{art:PDG96}.  The second
Weinberg sum rule~\cite{art:Weinberg} is given in
equation~(\ref{eq:i2}). The Das--Mathur--Okubo (DMO) sum
rule~\cite{art:DMO} is given by equation~(\ref{eq:i3}). Its asymptotic
prediction is a function of the pion decay constant $f_\Pgp$, the mean
square of the pion charge radius $\langle r_\Pgp^2\rangle = (0.439 \pm
0.008)\,{\rm fm}^2$~\cite{art:Amendolia} and the axial-vector form
factor of the pion $F_{\rm A} = 0.0058 \pm
0.0008$~\cite{art:PDG96}\footnote{Our definitions of $F_{\rm A}$ and
  $f_\Pgp^2$ differ by a factor of $1/2$ from those given
  in~\cite{art:PDG96}}, and equation~(\ref{eq:i4}) gives the
electromagnetic mass difference of pions~\cite{art:DGM}. Note that
equation~(\ref{eq:i4}) does not depend on the cut-off value $\lambda$
by virtue of the second Weinberg sum rule.

The saturation of these four sum rules is tested taking into account
the full correlations between the measured spectral functions.  The
plots of figure~\ref{fig:sum_rules} show the measured values of the
integrals~$I_1$-$I_4$ as error bands including all experimental
uncertainties versus the upper integration limit.  The asymptotic
predictions are drawn as thin lines denoting the present $\pm
1\,\sigma$ ranges.

All four sum rules appear to be saturated at the \Pgt-mass scale
within their errors. However, due to the small phase space near the
\Pgt-mass which appears in the denominator of the spectral functions
these errors are very large except for the DMO sum rule where the
$1/s$~factor suppresses the high energy tail.  The value of
equation~(\ref{eq:i3}) at $s_0 = m_\Pgt^2$ is:
\begin{equation}
  \label{eq:i3_result}
  I_3(m_\Pgt^2) = ( 26.3 \pm 1.8) \cdot 10^{-3},
\end{equation}
where the error covers all experimental uncertainties.

\subsection{Pion polarizability}
\label{sec:pion_polarizability}
Assuming that the DMO sum rule shown in figure~\ref{fig:sum_rules}~(c)
is already saturated at the \Pgt-mass scale, its value can be used to
predict the electric polarizability of the charged pion as proposed
in~\cite{art:Vato97}:
\begin{equation}
  \label{eq:alpha_E}
  \alpha_{\rm E} = \frac{\alpha}{m_{\pi^\pm}}\left(
    \frac{\langle r_\pi^2\rangle}{3} - 
    \frac{I_3}{f_\pi^2}\right).
\end{equation}
Using the result from the previous section for the DMO sum rule 
(equation~(\ref{eq:i3_result})) one gets:
\begin{equation}
  \label{eq:alpha_E_result}
  \alpha_{\rm E} = (2.71 \pm 0.88) \cdot 10^{-4}\,{\rm fm}^3,
\end{equation}
which is in good agreement with the value $\alpha_{\rm E} = (2.64 \pm
0.36) \cdot 10^{-4}\,{\rm fm}^3$, derived in~\cite{art:Vato97}.


\section{Summary}\label{sec:conclusions}
Measurements of the spectral functions of the vector current and the
axial-vector current and their applications in QCD have been
presented.  Within the framework of the Operator Product Expansion, a
simultaneous determination of the strong coupling constant \alphas\ 
and non-perturbative correction terms has been performed.  The sum of
$R_{\Pgt,{\rm V}}$ and $R_{\Pgt,{\rm A}}$ was found to involve a large
cancellation of the non-perturbative terms and thus has been used
together with the $\Pgt$ lifetime and the branching ratio $B(\Pgt
\rightarrow \Pgm \Pgngm \Pgngt)$ to give a precise measurement of the
strong coupling constant.  CIPT has led to the value
$$\alphas(m_\Pgt^2) = \MEANASVLMT \pm \MEANASXMMT_{\rm exp} \pm
\MEANASMTMT_{\rm theo},$$ at the \Pgt-mass scale and $$\alphas(\mzsq) =
\MEANASVLMZ \pm \MEANASXMMZ_{\rm exp} \pm \MEANASMTMZ_{\rm theo}$$ at
the \PZz-mass scale, where the first error stems from the experimental
uncertainties and the second error originates from the theoretical
uncertainties.  The values obtained for $\alphas(\mzsq)$ using FOPT or
RCPT are $2.3\,\%$ and $4.1\,\%$ smaller, respectively.

The total amount of non-perturbative corrections to $R_{\Pgt,{\rm V}}$
($R_{\Pgt,{\rm A}}$) was found to be $(1.6 \pm 0.4)\,\%$ ($(-2.3 \pm
0.4)\,\%$), while the correction on the sum of $R_{\Pgt,{\rm V}}$ and
$R_{\Pgt,{\rm A}}$ due to non-perturbative QCD is found to be only
$(-0.3 \pm 0.4)\,\%$.  Here the errors include all experimental and
theoretical uncertainties.

Assuming the validity of the Operator Product Expansion for energy
scales below the \Pgt\ mass a test of the \lq running\rq\ of the
strong coupling between $s_0 \simeq 1.3\,{\rm GeV}^2$ and $s_0 =
m_\Pgt^2$ has been performed. A good agreement between the predicted
\lq running\rq\ from the $4$-loop $\beta$-function and the fitted
$\alphas$ values has been observed for CIPT.

The saturation of QCD sum rules at the \Pgt-mass scale has been
tested, yielding a measurement of the pion polarizability of
$\alpha_{\rm E} = (2.71 \pm 0.88) \cdot 10^{-4}\,{\rm fm}^3$ as
determined from the DMO sum rule.


\section*{Acknowledgements}
We particularly wish to thank the SL Division for the efficient
operation of the LEP accelerator at all energies and for their
continuing close cooperation with our experimental group.  We thank
our colleagues from CEA, DAPNIA/SPP, CE-Saclay for their efforts over
the years on the time-of-flight and trigger systems which we continue
to use.  In addition to the support staff at our own
institutions we are pleased to acknowledge the  \\
Department of Energy, USA, \\
National Science Foundation, USA, \\
Particle Physics and Astronomy Research Council, UK, \\
Natural Sciences and Engineering Research Council, Canada, \\
Israel Science Foundation, administered by the Israel
Academy of Science and Humanities, \\
Minerva Gesellschaft, \\
Benoziyo Center for High Energy Physics,\\
Japanese Ministry of Education, Science and Culture (the Monbusho) and
a grant under the Monbusho International
Science Research Program,\\
German Israeli Bi-national Science Foundation (GIF), \\
Bundesministerium f{\"u}r Bildung, Wissenschaft,
Forschung und Technologie, Germany, \\
National Research Council of Canada, \\
Research Corporation, USA,\\
Hungarian Foundation for Scientific Research, OTKA T-016660,
T023793 and OTKA F-023259.\\
We also wish to thank A.~H{\"o}cker, J.H.~K{\"u}hn, C.J.~Maxwell and
M.~Neubert for numerous helpful discussions about the theoretical
peculiarities of the subject.

\baselineskip=11pt
\clearpage 


\end{document}